%% file: RI_network.tex
\documentclass[12pt]{article}
\usepackage{amssymb, amsmath, amsthm}
\usepackage{lscape} 
\usepackage{mathdots}
\usepackage{thmtools}
\usepackage{mathtools}
\declaretheorem[style=definition]{example}
\usepackage[makeroom]{cancel}
\usepackage{appendix}
\usepackage{bm}
\usepackage{bbm}
\usepackage{caption}
\usepackage{subcaption}
\usepackage{enumerate} 
\usepackage{graphicx}
\usepackage{natbib}
\usepackage{setspace}
\usepackage{hyperref} 
\usepackage{color}
\usepackage{tikz}
\usepackage{mathrsfs}
\usetikzlibrary{arrows.meta,bending,calc}
\newenvironment{proofsketch}{%
  \begin{proof}%
}{%
  \end{proof}%
}

\usepackage[top=.9in, bottom=.9in, left=.9in, right=.9in]{geometry}

\DeclareMathOperator*{\supp}{supp}

\newtheorem{lem}{Lemma}
\newtheorem*{lem2}{Lemma 2}
\newtheorem{prop}{Proposition}
\newtheorem{thm}{Theorem}

\newtheorem{cor}{Corollary}

\newtheorem{defn}{Definition}
\newtheorem{rem}{Remark}
\newtheorem{obs}{Observation}

\newtheorem{lesson}{Lesson}

\hypersetup{
    colorlinks,
    linkcolor={red!50!black},
    citecolor={blue!50!black},
    urlcolor={blue!80!black}
}

\input{macros.tex}

\begin{document}

\renewcommand{\thefootnote}{\fnsymbol{footnote}}
\renewcommand\thmcontinues[1]{Continued}
\renewcommand{\thefootnote}{\arabic{footnote}}
\renewcommand\thmcontinues[1]{Continued}
\title{Rationally Inattentive Echo Chambers}
\author{
Lin Hu\footnote{Research School of Finance, Actuarial Studies and Statistics, Australian National University, lin.hu@anu.edu.au. }
\and Anqi Li\footnote{Department of Economics, University of Waterloo, angellianqi@gmail.com.} 
\and Xu Tan\footnote{Department of Economics, University of Washington, tauxu@uw.edu. } 
}

\date{}
\maketitle

 \thispagestyle{empty}
\begin{abstract}
Rationally inattentive players allocate limited attention capacities to biased primary sources and to other players as secondary sources to acquire information about an uncertain state. The resulting Poisson attention network stochastically transmits information from primary sources to a recipient either directly or indirectly through other players. We give conditions for the rise of echo-chamber equilibria in which players restrict attention to their own-biased primary source and same-type peers. We characterize the peer attention networks within echo chambers and analyze their comparative statics. Our results explain why modern information environments foster echo-chamber formation, how small differences in attention capacities can magnify into large disparities in allocations, and why regulatory interventions such as altering user visibility on social media or mandating exposure to opposing views may entail unintended consequences.

\bigskip


\bigskip


\end{abstract} 
\vfill

\renewcommand{\thefootnote}{\arabic{footnote}}
\pagebreak

\clearpage
\pagenumbering{arabic}

\section{Introduction}\label{sec_intro}

The Cambridge English Dictionary defines echo chambers as ``environments in which people encounter mostly beliefs or opinions that coincide with their own, so that their existing views are reinforced and alternative ideas are seldom considered.'' Such environments have recently proliferated on the Internet and social media, sparking intense debate about their social impacts \citep{bakshyetal, barbera, cossardetal}. Most ongoing discussions about echo chambers focus on their behavioral origins. This paper provides an account of echo chambers based on rational inattention (RI) alone, that is, the rational and flexible allocation of limited attention capacity across information sources.

The concept of RI has become increasingly relevant in today’s digital age where information is abundant while attention remains scarce. Personalization technologies allow individuals to selectively follow information sources of interest, including both traditional outlets and newer secondary sources  on social media. Since \cite{sunstein} and \cite{filterbubble}, it has been postulated---and recently  formalized by \cite{che} (CM19) in single-agent decision problems---that RI may foster a selective exposure to content. This paper studies a game in which multiple RI players make strategic attention allocations between biased information sources and among one another. We examine how RI may foster echo chambers among like-minded people, analyze equilibrium comparative statics, and use our results to inform policy debates and reforms.

We analyze a model of decision under uncertainty where the true state of the world is $+1$ or $-1$ with equal probability. Each of finitely many players makes a decision of $+1$ or $-1$, which, if differs from the true state, yields a loss. The loss is smaller under the default, defined as the player's preferred decision under the common prior, also called his type.

There are two primary sources, each associated with one state. The $\omega$-revealing source broadcasts the message “$\omega$” in state $\omega$ and is silent in state $-\omega$. Each player can acquire information about the state by allocating a fixed attention capacity across the primary sources and other players as potential secondary sources. Given a joint attention allocation, 
messages about the true state are transmitted first from the primary source to players and then from informed players to others.  The success of each transmission is governed by an independent Poisson process, with rate and time given by the source’s exogenous visibility and the recipient's endogenous attention to the source, respectively. 
After that, players update beliefs and make terminal decisions. We analyze the perfect Bayesian equilibria of this game.

Our game captures many real-world situations. For instance, our players may represent busy parents choosing between traditional and modern parenting approaches. Primary sources are scientific experiments aimed to test different null hypotheses, which parents can access directly through pediatric journals and  indirectly through online groups. The $+1$-revealing source tests the null hypothesis that approach $-1$ is better. A message from this source rejects the null, while silence leads parents to update in favor of $-1$. Another example is voters voting instrumentally between Democratic and Republican candidates of uncertain qualities. Traditional news outlets produce original reporting on candidate quality, and voters follow these outlets and other voters on social media.  The $+1$-revealing source represents a liberal outlet broadcasting a newsworthy message that endorses the Republican candidate when the latter is truly of higher quality. When the Democratic candidate is better, the source remains silent, or equivalently broadcasts an ideological message.

In our model, silence by the $\omega$-revealing source leads to belief update in favor of decision $d=-\omega$. Following CM19, we refer to the $\omega$-revealing source as biased toward decision $d=-\omega$ (or simply $d$-biased), and define a player's own-biased source as one biased toward his default. An echo-chamber equilibrium (ECE) is one in which players restrict attention to their own-biased sources and other same-type players.\label{page:ece_defn}  After playing an ECE, a majority of players receive no message and update toward their defaults, while a minority receive conclusive evidence against their defaults and reverse toward the alternative.

In single-agent decision problems with limited attention and Poisson learning, attending to one's own-biased source and choosing the default following no message is optimal  (CM19).\label{page:CM19} This is because under Poisson learning, even paying full attention to a source may still fail to yield a message. Ex ante, a player foresees in the event of receiving no message, his best decision may still be wrong.  Fixing an intended decision $d$ in that event, only $d$-rejecting messages improve decision-making relative to choosing $d$ unconditionally. Attention scarcity then implies seeking only $d$-rejecting messages from the $d$-biased primary source.  Moreover, setting $d$ to the default is optimal under a symmetric prior. This is because, while attending to the two primary sources leads to accurate decisions with the same probability, choosing the default incurs a smaller loss if wrong.

In our multi-agent environment, it remains true that players seek only information that rejects  their intended decisions given no message.\label{page:intro_multi} In equilibrium,  players with the same intended decision attend to the same primary source and one another, thereby forming closed chambers. However, it is not immediate that the intended decision must equal the default. For instance, when many opponents of a player attend to his opposite-biased source, the player may want to join them so that with  high probability, he hears from at least one source and learns the true state. Absent a message, the player updates in favor of and chooses the alternative. While this decision incurs a significant loss if wrong, the expected loss may nevertheless be smaller than if the player had planned to choose the default given no message and gathered information accordingly. The equilibrium implication of this best-response argument can be nuanced. Depending on model parameters, ECE may fail to emerge, or other equilibrium forms may arise.

 Theorem \ref{thm_ece} prescribes conditions that foster echo-chamber formation in a symmetric society, where both types have equal population and all players share identical characteristics.\label{page:intro_ece}  We show that the ECE arises as the unique equilibrium when players' preference for default is strong,  attention is scarce, or  population is large. These conditions are consistent with several ongoing trends, such as increasingly polarized  preferences across political and science domains \citep{pew2017, gentzkow2019measuring}; attention crowded out by abundant entertainment and eye-grabbing clickbait \citep{priorbook,pewscience}; and tech-enabled large-scale virtual connections among individuals who rarely meet in person. 

Theorem \ref{thm_ece} can be understood via bounding the value of attending to peers; the latter refer to other members of the same chamber.  The large-population limit is the most illuminating.\label{intro_large_population} In a large chamber of size $n$, each member pays barely enough attention to the primary sourc required for disseminating information to others and divides the remaining attention evenly among $n-1$ peers. As information transmits from the primary source to a recipient via a peer, both the peer's advantage in disseminating information relative to the primary source and the recipient's capacity of receiving information are $O(1/n)$. Compared to a direct transmission from the primary source, an indirect transmission via a peer increases the success rate by  $O(1/n^2)$.  Multiplying by the number of peers, the total enhancement still vanishes at rate $O(1/n)$,  too small compared to the preference bias for the default. Hence no player would join a large chamber against his default. Moreover, we show that chambers must be large, which leaves the ECE as the only asymptotic possibility.

Theorem \ref{thm_tau} analyzes equilibrium comparative statics. \label{page:intro_comparative} Part A shows that starting from a symmetric society and ECE, slightly raising a player's bandwidth leads him to pay more attention to the primary source and attract more attention from each peer. By contrast, each of his peers pays less attention to the primary source and attracts less peer attention. This result has empirical support. For instance, political scientists document that nowadays, while most people have little time for serious news consumption, a small group of ``news junkies'' is still willing to devote substantial time to consuming hard news \citep{priorbook,pewscience}.  Theorem \ref{thm_tau} shows how such difference in attention capacity can trigger a feedback loop that leads to ``the law of the few''\footnote{The law of the few refers to a phenomenon that the vast majority relies on a small number of key individuals for information dissemination. \citep{katzlazarsfeld}.  }  and fat-tailed opinion distributions, increasingly observed on social media \citep{luetal, facebookshare}.

Part B of Theorem \ref{thm_tau} 
shows that starting from the same initial conditions, slightly raising a player's visibility has, in general, ambiguous effects on equilibrium attention.\label{page:intro_lambda} Among several possibilities is a  counterintuitive one in which the player with enhanced visibility cuts back primary-source attention and attracts less peer attention. This is because, while greater visibility lowers the threshold one must clear to disseminate information to others, it also lowers the peers' responsiveness to  his  primary-source attention.  Which of these effects dominates determines  equilibrium comparative static, whereas utilitarian welfare inside the chamber always improves. In response to the spread of misinformation culminating in the 2021 U.S. Capitol attack, reforms aimed at controlling user visibility on social media have been proposed by policymakers and voluntarily undertaken by platforms such as Meta \citep{romm}. We caution that such interventions may have nuanced equilibrium consequences and undermine welfare.

Proposition \ref{prop_population} shows that in a symmetric society, increasing population size leads each player to reduce primary-source attention. Eventually, the total primary-source attention inside a chamber declines,  and each member becomes worse off. This result speaks to recent anti-polarization efforts, including platforms such as \href{https://www.allsides.com/about}{Allsides.com} that mandatorily expose users to viewpoints from all sides. We model such a platform as a ``mega source'' that broadcasts about both states. 
When a mega source is in use, each player still finds default-rejecting messages decision-improving, but can now disseminate both types of messages after exposure to the source. Consequently, equilibrium attention is the same as in an ECE with twice the chamber size, and equilibrium welfare is lower than under biased primary sources.

While it is tempting to attribute Theorem \ref{thm_tau} to the strategic substitutability between players' primary-source attention, this intuition is incomplete. As a player directs more attention to the primary source, he becomes more informed and attracts more attention from each peer. How these peers adjust their own primary-source attention and how these adjustments ripple across the population, are less immediate. Our analysis leverages an novel model feature: each player attracts the same attention from all peers. Consequently, the equilibrium effects of perturbing a player's primary-source attention can be analyzed using spectral results on low-rank perturbation, a method broadly applicable to games with positive or negative externalities.

Additionally, we develop a dual representation of hybrid network games in which players first form a strategic network and then play a game on it. This approach yields insights into equilibria in general asymmetric societies and reveals a deep connection between our equilibrium program and the program of  maximizing utilitarian welfare. We explore these extensions of the baseline model  in Section \ref{sec_extension}.

\section{Model}\label{sec_model}

\subsection{Setup}\label{sec_setup}

A finite set $I$ of players faces a random state $\omega$ that equals $\pm 1$ with equal probability. Each player $i \in I$ has a type $t_i$ (also called his \emph{default}) and makes a decision $d_i$, both in $\{-1,+1\}$. If the decision matches the true state, the player earns zero payoff.  Otherwise, he incurs a loss equal to $\beta_i\in(0,1)$ by choosing the default, and to $1$ if by choosing the alternative.  Let $l_i(d_i,\omega)$ denote player $i$'s loss from choosing $d_i$ in state $\omega$, 
\[
l_i(d_i, \omega)=\begin{cases}
0 &\text{ if } d_i=\omega,\\
\beta_i & \text{ if } d_i \neq \omega \text{ and } d_i=t_i,\\
1 & \text{ if } d_i \neq \omega \text{ and } d_i \neq t_i.
\end{cases}
\]
The assumption $\beta_i\in (0,1)$ implies that given only the prior, player $i$ strictly prefers the default to the alternative, and more so as $\beta_i$ decreases.\footnote{Our model is equivalent to a heterogeneous-prior model in which each player $i$ assigns probability $\frac{1}{1+\beta_i}$ to  $\omega=t_i$ and $\frac{\beta_i}{1+\beta_i}$ to  $\omega \neq t_i$. Payoff is zero if $d_i =\omega$ and $-1$ otherwise. } Let $I_+$ and $I_-$ denote the sets of players of positive and negative types, respectively, with $|I_+|$, $|I_-| \geq 2$.

There are two primary sources, each associated with one state. The $\omega$-revealing source broadcasts the  message ``$\omega$'' in state $\omega$ and is silent in state $-\omega$. Therefore, the message ``$\omega$'' fully reveals that the state is $\omega$. To gather information about the state, each player can attend to the primary sources and other players, who may serve as secondary sources. Let 
$C_i\coloneqq \{\omega\text{-revealing}: \omega =\pm 1\} \cup I \setminus \{i\}$ denote the set of sources player $i$ can attend to, and let $a_i^c$ denote his attention allocation to source $c \in C_i$. The attention allocation vector $\bm{a}_i\coloneqq (a_i^c)_{c \in C_i}$ must belong to the simplex $A_i\coloneqq \left\{\bm a_i \in \mathbb{R}_+^{|C_i|}: \sum_{c \in C_i} a_i^c \leq \tau_i\right\}$, meaning that attention to each source is nonnegative and total attention does not exceed the player's \emph{bandwidth} $\tau_i>0$. We refer to the latter restriction as the \emph{bandwidth constraint}.

The game proceeds as follows. \label{page:sequence}
\begin{enumerate}
    \item Players simultaneously and independently choose attention allocations, possibly using mixed strategies in $\prod_{i\in I}\Delta(A_i)$. Independently, the true state $\omega$ is realized. 
    \item Given $\omega$ and a realized profile $\bm{a}\coloneqq(\bm{a}_i)_{i\in I}$, the message ``$\omega$'' is transmitted in society for two rounds:
    \begin{itemize}
        \item First, the $\omega$-revealing source broadcasts the message, and each player $i$ receives it independently with probability $1-\exp(-a_i^{\omega\text{-revealing}})$. 
        \item Next, informed players disseminate the message to others. Fixing an informed player $i$, each player $j \in I \setminus \{i\}$ receives the message from him independently with probability $1-\exp(-\lambda_i a_j^i)$. 
    \end{itemize}
\item Players update beliefs and make terminal decisions. 
\end{enumerate}
 
The parameter $\lambda_c>0$ represents source $c$'s \emph{visibility}, treated as exogenous and normalized to one for the primary sources, i.e., $\lambda_c=1$ for $c \in \{\omega\text{-revealing}: \omega  =\pm 1\}$. Between a source $c$ and a recipient $i$, the probability of a successful information transmission, $1-\exp(-\lambda_c a_i^c)$, increases with the former's visibility and the latter's attention.  This probability is strictly less than one, capturing the reality that attention is scarce relative to information.

The solution concept is perfect Bayesian equilibrium (PBE), or equilibrium for short. If each player $i$'s strategy assigns probability one to a single attention allocation in $A_i$, then the equilibrium is a pure strategy perfect Bayesian equilibrium (PSPBE). 

\subsection{Illustrative examples}

\begin{example}\label{exm_parenting}
In the parenting example described in Section \ref{sec_intro},  \( -1 \) and $+1$ represent traditional spoon-feeding and baby-led weaning, respectively. The state \( \omega \) indicates which approach is best for infant development. While all parents prefer to adopt the best approach if fully informed, they differ in their preferences under uncertainty. Information is produced by scientific experiments aimed to test different approaches. Parents can access them directly through scientific journals or indirectly through other parents. $\tau_i$ captures parent $i$'s  information-gathering capacities, while $\lambda_i$ captures their visibility on online parent groups. Both parameters may vary across parents, depending on availability, education levels, communication skills, and willingness to help.
\end{example}

\begin{example}\label{exm_pe}
The state $\omega$ indicates whether the Democratic candidate $-1$ or the Republican candidate $+1$ is of higher quality. Each player is a partisan voter voting instrumentally between the two candidates \citep{pratstromberg}.  Original reporting on candidate quality is produced by a liberal outlet and a conservative outlet. In state $+1$, the liberal outlet broadcasts a newsworthy message ``$\omega=+1$'' endorsing the Republican candidate \citep{chiangknight}, whereas the conservative outlet broadcasts an ideological message equivalent to silence. Voters can follow the primary sources directly and also one another on social media. Their bandwidths $\tau_i$ reflect the opportunity costs of consuming political news, such as the abundance of entertainment opportunities \citep{priorbook}, while visibility $\lambda_i$ may depend on public prominence or ability to decipher and communicate complex information to the audience \citep{oprah}.
\end{example}

\section{Preliminaries}\label{sec_preliminary} 

\subsection{Individual player's problem}

This section formalizes the problems faced by individual players.\label{page:problem_start}

Fix an attention profile $\bm{a}$. In state $\omega$, player $i$ fails to receive the message ``$\omega$'' if all the information transmission paths from the $\omega$-revealing source to him are disrupted. These include the direct transmission path $\omega\text{-revealing}\longrightarrow i$, which is disrupted with probability 
\[
\delta_i^{\omega\text{-revealing}}\coloneqq \exp\left(-a_i^{\omega\text{-revealing}}\right),
\]
and each indirect transmission path $j\longrightarrow i$, $j \in I \setminus \{i\}$, which is disrupted with probability 
\[
\delta_i^j\coloneqq \exp\left(-\lambda_j a_i^j\right).\]
The joint \emph{disruption probability} equals 
\[D_{i}^{\omega\text{-revealing}}(\bm{a})\coloneqq 
 \delta_i^{\omega\text{-revealing}}\prod_{j \in I \setminus \{i\}} \left[\delta_j^{\omega\text{-revealing}}+(1-\delta_j^{\omega\text{-revealing}})\delta_i^j \right].
\]
When we describe this probability in the sequel, we do so without reference to the primary source whenever the context is clear.  

Fix the (possibly mixed) attention strategies $\sigma_{-i} \in \prod_{j \neq i}\Delta(A_j)$ of player $i$'s opponents. In Stage 3 of the game, player $i$ earns zero payoff if he receives a message about $\omega$ and sets $d_i=\omega$. Upon receiving no message, he updates his belief and chooses $d_i$ to minimize the expected loss. 
For a realized attention allocation $\bm{a}_i \in \supp(\sigma_i)$, the minimized expected loss equals 
\begin{align*}
L(\bm{a}_i, \sigma_{-i})\coloneqq \frac{1}{2}
\min \left\{\beta_i \mathbb{E}_{\bm{a}_{-i}\sim \sigma_{-i}} D_i^{-t_i\text{-revealing}}(\bm{a}_i, \bm{a}_{-i}),\; \mathbb{E}_{\bm{a}_{-i}\sim \sigma_{-i}} D_i^{t_i\text{-revealing}}(\bm{a}_i, \bm{a}_{-i})\right\},
\end{align*}
where the factor $1/2$ represents the prior, the first term inside the minimization operator is the expected loss from choosing the default $t_i$ when the true state is $-t_i$, and the second is the expected loss from choosing the alternative $-t_i$ when the true state is $t_i$. Expectations are taken over the opponents' (possibly random) attention allocations. In Stage 1 of the game, player $i$ specifies an attention strategy $\sigma_i \in \Delta(A_i)$ such that each $\bm{a}_i \in \supp(\sigma_i)$ solves\footnote{The analysis is unchanged if players instead move sequentially, first allocating attention across the primary sources and, after receiving no message, allocating the remaining attention across secondary sources. \label{page:problem_end}
}
\[
\min_{\bm{a}_i \in A_i} L(\bm{a}_i, \sigma_{-i}).\]

\subsection{Concepts and notation}
 This section introduces the key concepts for the main analysis. 

We first define primary-source bias. Following CM19, we interpret the $\omega$-revealing source as an experiment designed to reject that the state is $-\omega$. In the absence of the message ``$\omega$'', players update in favor of state $-\omega$ and are more inclined to make decision $d=-\omega$ than under the prior. For this reason,

\begin{defn}\label{defn_bias}
    Call the $\omega$-revealing source \emph{biased toward decision $d=-\omega$}, or simply \emph{$(-\omega)$-biased.} A player's \emph{own-biased} primary source is one biased toward his default. 
\end{defn}

We next introduce the concept of echo-chamber equilibrium:

\begin{defn}\label{defn_ec}
    An equilibrium is an \emph{echo-chamber equilibrium (ECE)} if each player restricts attention to his own-biased source and other same-type players on the equilibrium path.  
\end{defn}

An ECE has two salient features. The first is a selective exposure to one's own-biased source and other same-type players. Second, after playing an ECE, a majority of players receive no message and update in favor of their defaults. Occasionally, however, a message from the own-biased source arrives and decisively rejects the default, triggering a drastic reversal toward the alternative. This combination of belief polarization and reversal is a hallmark of ECE; suggestive evidence is discussed in Section~\ref{sec_empirical}.

We focus on symmetric societies as a benchmark: 

\begin{defn}\label{defn_symmetry}
A society is \emph{symmetric} if both types have equal population, i.e., \( |I_+| = |I_-|=N \), and all players share identical characteristics, i.e., \( (\beta_i, \tau_i, \lambda_i)\equiv (\beta, \tau, \lambda) \) across \( i \in I \).
\end{defn}

Throughout, we use $f^+$ to denote the positive part of a real-valued function $f$, i.e., $f^+(x)=\max\{f(x),0\}$, and use Landau notation for asymptotics. Additional notation is introduced when first used.

\section{Results}\label{sec_symmetric}

\subsection{Preliminary lemmas}
This section presents two introductory lemmas that facilitate later analysis.

The first lemma clarifies the overlap and distinction between our multi-agent framework and the single-agent decision-making framework studied in CM19. 

\begin{lem}
\label{lem_partition}
In every PSPBE, society is partitioned into at most two \emph{chambers} $S_d$, $d \in \{-1,+1\}$. Each member of $S_d$ restricts attention to the $d$-biased primary source and other members of $S_d$ (henceforth referred to as  \emph{peers}), and chooses  $d$ upon no receiving no message.  
\end{lem}

Lemma \ref{lem_partition} builds on the single-agent, single-period example studied in CM19.\label{page:single-agent} A well-known feature of Poisson learning is that it captures attention scarcity: even paying full attention to a source may fail to yield a message. Consider the single-agent benchmark. Ex ante, the agent foresees that in the event of receiving no message,  even his best decision may still be wrong.  For a fixed intended decision $d$ in that event, only $d$-rejecting messages improve decision-making relative to choosing $d$ unconditionally. Attention scarcity then implies paying full attention to the $d$-biased source is optimal. Moreover, setting $d$ to the default is optimal under a symmetric prior. This is because while attending to the two primary sources leads to accurate decisions with the same probability, choosing the default incurs a small loss if wrong.

In our multi-agent environment, it remains true that players seek only information that rejects their intended decisions given no message.\label{page:multi-agent} In equilibrium, players with the same intended decision restrict attention to the same primary source and to one another, thereby forming closed chambers. However, it is not immediate that the intended decision must equal the default. For instance, when many opponents attend to his opposite-biased source, player \(i\) may follow suit. With high probability, he then hears from at least one source and learns the true state. Absent a message, he updates toward and chooses the alternative, incurring a significant loss if wrong. Nevertheless, the  expected loss may still be smaller than if he had planned to choose the default given no message and gathered information accordingly. This trade-off between information and preference is intrinsic to players' best responses, and its equilibrium consequences can be nuanced. In Section \ref{sec_proofsketch}, we show that depending on model parameters, ECE may fail to arise, or alternative equilibrium forms may arise. 

The second lemma provides preliminary characterizations of the peer attention network inside a chamber. To begin, we define two useful functions and describe their properties: 

\begin{defn}
\label{df-two functions}
For $\lambda>1$ and $x>0$, define
\[
\phi(\lambda)\coloneqq \log\left(\frac{\lambda}{\lambda-1}\right),
\qquad
h(x,\lambda)\coloneqq \frac{1}{\lambda}
\log\left((\lambda-1)(\exp(x)-1)\right).
\]
We then set $\phi(\lambda)=+\infty$ for
$\lambda\in[0,1]$ and $h(x,\lambda)=-\infty$ for $x=0$ or $\lambda \in [0,1]$. 
 \end{defn}

 \begin{obs}\label{obs_function}
\begin{enumerate}[(i)]
\item $\phi'(\lambda)<0$ on $\left(1,+\infty\right)$ and $\lim_{\lambda \downarrow 1} \phi(\lambda)=+\infty$. 
\item For each $\lambda>1$, $h(x,\lambda)$ is increasing and concave in $x$, with $h(\phi(\lambda), \lambda)=0$, $h_x(\phi(\lambda),\lambda)=1$, and $h_{x}\left(x, \lambda\right) \in \left(0,1\right)$ on $(\phi\left(\lambda\right), +\infty)$. 
\item For each $x >0$, $h(x,\lambda)$ is non-monotonic in  $\lambda$ over $(1,+\infty)$. 
\end{enumerate}
\end{obs}

\begin{figure}[htbp]
    \centering
    \begin{minipage}[t]{0.32\textwidth}
        \centering
        \includegraphics[width=\linewidth]{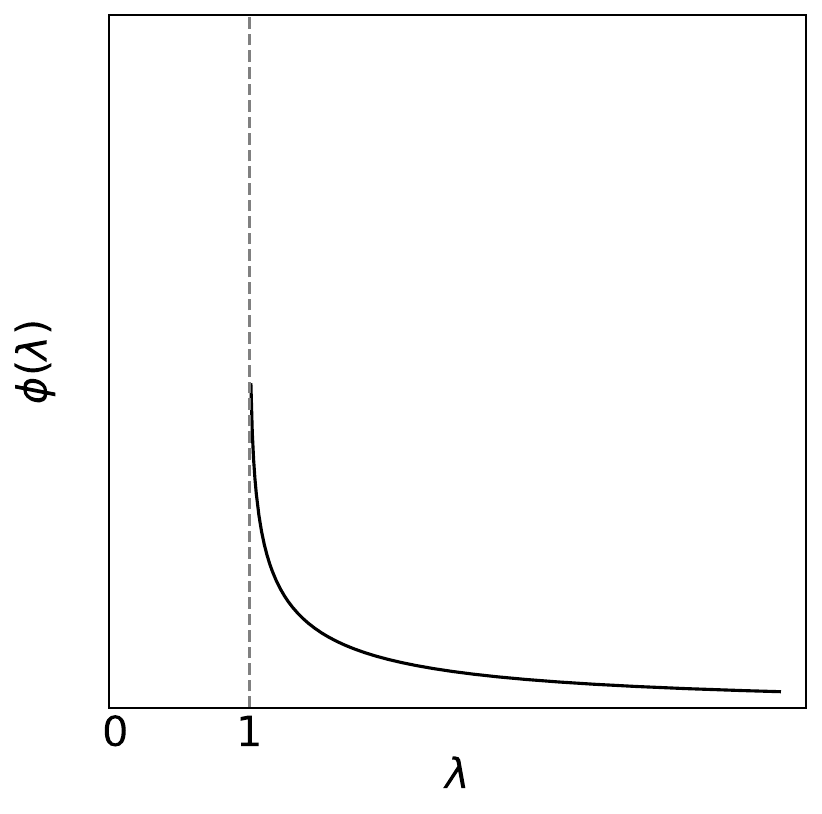}
        \caption*{(a) $\phi(\lambda)$}
        \label{fig:phi}
    \end{minipage}%
    \hfill
    \begin{minipage}[t]{0.32\textwidth}
        \centering
        \includegraphics[width=\linewidth]{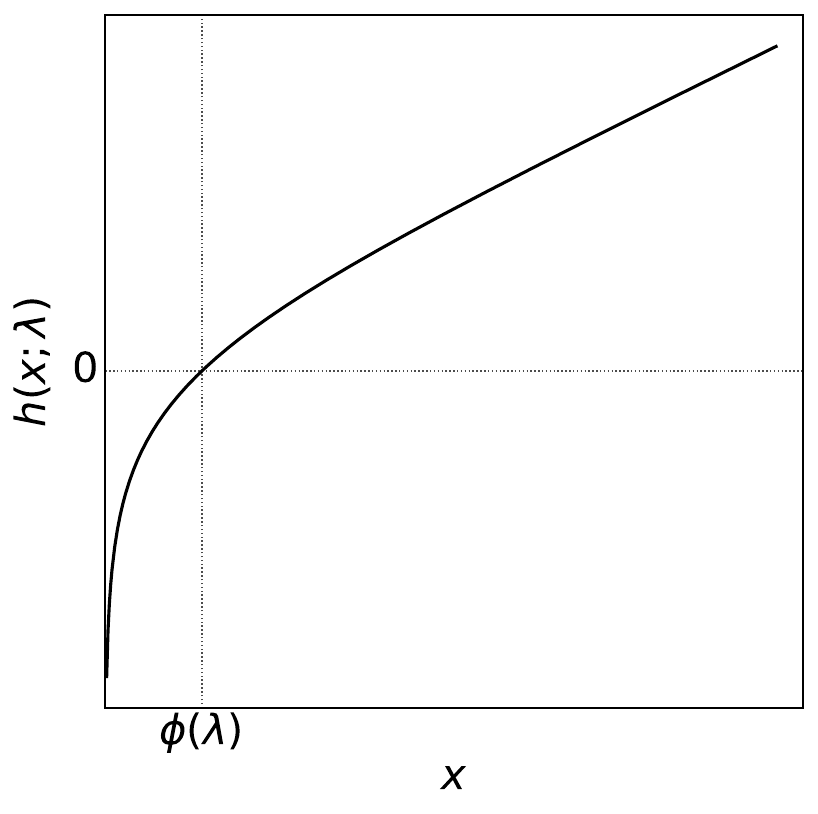}
        \caption*{(b) $h(x, \lambda)$ for fixed $\lambda>1$}
        \label{fig:h}
    \end{minipage}%
    \hfill
    \begin{minipage}[t]{0.32\textwidth}
        \centering
        \includegraphics[width=\linewidth]{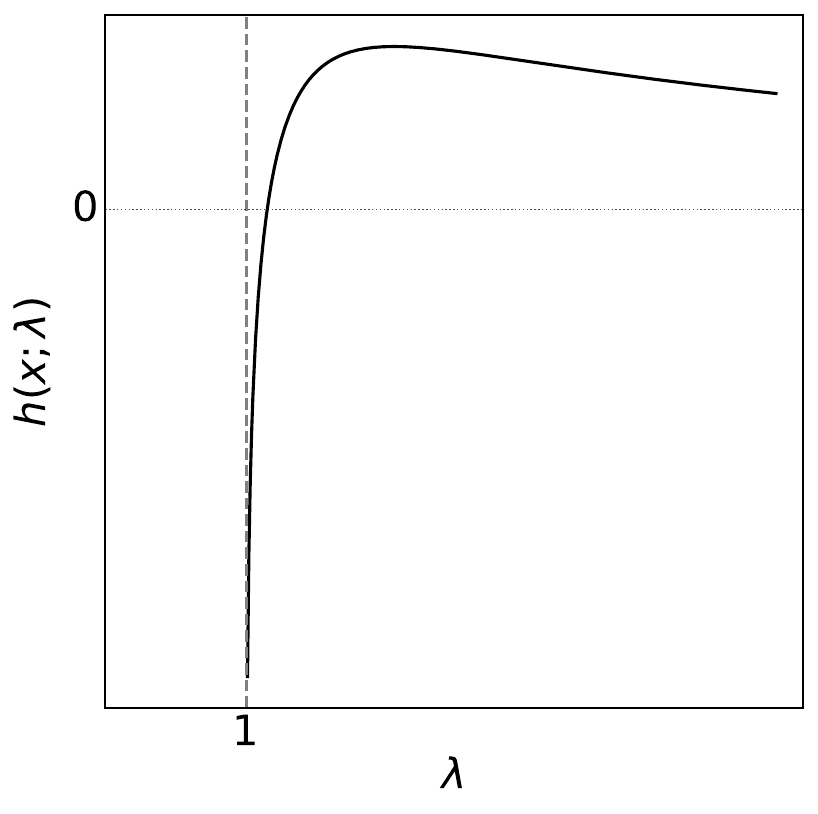}
        \caption*{(c) $h(x; \lambda)$ for fixed $x>0$}
        \label{fig:h_lambda}
    \end{minipage}
    \caption{Plots of $\phi(\lambda)$ and $h(x;\lambda)$}
    \label{fig:phi_h_combined}
\end{figure}

\begin{lem}\label{lem_general}
Consider a PSPBE in which $|S_d|\geq 2$. For each $i\in S_d$, let $x_i$ denote player $i$'s attention  to the $d$-biased primary source. The following hold when $x_i>0$: 
        \begin{enumerate}[(i)]
            \item Player $i$'s attention to a peer $j \in S_d\setminus \{i\}$ equals \[a_i^j= h^+\left(x_j,\lambda_j\right)\coloneqq\begin{cases}
h\left(x_j, \lambda_j\right) &\text{ if } x_j>\phi(\lambda_j),\\
0 &\text{ otherwise}.
\end{cases}\]
\item The log-disruption probability faced by player $i$ equals 
\[-\tau_i-\sum_{j \in S_d \setminus \{i\}}\left[x_j-\phi(\lambda_j)-h(x_j,\lambda_j)\right]\mathbbm{1}_{x_j>\phi(\lambda_j)}.\]
        \end{enumerate}
\end{lem}

Lemma \ref{lem_general}  says that in a chamber of size at least two,  a member $j$ receives the same attention $h^+(x_j,\lambda_j)$ from all peers who pay positive attention to the primary source.  This amount depends only on $j$'s primary-source attention $x_j$---which measures his informedness as a secondary source---and visibility $\lambda_j$.  For $j$ to attract positive peer attention, $x_j$ must exceed a threshold $\phi(\lambda_j)$. For this to happen, two necessary conditions  are (1) $\lambda_j>1$ and (2) $\tau_j>\phi(\lambda_j)$. Intuitively, if $\lambda_j\leq 1$, then $j$ is no more visible than the primary source and so should be ignored by his peers. Correspondingly, $\phi(\lambda_j)$ diverges by  Observation \ref{obs_function}.  As $\lambda_j$ rises above one, $\phi(\lambda_j)$ becomes finite. The condition  $\tau_j > \phi(\lambda_j)$ then ensures that $j$ has enough bandwidth to clear the threshold.\label{page:twofun_end}

After clearing the threshold, $j$ starts to disseminate information to peers, with capability $x_j-\phi(\lambda_j)$ relative to the primary source. A peer, say $i$, directs $h(x_j,\lambda_j)$ units of attention to $j$. Compared to paying direct attention to the primary source, this lowers $i$'s log-disruption probability by $x_j-\phi(\lambda_j)-h(x_j,\lambda_j)$. By Observation \ref{obs_function}, this amount is positive and, as $x_j\downarrow \phi(\lambda_j)$, vanishes at rate\label{page:taylor}
\begin{align*}
&  \ x_j - \phi(\lambda_j) - h(x_j, \lambda_j) \\
= &\ x_j - \phi(\lambda_j) - \bigg[ \underbrace{h(\phi(\lambda_j), \lambda_j)}_{= 0} + \underbrace{h_x(\phi(\lambda_j), \lambda_j)}_{= 1} (x_j - \phi(\lambda_j)) 
  + \frac{1}{2} \underbrace{h_{xx}(\zeta, \lambda_j)}_{<0} (x_j - \phi(\lambda_j))^2 \bigg] \\
 \tag{second-order Taylor expansion with $\zeta\in (\phi(\lambda_j),x_j)$}= & \ O \left((x_j-\phi(\lambda_j))^2\right).
\end{align*}
This observation plays an integral role when we analyze the game's asymptotics in the next section. 

The current version of the lemma focuses on the case where player $i$'s primary-source attention is positive, i.e., $x_i>0$. The case $x_i=0$ is deferred to Appendix~\ref{sec_proof_thm1}. 

Armed with Lemmas \ref{lem_partition} and \ref{lem_general}, we now turn to the main analysis: Sections \ref{sec_ece}--\ref{sec_empirical} state the main results and discuss their intuitions and implications; Sections \ref{sec_proofsketch} and \ref{sec_proofsketch2} sketch the proofs.

\subsection{Echo-chamber formation}\label{sec_ece}
Our first theorem prescribes asymptotic conditions for an ECE to arise as the unique equilibrium in a symmetric society.  To focus on nontrivial environments in which peer attention can arise, we maintain that $\lambda>1$ and $\tau>\phi(\lambda)$. 

\begin{thm}\label{thm_ece}
Starting from a symmetric society with $\beta\in(0,1)$, $\lambda>1$, $\tau>\phi(\lambda)$, and $N\geq2$, an ECE eventually emerges as the unique equilibrium in each of the following limits, holding all else constant:
\begin{enumerate}[(i)]
\item preference for default becomes sufficiently strong, $\beta\downarrow0$;
\item attention becomes sufficiently scarce, $\tau\downarrow\phi(\lambda)$, or equivalently visibility falls to $\lambda\downarrow\phi^{-1}(\tau)$;
\item population becomes sufficiently large, $N\to\infty$.
\end{enumerate}
\end{thm}

The limiting conditions in Theorem~\ref{thm_ece} capture four salient features of modern decision environments. First, there is substantial evidence that preferences are deeply ingrained and polarized in domains such as politics and science \citep{pew2017, gentzkow2019measuring}. Second, the volume of information available through the Internet and social media far exceeds what any individual can process, while constant distractions---from entertainment to attention-grabbing clickbait---leave limited time for serious news consumption \citep{priorbook, pewscience}. 
Third, platform algorithms amplify the visibility of a small number of high-profile users, while leaving the vast majority barely visible \citep{biega2018equity, cakmak2024shorts}. 
Finally, online platforms enable large-scale interactions among individuals who may rarely, if ever, meet in person.
These trends---polarized preferences, scarce attention, limited visibility for most individuals, and large-scale interaction---motivate our focus on the limits $\beta\downarrow0$, $\tau\downarrow\phi(\lambda)$ (equivalently $\lambda\downarrow\phi^{-1}(\tau)$), and $N\to\infty$. Theorem~\ref{thm_ece} shows that an ECE arises as the unique equilibrium in each of these limits.

The $\beta\downarrow0$ limit operates through a simple channel: when preference for default is sufficiently strong, the loss saving from choosing the default over the alternative exceeds any informational gain from joining a chamber against one's default. Attending to the own-biased source becomes a dominant strategy, hence every equilibrium is an ECE. The uniqueness of ECE  follows from its construction in  Section \ref{sec_proofsketch}.

The limit $N\rightarrow \infty$ can be understood via bounding the value of attending to peers.\label{page:largeN} In Section \ref{sec_proofsketch}, we show that members of a chamber allocate attention in a symmetric manner. 

In a large chamber of size $n$, each member pays $h(x(n),\lambda)=O(1/n)$ units of attention to each peer. Consequently, his primary-source attention barely exceeds the threshold,  $x(n)-\phi(\lambda)=O(1/n)$. As information trickles from the primary source to a recipient via a peer, both the peer's advantage in disseminating information relative to the primary source, and the recipient's capacity of receiving information---as captured by $x(n)-\phi(\lambda)$ and $h(x(n),\lambda)$, respectively ---are $O(1/n)$. Compared to a direct transmission from the primary source, an indirect  transmission via a peer increases the success rate only by $O(1/n^2)$---an intuition confirmed by letting $x(n)-\phi(\lambda)=O(1/n)$ in the second-order Taylor expansion on page \pageref{page:taylor}. 
Multiplying by the number $n-1$ of peers, the total enhancement still vanishes at rate $O(1/n)$, too small compared to the preference bias for the default.  Hence no player would join a large chamber against his default. Additionally, we show that chambers must be large (see Section \ref{sec_proofsketch}), which leaves the ECE as the only asymptotic possibility. 

The limit $\tau\downarrow \phi(\lambda)$ is analogous: denote $\Delta=\tau-\phi(\lambda)$ and replace $1/n$ with $\Delta$ in the preceding argument.

\subsection{Comparative statics}\label{sec_cs}

This section examines equilibrium comparative statics. 

Our second theorem begins with a symmetric society and perturbs a player's bandwidth or  visibility.\label{page:thm_tau_motivation} This exercise is motivated by a few outliers with large bandwidths or high visibility, who---despite their small population---nevertheless occupy central positions in the modern information landscape. For instance, political scientists are interested in a small group of individuals known as  ``news junkies,'' who---unlike their contemporaries---are willing to devote substantial time to consuming hard news \citep{priorbook, pewscience}. Recently, an important subject of system design concerns how to concentrate traffic and visibility on a few prominent high-profile users in order to maximize platform profits \citep{mousavi2024tiktokexplanations, cakmak2024shorts}. We examine how the  presence of such outliers affects equilibrium attention and welfare.

\begin{thm}\label{thm_tau} 
Start from a symmetric society under one of the limiting conditions in Theorem \ref{thm_ece}, i.e., strong preference, scarce attention, or large population. After each perturbation stated below, the game has a unique equilibrium, which remains an ECE. 

\medskip  
\noindent\emph{(Part A)}
The following happen after slightly increasing $\tau_i$ for some $i\in I_+$: 
\begin{enumerate}[(i)]
\item Player $i$ pays strictly more attention to the primary source and attracts strictly more attention from each peer. 
\item Each player $j \in I_+\setminus \{i\}$ pays strictly less attention to the primary source and attracts strictly less attention from each peer (including $i$). 
\item Each member of $I_+$ faces a lower expected loss. 
\end{enumerate}

\medskip 
\noindent\emph{(Part B)} Slightly increasing $\lambda_i$ for some $i\in I_+$ gives rise to one of the following situations: 
\begin{enumerate}[(i)]
    \item Player $i$ pays strictly more attention to the primary source and attracts strictly more attention from each peer. The opposite is true for each player $j\in I_+\setminus \{i\}$.
    \item Player $i$ pays strictly less attention to the primary source and attracts strictly less attention from each peer. The opposite is true for each player $j\in I_+\setminus \{i\}$. 
    \item Equilibrium attention allocations remain unchanged.
\end{enumerate}
In all three cases, utilitarian welfare inside $I_+$ (measured by negative expected total loss) strictly increases.\label{page:thm_tau_end}
\end{thm}

Part A of Theorem \ref{thm_tau} establishes a reinforcement mechanism that turns the bandwidth  inequality between a few outliers and the vast majority into uneven attention choices. In particular, \emph{all} the other chamber members increase attention to $i$ while reducing attention to the primary source and one another after the perturbation. Patterns consistent with this result include the vast majority relying on a few news junkies for second-hand news (also known as ``the law of the few'') and fat-tailed opinion distributions increasingly observed on social media \citep{luetal, facebookshare}. Such patterns may indicate high welfare in the Pareto sense. As shown in Section \ref{sec_proofsketch2},  the increase in $i$'s primary-source attention is large enough to offset the declines in his peers' primary-source attention combined. As more information flows into the chamber, each member becomes better off.

\begin{figure}[h!]
\centering
\begin{subfigure}{.4\textwidth}
	\captionsetup{width=.9\linewidth}
\begin{center}
 		\includegraphics[width=\linewidth]{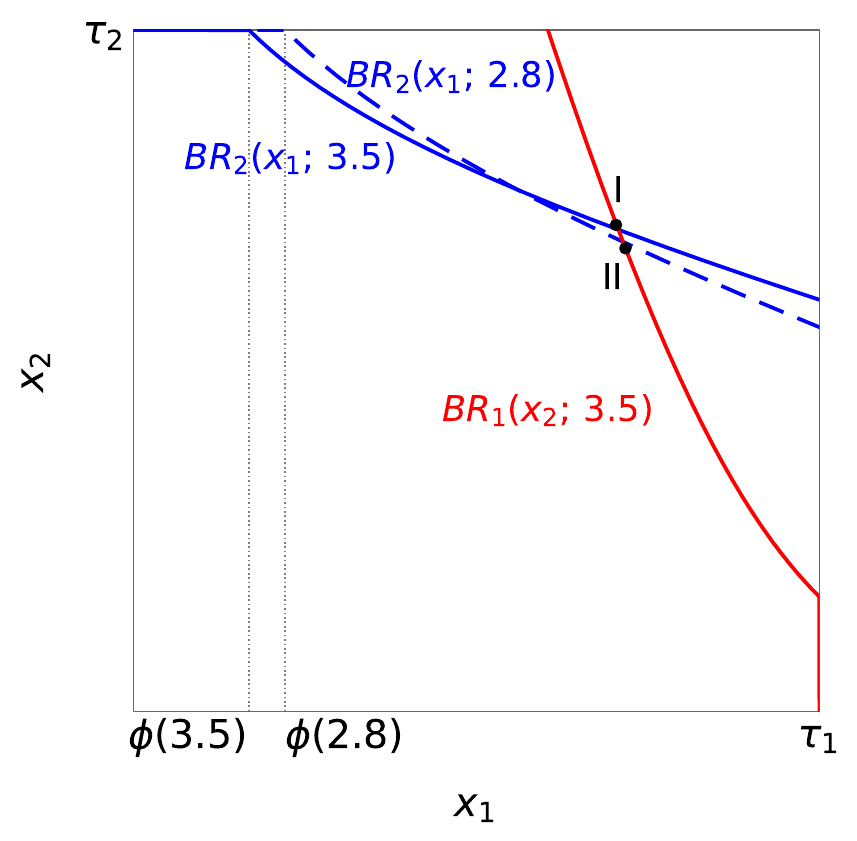}
  \caption{Decrease $\lambda_2$ from $3.5$ to $2.8$.}\label{figure2a}
\end{center}
\end{subfigure}%
\hspace{0.1\textwidth}
\begin{subfigure}{.4\textwidth}
		\captionsetup{width=.9\linewidth}
\begin{center}
 		\includegraphics[width=\linewidth]{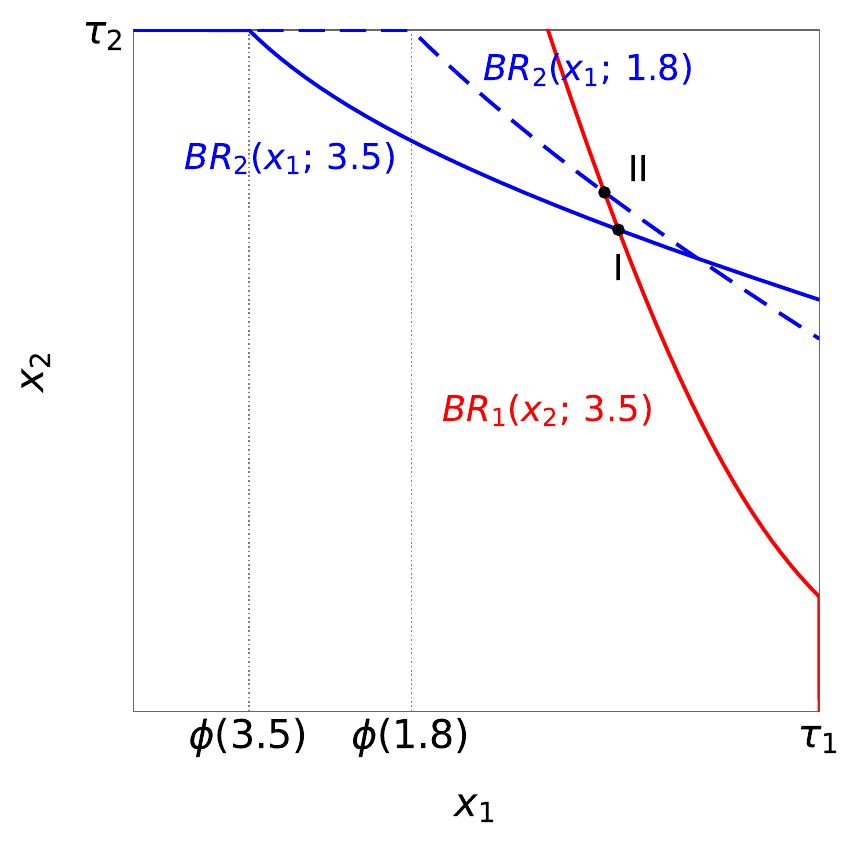}
  \caption{Decrease $\lambda_2$ from $3.5$ to $1.8$.}\label{figure2b}
\end{center}
\end{subfigure}
\caption{A two-player example. The solid and dashed blue curves depict player $2$'s best-response primary-source attention as a function of player $1$'s primary-source attention, for different values of $\lambda_2$.}
\label{figure2}
\end{figure}

Two countervailing forces underlie the nuanced results in Part B. \label{page:lambda} On the one hand, enhanced visibility lowers the threshold required for disseminating information to peers. However, it also makes the peers less responsive to changes in one's primary-source attention.  These two effects, corresponding to the intercept and slope of players' best response functions, work in opposite directions, as illustrated in Figure \ref{figure2}. 
Their net effect on equilibrium attention is succinctly summarized by the sign of $h_{\lambda}(x_i,\lambda_i)$, which, by Observation \ref{obs_function}, is generally indeterminate. Despite this ambiguity, utilitarian welfare inside the chamber always improves.\footnote{We cannot rank equilibrium welfare before and after the perturbation in the Pareto sense. For instance, in Case (i), expected loss increases for player $i$ but decreases for every other chamber member.  }

These results have policy implications. In response to the spread of misinformation and fake news  culminating in the 2021 U.S. Capitol attack, reforms aimed to control user visibility on social media have been proposed by policymakers and voluntarily undertaken by platforms \citep{romm}.  For instance, Meta caps individual users' daily posts to 25 articles, beyond which their visibility is substantially reduced. 
We caution that such interventions may entail nuanced equilibrium consequences and undermine welfare.

The next proposition increases the population size of a symmetric society and examines the comparative static of the ECE:   

\begin{prop}\label{prop_population}
In a symmetric society, increasing the population size $N$, holding all else constant, eventually increases the expected loss faced by each player in the ECE. 
\end{prop}

As chamber size increases, each member gains access to more secondary sources and reduces primary-source attention.  Eventually, the total primary-source attention inside a chamber declines, and each member faces a greater loss.  

Proposition \ref{prop_population} informs recent anti-polarization efforts, such as \href{https://www.allsides.com/about}{AllSides.com}, which mandatorily exposes users to diverse viewpoints across the political spectrum. \label{page:mega} We model such a platform a \emph{mega primary source} that broadcasts about both states. Each player allocates attention among this source and other players, $C_i = \{\text{mega}\}\cup I\setminus \{i\}$, $\bm{a}_i \in \left\{\mathbb{R}_+^{2N}: \sum_{c \in C_i} a_i^{c}\leq \tau\right\}$. In state $\omega$, the message ``$\omega$'' first transmits from the mega source to each player $i$ independently with probability $1-\exp(-a_{i}^{\text{mega}})$, and then from each informed player $i$ to each player $j\in I\setminus \{i\}$ independently with probability $1-\exp(-\lambda a_j^i)$. 

In this augmented model, the same attention network transmits messages for both states. If a player receives no message, his posterior belief remains at the prior, leading him to choose the default. Consequently, only the default-rejecting message remains decision-improving, even though each player is capable of transmitting both message types after exposure to the mega-source. The equilibrium attention allocation is therefore identical to that of the ECE with twice the chamber size. Proposition \ref{prop_population} implies a perverse welfare consequence for both individual players and society as a whole:

\begin{cor}\label{cor_merger}
In a large symmetric society, the use of a mega source increases each player's expected loss and, hence, undermines utilitarian welfare, compared to biased primary sources. 
\end{cor}\label{page:mega_end}

\subsection{Empirical implications}\label{sec_empirical}
After playing an ECE, most people update toward their default, while a minority experiences a sharp reversal toward the alternative. This combination of belief polarization and occasional, drastic belief reversal is a hallmark of Bayesian rationality. In traditional media studies, outcomes  consistent with Bayesian updating have been documented by economists and political scientists even in emotionally contentious settings.\footnote{Bayesian rationality is a recurring theme in two branches of media studies: (1) media persuasion, surveyed by \cite{dellavignagentzkow}, and (2) filtering bias, reviewed in Section \ref{sec_literature} of this paper. A prominent study of media persuasion conducted by  \citet{dellavigna2007fox} exploits the staggered rollout of Fox News between 1996 and 2000. The authors find  that access increased the Republican presidential vote share by 0.4--0.7 percentage points, implying a persuasion rate of $11.6\%$ among non-Republicans. } With the rise of social media, secondary sources have gained prominence over traditional media  \citep{allcott2017social}, yet existing social-media studies focus mainly on belief polarization and sometimes on belief reversal, but rarely on their coexistence.\footnote{A prominent study of the polarization effect of social media is conducted by \cite{allcottetal}. The authors show that deactivating Facebook before the 2018 U.S. midterm elections reduced polarization in users' political beliefs and voting behavior. Evidence for belief reversal comes from \cite{balsamoetal}, who find that exposure to cross-cutting content through one's inner circle on Twitter lowered belief coherence within ideological groups during Brazil's 2016 presidential impeachment debate. Similarly, \cite{barbera2015} finds that although political discussion networks on Twitter are ideologically segregated, most users encounter cross-cutting content through weak ties that moderate their political views. Both studies emphasize the robustness of their findings in the absence of explicit cross-partisan ties.} Our analysis shows that both phenomena can arise simultaneously, thereby suggesting new research avenues in social-media studies.

In addition, our analysis yields novel predictions about rationally inattentive echo chambers, including conditions that foster echo-chamber formation and properties of the  peer attention network within echo chambers. Our comparative-static results further link equilibrium outcomes to primitive objects such as attention capacity, visibility, and population size. Since these objects can be manipulated in controlled environments, one direction for future research is to run a controlled experiment that tracks subjects' attention allocation, belief evolution, and subsequent behavior. This experimental approach of studying social media was introduced to economics by \citet{allcottetal}.

\subsection{Proof sketch for Theorem \ref{thm_ece}}\label{sec_proofsketch}

In this section, we outline the proof of Theorem \ref{thm_ece}, along the way unlocking structural insights into \emph{all} equilibria of the game.  Throughout, we consider a symmetric society with $\lambda>1$ and $\tau>\phi(\lambda)$.  

We begin by characterizing PSPBEs, conditional on their existence. The next lemma establishes an attention symmetry between members of a chamber and quantifies the value of paying peer attention. 

\begin{lem}\label{lem_peersymmetry}
Consider a PSPBE in a symmetric society. In a chamber of size $n\ge2$, each member allocates $x(n)>\phi(\lambda)$ units of attention to the primary source and $h(x(n),\lambda)>0$ units to each peer. The log-disruption probability he faces equals 
\[-\tau-(n-1)\underbrace{\left[x(n)-\phi(\lambda)-h(x(n),\lambda)\right]}_{\coloneqq \xi(n)},\]
where $\xi(n)>0$ represents the reduction in the log-disruption probability from attending to a peer rather than to the primary source. 
\end{lem}

\begin{proofsketch}
When $n=2$, the result follows by inspection of Figure \ref{fig:best_response_diagram}: \begin{figure}[htbp]
\centering
\includegraphics[width=0.4\textwidth]{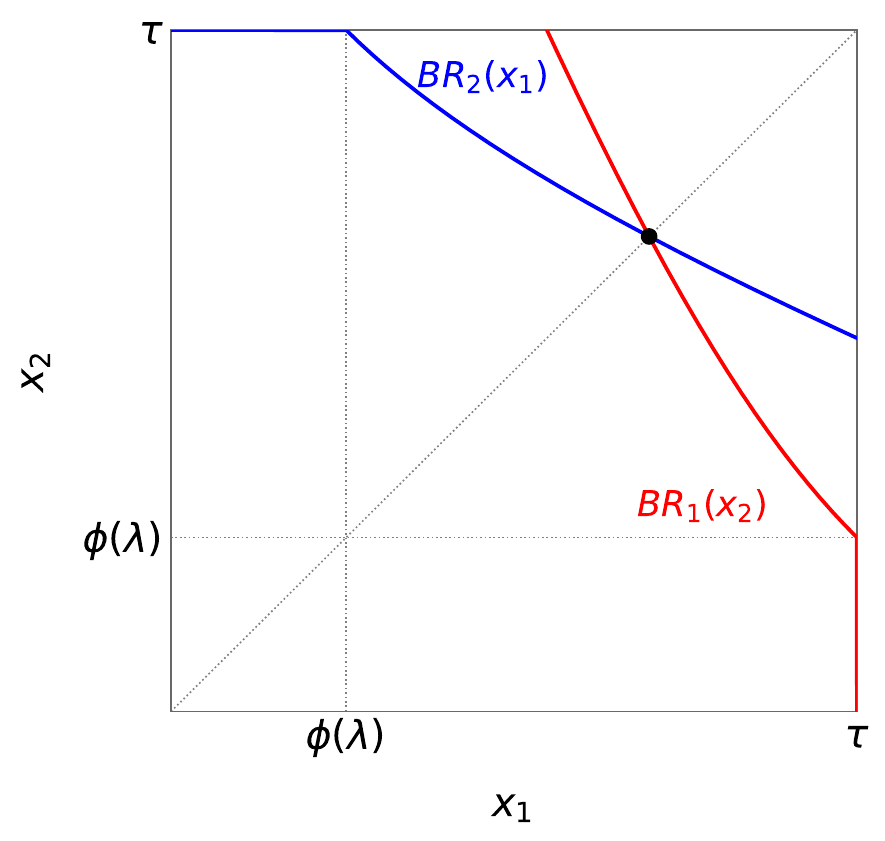}
\caption{The red curve depicts player 1's optimal attention allocation to the primary source as a function of player 2's, and the blue curve vice versa.} 
\label{fig:best_response_diagram}
\end{figure}
Until his peer’s primary-source attention crosses the threshold, a player pays full attention to the primary source. After that, he directs more attention to the peer as the latter's primary-source attention increases, though by less than one-for-one. The two best response functions have a unique intersection $(x(2),x(2))$, where $x(2) \in (\phi(\lambda),\tau)$. Thus in equilibrium, each player directs $x(2)$ units of attention to the primary source and the remaining attention to his peer. In Lemma \ref{lem_general} (ii), letting $x_j=x(2)$ yields $\xi(2)$ as the reduction in $i$'s log-disruption probability from attending to $j$ relative to the primary source. $\xi(2)>0$ holds by Observation \ref{obs_function}. 

The case $n>2$ requires more work and is deferred to Appendix \ref{sec_proof_thm1}.
\end{proofsketch}

The next lemma establishes an attention symmetry between same-type players and uses it to simplify equilibrium forms: 
\begin{lem}\label{lem_typesymmetry}
    In a PSPBE of a symmetric society, same-type players must join the same chamber. Consequently, the equilibrium takes one of the following forms:
  \begin{enumerate}[(i)]
        \item ECE, i.e., $S_+=I_+$ and $S_-=I_-$;
        \item One-sided attention, i.e., either $S_+=I$ and $S_-=\emptyset$, or $S_+=\emptyset$ and $S_-=I$.
    \end{enumerate}
\end{lem}

\begin{proofsketch}
Suppose, toward a contradiction, that two positive-type players $i$ and $j$ join different chambers $S_+$ and $S_-$, respectively. If $i$  were to leave $S_+$ and join $S_-$, he would have access to $|S_-|$ peers that include $j$,  each paying $x(|S_-|)$ units of attention to the primary source.  This is one more peer than $j$ currently faces. After some technical work postponed to Appendix \ref{sec_proof_thm1}, it can be shown that $i$'s log-disruption probability after deviation is $\xi(|S_-|)$ units lower than $j$'s current log-disruption probability.

Likewise, if $j$ were to leave $S_-$ and join $S_+$, then his log-disruption probability after  deviation is $\xi(|S_+|)$ units lower than $i$'s current log-disruption probability.

Now suppose $i$ finds the deviation weakly unprofitable. Writing ``LL'' for log-expected loss and ignoring the common constant $-\tau+\log 1/2$, 
\begin{align*}
\underbrace{\log \beta -|S_+|\xi(|S_+|)}_\text{j's LL after deviation}
<\underbrace{\log\beta -(|S_+|-1)\xi(|S_+|)}_\text{i's current LL}
\leq \underbrace{-|S_-|\xi(|S_-|)}_\text{i's LL after deviation}
<\underbrace{-(|S_-|-1)\xi(|S_-|)}_\text{j's current LL}.
\end{align*}
Thus $j$ strictly benefits from leaving $S_-$ and joining $S_+$, contradicting equilibrium. 
\end{proofsketch}

In general, both echo chambers and one-sided attention can arise in equilibrium, as illustrated in the next table. 
\begin{table}[ht]
\centering
\begin{tabular}{c|c|c|c}
\hline
$N$
& ECE equilibrium
& One-sided-attention equilibrium
& Intermediate region
\\ \hline
$2$
& $\beta\leq 0.855$
& $\beta\geq 0.864$
& \begin{tabular}[c]{@{}c@{}}
  $0.855<\beta<0.864$: \\
  no PSPBE
  \end{tabular}
\\[6pt]
$3$
& $\beta\leq 0.922$
& $\beta\geq 0.893$
& \begin{tabular}[c]{@{}c@{}}
  $0.893\leq\beta\leq0.922$: \\
  both equilibria coexist
  \end{tabular}
\\ \hline
\end{tabular}
\caption{PSPBE when $\lambda=2$ and $\tau=2$.}
\end{table}
However, 
\begin{lem}\label{lem_ece}
In a symmetric society, the ECE emerges as the unique PSPBE under each of the limiting conditions stated in Theorem \ref{thm_ece}. 
\end{lem}

\begin{proofsketch}
  We only sketch the proof for the limit $N\rightarrow \infty$. 
  
  Inside a chamber of a large size $n$, each player directs $h(x(n),\lambda)=O(1/n)$ units of attention to each peer. Consequently, their primary-source attention satisfies $x(n)=\phi(\lambda)+O(1/n)$. Substituting into Lemma \ref{lem_general} (ii), a second-order Taylor expansion as on page \pageref{page:taylor} yields $\xi(n)=O\left((x(n)-\phi(\lambda))^2\right)=O(1/n^2)$. The economic intuition behind this result was already discussed in Section \ref{sec_ece}. 
  
  It follows that, for sufficiently large $N$,
\[\log\beta-(N-1)\xi(N) < -N\xi(N), \quad -(2N-1)\xi(2N)<\log\beta.\]
The first inequality says that starting from the ECE,  nobody benefits from joining the opposite-biased chamber to gain one more peer. The second inequality says that starting from one-sided attention, anyone who currently chooses the alternative following no message strictly prefers to leave the chamber and attend to his own-biased source alone. Thus for sufficiently large $N$, the ECE arises as the unique PSPBE, whereas one-sided attention cannot arise in equilibrium.
\end{proofsketch}

The next lemma rules out mixed-strategy equilibria in the limits, thereby completing the argument.
\begin{lem}\label{lem_mse}
In a symmetric society, mixed strategy equilibria do not exist under each of the limiting conditions stated in Theorem \ref{thm_ece}.
\end{lem}

\subsection{Proof sketch for Theorem \ref{thm_tau}}\label{sec_proofsketch2}
The comparative-static technique developed in this section does \emph{not} rely on environmental or equilibrium symmetry; Remark \ref{rem_cs} clarifies the role of symmetry in Theorem \ref{thm_tau}. Below we sketch the proof of Part A, adopting the following matrix conventions: for $n\in\mathbb{N}$, $\bm{1}_n$ denotes the $n$-dimensional vector of ones, and $I_n$ denotes the $n\times n$ identity matrix. For a vector $\boldsymbol{v}$, $\diag(\boldsymbol{v})$ denotes the diagonal matrix with diagonal entries given by $\boldsymbol{v}$, and $v_i$ denotes the $i$th coordinate.

In a chamber of size two,  Part A follows from the strategic substitutability between players' primary-source attention of degree less than one. As player $1$ raises his primary-source attention by $\epsilon>0$, he becomes more informative as a secondary source and attracts $h_x(x_1,\lambda_1) \epsilon$ more units of attention from player $2$,  where $h_x(x_1,\lambda_1) \in (0,1)$. In turn, player $2$ reduces primary-source attention and attracts less attention from player $1$.  This feedback loop continues until the game settles at the new equilibrium. 

With more than two players, strategic substitutability alone is insufficient. For instance, among three players $i$, $j$ and $k$, it is not immediate how an increase in $i$'s primary-source attention may lead $j$ and $k$ to adjust their primary-source attention. In the case where $j$ significantly cuts back primary-source attention, $k$ may raise primary-source attention, which leads to further adjustments by $i$ and $j$, and so forth. Establishing Part A  requires additional model structures.

A key observation is that when all players are above their thresholds, each of them attracts the same attention from all peers. Starting from an ECE where $x_i>\phi(\lambda_i)$ for all $ i \in I_+$, the following holds: 
\[
x_i+\sum_{j \in I_+ \setminus \{i\}} h(x_j, \lambda_j)=\tau_i \quad \forall i \in I_+. 
\]
Differentiating this system of equations with respect to $\tau_1$ gives, in the matrix form: 
\begin{align*}
\underbrace{
\begin{bmatrix}
1 & g_2 & \cdots & g_N \\
g_1 & 1 & \cdots & g_N \\
\vdots & \vdots & \ddots & \vdots \\
g_1 & g_2 & \cdots & 1
\end{bmatrix}
}_{M}
\underbrace{
\begin{bmatrix}
\partial x_1/\partial \tau_1 \\
\partial x_2/\partial \tau_1 \\
\vdots \\
\partial x_N/\partial \tau_1
\end{bmatrix}
}_{\partial \boldsymbol{x}/\partial \tau_1}
\; = \;
\underbrace{
\begin{bmatrix}
1 \\
0 \\
\vdots \\
0
\end{bmatrix}
}_{\boldsymbol{e}_1}.
\end{align*}
On the left-hand side, the matrix $M$ is the Jacobian of the system at the equilibrium before perturbation. 
The term $g_i \coloneqq h_x(x_i,\lambda_i)$ captures the marginal effect of perturbing $i$'s primary-source attention on the peer attention he attracts. By Observation \ref{obs_function}, we have $g_i \in (0,1)$, where $g_i>0$ reflects strategic substitutability and $g_i<1$ bounds its degree. Moreover, $g_i$ is constant across all peers of $i$. Consequently, we may express $M$ as a rank-one update of a diagonal matrix:
\[
M = \operatorname{diag}(1-g_1,\dots,1-g_N) + \boldsymbol{1}_N \boldsymbol{g}^{\top}, 
\quad  \boldsymbol{g} \coloneqq [g_1,\dots,g_N]^{\top}.
\]
Applying the Sherman-Morrison formula \citep{sherman1950adjustment},\footnote{Let $A$ be a non-singular matrix and $u$ and $v$ be vectors. Whenever $1+v^{\top}A^{-1}u \neq 0$, $A+u v^{\top}$ is invertible with 
\[(A+u v^{\top})^{-1}=A^{-1}-\frac{A^{-1} u v^{\top} A^{-1}}{1+v^{\top} A^{-1} u}.\]
} 
\[
M^{-1}=\operatorname{diag}\left(\frac{1}{1-g_1},\dots, \frac{1}{1-g_N}\right)\left\{I_N-\frac{1}{1+\sum_{i=1}^N \frac{g_i}{1-g_i}} 
\boldsymbol{1}_N \begin{bmatrix}
\dfrac{g_1}{1-g_1}, & \dots, & \dfrac{g_N}{1-g_N}
\end{bmatrix}
\right\}. 
\]
Therefore 
\[
\frac{\partial \boldsymbol{x}}{\partial \tau_1}=M^{-1}\boldsymbol{e}_1=\operatorname{diag}\left(\frac{1}{1-g_1},\dots, \frac{1}{1-g_N}\right)\left(\boldsymbol{e}_1-\frac{\frac{g_1}{1-g_1}}{1+\sum_{i=1}^N \frac{g_i}{1-g_i}} \boldsymbol{1}_N\right),
\]
where the right-hand-side vector is strictly positive along the first coordinate  and strictly negative along the remaining coordinates, i.e., $\partial x_1/\partial \tau_1>0$ and $\partial x_j /\partial \tau_1<0$, $j\neq 1$. Substituting into $h$ gives $\partial h(x_1,\lambda_1)/\partial \tau_1>0$ and $\partial h(x_j,\lambda_j)/\partial \tau_1<0$, $j\neq 1$.  Parts (i) and (ii) of the theorem are established. 

Turning to Part (iii), recall from Lemma~\ref{lem_general}(ii) that, when all players exceed their thresholds, the log-disruption probability for player $i$ equals 
\[-\tau_i-\sum_{j \in I_+\setminus \{i\}}\left[x_j-\phi(\lambda_j)-h(x_j,\lambda_j)\right]=-\sum_{j\in I_+} x_j+\sum_{j \in I_+\setminus \{i\}}\phi(\lambda_j).\]
Therefore, it suffices to sign the derivative of the total primary-source attention inside $I_+$ w.r.t. $\tau_1$:  
\[\sum_{i \in I_+}\frac{\partial x_i}{\partial \tau_1}=\boldsymbol{1}_N^{\top} \frac{\partial \boldsymbol{x}}{\partial \tau_1}.\]
In general, the sign is indeterminate.  However, when the original environment and equilibrium are  symmetric,  $g_i\equiv g$, so  the expression simplifies to
\[\boldsymbol{1}_N^{\top} \frac{\partial \boldsymbol{x}}{\partial \tau_1}=\frac{1}{1+(N-1)g}>0.\]
Thus  the perturbation increases the total primary-source attention inside the chamber and makes each member better-off.  

\begin{rem}\label{rem_tau}
\emph{A related comparative-static exercise considers a uniform increase in all players' bandwidths in a symmetric society. Let $\tau$ denote the common bandwidth. Then
    \begin{align*}
    M \frac{\partial \boldsymbol{x}}{\partial \tau}=\boldsymbol{1}_N \implies &\frac{\partial \boldsymbol{x}}{\partial \tau}=\sum_{i=1}^N M^{-1}\bm{e}_i=\frac{1}{1+(N-1)g}\boldsymbol{1}_N\\
\implies  & \mathbf{1}_N^{\top} \frac{\partial \bm x}{\partial \tau}
=\frac{N}{1+(N-1)g}.
    \end{align*}
Thus, the effect of uniformly increasing \(\tau\) on each player's primary-source attention is the same as that of increasing one member's bandwidth on the chamber's total primary-source attention. The effect on total primary-source attention and, hence, individual welfare, is \(N\) times as large.}

\emph{In Appendix \ref{sec_proof_thm2}, we use the same approach to analyze the comparative static of  perturbing players' visibility. }
\end{rem}

\begin{rem}\label{rem_cs}
   \emph{Our comparative-static analysis only requires that the equilibrium before perturbation is locally unique and strict. The limiting conditions imposed in the theorem ensure that the equilibrium is globally unique and strict before perturbation. As shown in  Appendix \ref{sec_proof_thm2},  these properties together imply that the equilibrium remains unique and an ECE after the perturbation.  }
\end{rem}

\begin{rem}
\emph{The low-rank perturbation structure used in our proof does \emph{not} depend on whether players' primary-source attention are strategic substitutes or complements. Thus our approach may apply broadly to (network) games with strategic externality. Recently, spectral methods for low-rank perturbations have been used in other areas of economics, such as misspecified Bayesian learning \citep{he2023learning}.}
\end{rem}

\section{Extensions}\label{sec_extension}
This section presents the main extensions of the baseline model. Additional robustness checks---including noisy primary sources and environments with multiple states, types, and sources---are conducted in the Online Appendix.

\subsection{Asymmetric society}\label{sec_asymmetric}

This section extends the baseline model to asymmetric societies. 

Society consists of finitely many groups collected in a set $G$. Each group $g\in G$ has $N \in \mathbb{N}$ members who share common profile $(t_g,\beta_g,\tau_g,\lambda_g)$ of characteristics, including default $t_g\in\{-1,+1\}$, preference strength $\beta_g\in(0,1)$, bandwidth $\tau_g>0$, and visibility $\lambda_g>0$. Different groups may share an identical characteristic profile.  Define
$I_+\coloneqq \cup_{g: t_g=1} g$ and $I_-\coloneqq \cup_{g: t_g=-1} g$. To focus on nontrivial environments, suppose that
$
\max_{g\in G} \tau_g-\phi(\lambda_g)>0,
$
and that both $I_+$ and $I_-$ are nonempty.

We first establish the analog of Theorem \ref{thm_ece} in an asymmetric society:  
\begin{prop}\label{prop_ece}
In each of the following limits, eventually a PSPBE exists, and every PSPBE is an ECE:
\begin{enumerate}[(i)]
    \item  all groups have sufficiently strong preferences for defaults, i.e., $\max_{g \in G} \beta_g \downarrow 0$;
    \item all groups' attention becomes sufficiently scarce, i.e.,  $\max_{g \in G}\tau_g-\phi(\lambda_g)\downarrow 0$; 
    \item  population becomes sufficiently large, i.e., $N\rightarrow \infty$.
\end{enumerate}
\end{prop}

We next provide a graph-based characterization of the attention network inside an echo chamber. For a fixed ECE, draw a directed link from $i$ to $j$, denoted by $i\longrightarrow j$, whenever $i$ pays positive attention to $j$. These links form a directed graph on members of a chamber $I_+$. Define the set of peers that  $i$ attends to as 
\[
\mathrm{COR}_i\coloneqq\{j\in I_+:i\longrightarrow j\}.
\]
Following \cite{sadler2021}, call the graph on $I_+$ a \emph{directed nested split graph} if, for any distinct $i,j\in I_+$, $|\mathrm{COR}_i|\leq|\mathrm{COR}_j|$ implies $\mathrm{COR}_i\subseteq\mathrm{COR}_j\cup\{j\}$. In words, if $i$ attends to fewer peers than $j$, then aside from $j$ (who cannot attend to himself), the set of peers attended to by $i$ is nested by that attended to by $j$. 

\begin{prop}\label{prop_cp}
In any ECE, the attention network within a chamber is a directed nested split graph.
\end{prop}

To develop intuitions, let us partition a chamber into 
    \[\mathrm{COR}\coloneqq \left\{i \in I_+: x_i>\phi(\lambda_i)\right\}, \quad \mathrm{PER}\coloneqq I_+\setminus \mathrm{COR},\] 
    read as ``core'' and ``periphery.'' 
By definition, COR (resp. PER) consists of players whose primary-source attention exceeds (resp. falls short of) their personal thresholds. Since these thresholds are positive whenever they are finite, COR members pay positive attention to the primary source. By Lemma \ref{lem_general}, they also pay positive attention to one another while ignoring PER, i.e.,
\[\mathrm{COR}_i=\mathrm{COR}-\{i\} \quad \forall i \in \mathrm{COR}.\]

In an asymmetric society, a PER player may pay zero attention to the primary source (e.g., because he has a small bandwidth). Adapting Lemma \ref{lem_general} to allow this possibility, the attention of each $i \in \mathrm{PER}$ to a peer $k$ becomes $\frac{1}{\gamma_i}h^+(x_k,\frac{\lambda_k}{\gamma_i})$, where $\gamma_i\geq1$. Consequently, player $i$ ignores PER while selectively attending to a subset of COR:
\[
\mathrm{COR}_i=\left\{k\in\mathrm{COR}:x_k>\phi(\lambda_k/\gamma_i)\right\}\quad \forall i \in \mathrm{PER}.
\]
The term $\gamma_i$ captures $i$'s selectivity. An increase in $\gamma_i$ raises $\phi(\lambda_k/\gamma_i)$ and causes the set to shrink, as if $i$ screens COR members more aggressively. Between two distinct PER members $i,j$, 
\[\text{either} \quad \mathrm{COR}_i\subseteq \mathrm{COR}_j\quad \text{or}\quad \mathrm{COR}_j \subseteq \mathrm{COR}_i.\]

Combining the above arguments yields a directed nested split graph.

\subsection{Efficient attention allocation}\label{sec_efficient}
This section analyzes the \emph{efficient allocations} that maximize utilitarian welfare in a symmetric society, highlighting how they differ from and relate to the ECE. 

Throughout, we refer to maximizing utilitarian welfare---or equivalently minimizing the total expected loss---as the \emph{efficiency program}, and to individual players minimizing expected losses, taking others' attention as given, as the \emph{equilibrium program}. The main difference between the two is that in the equilibrium program, individuals do not internalize the potential social benefits generated by their primary-source attention. Increasing a player's primary-source attention above the equilibrium level makes him a more informative secondary source. To the extent that this change benefits his peers but not the player himself,  it wouldn't be internalized by him in the equilibrium program.

As a social planner attempts to unlock these social benefits, a technical challenge emerges: the efficiency program may be neither convex nor concave in the attention allocation vector. By contrast, the equilibrium program is always convex in individual players' own attention allocation. The next example illustrates this distinction:

\begin{example}\label{exm:saddlepoint}
In a society of two same-type players, minimizing total expected loss is equivalent to  minimizing the sum of disruption probabilities across players: 

\[
\min_{(x_1, x_2) \in [0,\tau]^2}
\sum_{i=1}^2 \exp(-x_i)
\left[
\exp(-x_j)
+(1-\exp(-x_j))\exp(-\lambda(\tau-x_i))
\right].
\]
In the objective, $x_i$ denotes player $i$'s attention to the primary source. Increasing $x_i$ lowers player $j$'s disruption probability---a form of positive externality. When $\lambda=2$ and $\tau=1$, the Hessian of the objective is everywhere indefinite, so the objective admits a global saddle-point structure.
There are two global minimizers $(x_1,x_2)=(1,.73)$ and $(.73,1)$, where one player pays full attention to the primary source, and the other splits attention between the primary source and his peer. 


By contrast, in the equilibrium program, player $1$ chooses $x_1$ to minimize his own log-disruption probability, taking $x_2$ as given: 
\[\min_{x_1 \in [0,\tau]} -x_1+\log\left[\exp(-x_2)+(1-\exp(-x_2))\exp(-\lambda (\tau-x_1))\right].\]
For any $x_2>0$, this objective is strictly convex in $x_1$. 
\end{example}

A lack of convexity renders standard methods (e.g., the first-order approach) ineffective at solving the efficiency program. Global properties of efficient allocations---such as those considered in the next theorem---are most  challenging to address. We overcome these challenges and obtain partial characterizations of efficient allocations under two limiting conditions: $\tau\downarrow \phi(\lambda)$ and $N\rightarrow \infty$. The next theorem reports our findings, focusing on  (1) whether efficient allocations exhibit echo chambers; (2) the \emph{efficiency gap} of the ECE, defined as the gap between the expected losses per capita under the ECE and efficient allocations.

\begin{thm}\label{thm:efficiency}
Consider symmetric environments with $\lambda>1$ and $\tau>\phi(\lambda)$.
\begin{enumerate}[(i)]
\item Denote $\Delta=\tau-\phi(\lambda)$. For sufficiently small $\Delta>0$, every efficient allocation is within $O(\Delta^2)$ of an echo-chamber allocation in the sup-norm, and the efficiency gap of the ECE is $O(\Delta^2)$.

\item For sufficiently large $N$,  the gap between the expected losses per capita under the ECE and the \emph{fully symmetric} efficient allocation is $\Theta(1)$,\footnote{An allocation is fully symmetric if all players pay identical attention to their own-biased primary source, the opposite-biased primary source, each other same-type player, and each opposite-type player. The ECE is fully symmetric. } and the gap between the latter and an unconstrained efficient allocation is also $\Theta(1)$. Consequently, the efficiency gap of the ECE is $\Theta(1)$. 
\end{enumerate}
\end{thm}

In the equilibrium analysis, we showed that for small $\Delta=\tau-\phi(\lambda)$, attending to a peer lowers one's disruption probability by $O(\Delta^2)$ compared to attending directly to the primary source. In the efficiency program, the same intuition and conclusion continue to  hold, implying that every unit of attention directed to one's opposite-biased source generates a social benefit of $O(\Delta^2)$, at a non-negligible opportunity cost on the individual (who could at least learn directly from his own-biased source).  Hence, the planner directs at most $O(\Delta^2)$ units of each individual's attention to the opposite-biased source.

Part~(ii) of Theorem~\ref{thm:efficiency} paints a distinctive picture in the large-population limit. We first compare the ECE with the fully symmetric efficient allocation, which is shown to exhibit echo chambers; see Remark~\ref{rem_dual}. The main difference, therefore, lies in the primary-source attention. As $N\rightarrow \infty$, primary-source attention collapses to $\phi(\lambda)$ under the ECE but tends to $x^{\infty} \in (\phi(\lambda),\tau)$ under the fully symmetric efficient allocation. The difference $x^{\infty}-\phi(\lambda)$ yields a non-vanishing gap in expected loss per capita.

Note, however, that $x^{\infty}<\tau$: under the full symmetry restriction, asking all individuals to pay full attention to the primary source and no attention to one another is inefficient. By relaxing this restriction, society as a whole can do better.\label{page:efficient_asy} 

To see why, consider a benchmark in which each individual has access to $N$ hypothetical peers who pay full attention $\tau$ to the primary source, while he himself allocates $\tau$ units of attention between these hypothetical peers and his own-biased source.  While infeasible, this benchmark provides a lower bound on expected loss. It can be shown that, when $N$ is large, the individual allocates $\tau/N$ units of attention to each hypothetical peer and zero attention to the primary source. 
The resulting disruption probability is 
\begin{align*}
\left(\exp(-\tau)+(1-\exp(-\tau))\exp\left(-\frac{\lambda \tau}{N}\right)\right)^N\rightarrow \exp(-\lambda \tau (1-\exp(-\tau))) \text{ at rate } O(1/N).
\end{align*}

We then construct a feasible asymmetric allocation to approximate this benchmark. Among $N$ players of the same type, let $K$ pay full attention to the primary source, just as the hypothetical peers in the benchmark; call them \emph{transmitters}. Each of the remaining $N-K$ players, called \emph{benefiters}, divides his bandwidth evenly across the transmitters. Replacing $N$ by $K$ in the preceding argument, each benefiter's disruption probability approaches the benchmark at rate $\Theta(1/K)$. Meanwhile, each transmitter's disruption probability is $\Theta(1)$ higher than under the benchmark. Across all players, the total expected loss exceeds that under the benchmark by 
$K\Theta(1)+(N-K)\Theta(1/K),$
whose order is minimized at $K=\Theta(\sqrt{N})$. Thus, by assigning a vanishing fraction $\Theta(1/\sqrt{N})$ of the players as transmitters, we approximate the benchmark loss per capita at rate  $\Theta(1/\sqrt{N})$.\label{page:efficient_asy_end}

\begin{rem}\label{rem_dual}
\emph{Appendix~\ref{sec_proof_thm3} develops a dual approach to obtain partial characterizations of efficient allocations. Here, we provide an overview of the approach. }

\emph{To begin, notice that in both the efficiency program and the equilibrium program, peer attention generates only private benefits but no social benefits. Once primary-source attention is fixed, the problem of peer attention allocation is analogous across both programs. This similarity has a succinct dual representation. Fixing a  terminal decision upon receiving no message, say $d_i=+1$, player $i$'s attention to player $j$ equals 
\[\frac{1}{\gamma_i}h^+\left(x_j^{+},\frac{\lambda}{\gamma_i}\right),\]
where $x_j^+$ denotes $j$'s attention to the $+1$-biased primary source, and $\gamma_i >0$ denotes the Lagrange multiplier on $i$'s bandwidth constraint. Since the latter binds, we have 
\[
\frac{1}{\gamma_i}\sum_{j \neq i} h^+\left(x_j^+, \frac{\lambda}{\gamma_i}\right)
= \tau-x_i^{+}-x_i^{-},\]
where the left-hand side is $i$'s total attention to peers,  and the right-hand side is the residual capacity after deducting primary-source attention.  The equilibrium analysis so far exploits the facts that $x_i^-=0$ and $\gamma_i=1$; the latter holds because in the equilibrium program, $i$ equalizes the marginal, private benefit of primary-source attention---which lowers his log-disruption probability by one unit---and the shadow cost $\gamma_i$. In the efficiency program, primary-source attention is chosen by the social planner to internalize its social benefits. Given these choices, player $i$ adjusts $\gamma_i$ to bind his bandwidth constraint.}

\emph{The dual approach is used to establish, among other results, that every efficient allocation satisfying a weaker symmetry requirement than full symmetry exhibits echo chambers. This result was invoked in the preceding argument to establish the efficiency gap of the ECE. }\label{page:rem_dual_end}
\end{rem}

\subsection{Multiple periods with costly waiting}\label{sec_cm}
Our model is multi-agent and static, building on the single-agent static example in CM19. The main analysis in CM19 studies a single-agent dynamic attention-allocation problem with costly waiting. This raises a natural question: how do our multi-agent insights change in a dynamic environment of the type studied in CM19? 
In this section, we shed light on this question using a simple two-player, two-period model.

\vspace{-10pt}

\paragraph{Setup.} There are two players, Alan and Bob. Following CM19, Alan initially assigns probability $p_0\in(1/2,1)$ to the state being $+1$, while Bob assigns the same probability to the state being $-1$. We therefore refer to Alan as $+1$-biased and Bob as $-1$-biased.\footnote{In static settings, biased preferences over terminal decisions are mathematically equivalent to heterogeneous priors. This equivalence breaks down in dynamic decision-making problems with costly waiting.}

The game unfolds over two periods. In period one, each player allocates $\tau>\phi(\lambda)$ units of attention between the two primary sources. At the end of period one, he either stops and makes a terminal decision of $-1$ or $+1$, or pays a waiting cost $c>0$ and continues to period two. In period two, each player again has $\tau$ units of attention to allocate, but may now attend to the other player as a secondary source, in addition to the two primary sources. With probability $1-\exp(-\lambda a_j^i)$, $j$ is successfully connected to $i$, and $i$ transmits all primary-source messages he has received in both periods to $j$. At the end of period two, each player makes a terminal  decision of $-1$ or $+1$, and the game ends. Payoff is zero if the terminal decision matches the true state and $-1$ otherwise.

\vspace{-10pt}
\paragraph{CM19's result.} In the single-agent static benchmark, the agent pays full attention to his own-biased source and chooses his default upon receiving no message. CM19 show that dynamic attention allocation with costly waiting may reverse this prediction. They identify a novel attention strategy, which we refer to as \emph{opposite-biased attention}. As illustrated in Panel (a) of Table \ref{table:comparison}, Alan keeps attending to the opposite-biased source and, if no message arrives by the end of period two, chooses the alternative. By contrast, \emph{own-biased attention}, illustrated in Panel (b), refers to Alan attending to the own-biased source and choosing the default if no message arrives.

\begin{table}[h!]
\centering
\begin{minipage}[t]{0.49\textwidth}
\centering\small
\begin{tabular}{@{}lc@{}}
\hline
& \textbf{Alan} \\
\hline
\textbf{Period 1} & $-1$-biased \\
\textbf{Period 2} & $-1$-biased \\
\shortstack[l]{\textbf{Period-2 decision}\\\textbf{if uninformed}} & $-1$ \\
\hline
\end{tabular}
\subcaption{Opposite-biased attention.}
\label{table:panel_a}
\end{minipage}\hfill%
\begin{minipage}[t]{0.49\textwidth}
\centering\small
\begin{tabular}{@{}lc@{}}
\hline
& \textbf{Alan} \\
\hline
\textbf{Period 1} & $+1$-biased \\
\textbf{Period 2} & $+1$-biased \\
\shortstack[l]{\textbf{Period-2 decision}\\\textbf{if uninformed}} & $+1$ \\
\hline
\end{tabular}
\subcaption{Own-biased attention.}
\label{table:panel_b}
\end{minipage}
\caption{Single agent attention dynamics over two periods.}
\label{table:comparison}
\end{table}

CM19 show that opposite-biased attention can dominate own-biased attention. The reason is that, ex ante, Alan believes that state $+1$ is more likely. He therefore expects to receive a conclusive message sooner from the opposite-biased source revealing his default,  than from the own-biased source revealing his alternative. Upon receiving a conclusive message after one period, Alan acts on it and stops, thereby saving the waiting cost. When the cost saving is significant and the prior is sufficiently close to neutral, opposite-biased attention is optimal. 

\vspace{-10pt}
\paragraph{Lessons from our model.} A full analysis of our model is deferred to Online Appendix~\ref{sec_online_cm}. Here we highlight two new insights. 

First, sustaining opposite-biased attention in equilibrium is harder in the presence of Bob. In Table \ref{table:deviate}, we start from opposite-biased attention (marked black) and consider Alan's unilateral deviation to own-biased attention (marked red). 
Compared to Panel (b) of Table \ref{table:comparison}, Alan can now attend to Bob in period two, in addition to the $+1$-biased source. The informational value of Bob makes the deviation more attractive than in CM19's single-agent setting. For instance, when $\tau=1.5$, $\lambda=1.5$, and $p_0=0.6$, opposite-biased attention is optimal in CM19 if 
$c\in[0.0641,\;0.1112]$ and is an equilibrium in our model if $c\in[0.1061,\;0.1112]$. For $c\in[0.0641,\;0.1061)$, opposite-biased attention is optimal in CM19 but cannot arise in equilibrium in our model.

\begin{table}[h!]
\centering
\begin{tabular}{lcc}
\hline
 & \textbf{Alan} & \textbf{Bob} \\
\hline
\textbf{Period 1} & $-1$-biased $\rightarrow$ {\color{red}{$+1$-biased}} & $+1$-biased \\
\textbf{Period 2} & $-1$-biased $\rightarrow$ {\color{red}{$+1$-biased \& Bob}} & $+1$-biased \\
\textbf{Period-two decision if uninformed} & $-1$ $\rightarrow$ {\color{red}{$+1$}} & $+1$\\
\hline
\end{tabular}
\caption{Alan deviates from opposite-biased attention to own-biased attention.}\label{table:deviate}
\end{table}

Second, a new form of attention strategy emerges, which we refer to as \emph{alternating attention}. As illustrated in Table \ref{table:alternate}, both players attend to the opposite-biased source in period one. If silence remains, then in period two, they attend to their own-biased sources and one another, and choose their defaults in case silence still remains. 
This strategy combines the early-stopping advantage of opposite-biased attention in period one with the informational advantage of own-biased plus peer attention in period two. The presence of Bob is key. Without Bob, Alan's period-two posterior conditional on silence would coincide with his prior. A better strategy would then be to focus on the own-biased source and stop unconditionally after one period.  Numerical analysis shows that when $\tau=2$, $\lambda=2$, $p_0=0.9$, and 
$c\in[0.0116,\;0.0193]$,  alternating attention emerges as an equilibrium in our model.
\begin{table}[h!]
\centering
\begin{tabular}{lcc}
\hline
 & \textbf{Alan} & \textbf{Bob} \\
\hline
\textbf{Period 1} & $-1$-biased  & $+1$-biased \\
\textbf{Period 2} & Bob \& $+1$-biased & Alan \& $-1$-biased \\
\textbf{Period-two decision if uninformed} & $+1$ & $-1$\\
\hline
\end{tabular}
\caption{Alternating-attention equilibrium.}\label{table:alternate}
\end{table}

\section{Related literature}\label{sec_literature}
\paragraph{Rational inattention.} A large literature on rational inattention (RI) studies how a single decision maker can flexibly acquire information about a payoff-relevant state, using signal structures that map the states into lotteries over final decisions \citep{mackowiak2023rational}. While this approach is well suited for studying single-agent decision problems, it does not address how strategic players acquire information from multiple sources and from one another.
\cite{hellwig}, \cite{dentinetwork, denti}, and \cite{hebertlao} show that, in coordination games, equilibrium outcomes can depend sensitively on whether players can attend to and coordinate on the information endogenously gleaned by other players. 
In our model, by contrast, a player attracts peer attention because he can disseminate information more effectively than the primary sources.

The idea that rational decision makers may prefer biased information when facing limited information processing capacities dates back to \cite{calvert} and was later developed by \cite{suen}, \cite{nimark2019inattention}, \cite{hulisegal} and \cite{novak2024status} among others, in single-agent decision-making contexts. In media and politics, studies of this phenomenon---often termed \emph{filtering bias}---are surveyed by \cite{gentzkow2015media}. CM19 and \cite{zhong} analyze single-agent dynamic information acquisition problems, under Poisson learning and posterior-separable attention costs, respectively. \cite{dessein} characterize the most efficient Poisson attention network in a non-strategic organization. By contrast, we study the equilibrium Poisson attention network among multiple strategic players with different objectives from that in \cite{dessein}.

\vspace{-10pt}

\paragraph{Social network.} We study a hybrid game in which players form a strategic peer attention network and play a game of primary-source attention allocation over it. The literature on strategic network formation pioneered by \citet{jackson1996} and \citet{balagoyal} is vast, with recent works such as \citet{hagenbach2010}, \citet{galeottietal}, \citet{calvo2015}, and \citet{herskovic2020} studying the formation of information transmission networks. Unlike our work, these studies take information sources as exogenously given and model information transmission either as a physical process or as a cheap-talk game.  While most prior works in this area study discrete link formation, we work with  divisible links in line with \citet{blochdutta} and related contributions. A handful of papers predict a core–periphery architecture, relying on the assumption that link values exhibit strategic substitutability. Recent work by \citet{herskovic2020} and \citet{dentinetwork} demonstrates that coordination motives can also sustain a core–periphery structure when combined with rational inattention.\footnote{\citet{peregoyuksel} study a dynamic learning model in which each player can either produce or search for information, but not both. In equilibrium, specialization emerges, although, unlike in our model, information producers (analogous to our core) do not interact with one another, while information consumers (analogous to our periphery) search randomly, without directions, to be matched with information  producers. } 
 
A vast literature studies games played over fixed networks, including seminal work on strategic substitutes \citep{bramoulle2007, kranton14} and complements \citep{ballester}. As noted by \cite{jackson2015}, ``equilibria in games with strategic substitutes are difficult to characterize, and multiple equilibria rather than a unique equilibrium are the rule.''  Existing methodological contributions in this area have largely focused on the existence, uniqueness, and stability of pure-strategy equilibria, often assuming linear best responses or symmetric influences between players. Our approach differs, exploiting  the game's asymptotics to establish the ECE as the unique equilibrium.

A growing literature, surveyed by \cite{bothsurvey}, studies hybrid models that combine  strategic network formation with games played over networks. The seminal contribution of \citet{galeottigoyal} (GG10) predicts the ``law of the few'' in a public-good contribution game, relying on two model features: (1) total equilibrium investment in the public good is independent of the number of players, and (2) discrete links form at a constant unit cost. Subsequent work extends GG10's framework to heterogeneous players \citep{merlino}, weighted link-formation \citep{kinateder2022local}, and multiple public goods \citep{allmis2023homophily}. 
Our comparative statics instead exploit the low-rank perturbation structure between  players' mutual influences. \cite{sadler2021} examine how endogenous action choices and link formation jointly shape the outcomes of hybrid games.\label{pagesadlergolub} In our framework, these correspond to primary-source and peer attention whose interactions have a succinct dual representation. To the extent that our approach relies only on the assumption of divisible links, it may be useful for future studies of hybrid games.

\vspace{-10pt}

\paragraph{Echo chamber and related phenomena.} The term ``echo chamber'' carries different meanings in economics. For instance, \citet{levyrazinsurvey} interpret ``echo'' as correlation neglect, while \cite{bowen2023learning}, \cite{acemoglu2024model} and \cite{kranton2022social} use it to describe the spreading of misinformation. In cognitive science, echo chambers have been associated with a range of behavioral biases, including cognitive dissonance, selective exposure, confirmation bias and motivated reasoning \citep{festinger1957, stroud2017, nickerson1998confirmation, kunda1990}. Economic models of these phenomena and their applications to media and politics are surveyed by \cite{benjamin2019errors} and \cite{anderson2016handbook}.

We examine the rational-inattention origin of echo chambers. Other  rational models of echo chambers exploit alternative mechanisms, including cheap talk \citep{jann}, Bayesian persuasion \citep{meng, innocenti}, and social/DeGroot learning \citep{yildiz, williams, polanski2023}.\footnote{We view rationality as a first principle in economics, a useful benchmark, and a relevant force as  decisions are increasingly delegated to technologies designed to de-bias human judgment \citep{kleinberg2018human}. Our rational approach yields distinctive predictions that can be tested apart from bounded-rationality models; see Section \ref{sec_empirical}. In practice, behavior likely reflects both rational and bounded-rational forces.}

Our work is broadly related to the theories of homophily and of club/group formation \citep{jacksonjep, demange2005group}. The closest paper is \citet{baccarayariv} (BY), which studies group formation followed by in-group production and sharing of public goods. While in-group sharing is automatic in BY, peer attention allocation is strategic in our model. Moreover, BY’s mechanism relies on sorting individuals with similar preferences into limited-size groups---an idea pioneered by \citet{tiebout1956} but not explored here.

\appendix

\section{Proof of Theorem \ref{thm_ece}}\label{sec_proof_thm1}
Theorem \ref{thm_ece} follows from combining Lemmas \ref{lem_partition}-\ref{lem_mse}. 

\subsection{Proofs of Lemmas \ref{lem_partition}-\ref{lem_ece}}

We complete Lemma \ref{lem_general} to allow for zero primary-source attention.

\begin{lem2}[Complete]
Consider a PSPBE in which $|S_d|\geq 2$. For each $i\in S_d$, let $x_i$ denote player $i$'s attention  to the $d$-biased primary source. The following are true: 
\begin{enumerate}[(i)]
 \item For all $j \in S_d\setminus \{i\}$, 
        \[
a_i^j=\frac{1}{\gamma_i} h^+\left(x_j, \frac{\lambda_j}{\gamma_i}\right)=\begin{cases}
\dfrac{1}{\gamma_i} h\left(x_j, \dfrac{\lambda_j}{\gamma_i}\right) & \text{ if  } \dfrac{\lambda_j}{\gamma_i}>1 \text{ and } x_j>\phi\left(\dfrac{\lambda_j}{\gamma_i}\right),\\
0 &\text{ otherwise},
\end{cases}
\]
where $\gamma_i$ denotes the Lagrange multiplier on $i$'s bandwidth constraint and satisfies $\gamma_i=1$ if $x_i>0$ and  $\gamma_i\geq 1$ if $x_i=0$. 

\item The log-disruption probability faced by player $i$ equals 
\[-\tau-\sum_{j \in S_d\setminus \{i\}}\left(\left[x_j^{+}-\phi\left(\frac{\lambda_j}{\gamma_i}\right)\right]^+- \frac{1}{\gamma_i}h^+\left(x_j^{+},\frac{\lambda_j}{\gamma_i}\right)\right)\]

\end{enumerate}
\end{lem2}

\paragraph{Proof of Lemmas \ref{lem_partition} and \ref{lem_general}.}
Let $S_+$ and $S_-$ denote the sets of players who choose $d=+1$ and $d=-1$ following no message, respectively. Consider the problem faced by $i\in S_+$.  Since $d_i=+1$ incurs a loss in state $\omega=-1$, minimizing expected loss is equivalent to minimizing the probability that all information transmission paths from the $+1$-biased primary source to $i$ are disrupted. For a given attention allocation  $\boldsymbol{a}_{-i} \in A_{-i}$ by  $i$'s opponents, this disruption probability equals 
\[
\delta_i^{+1\text{-biased}} \prod_{j \in I\setminus \{i\}} \left[ \delta_j^{+1\text{-biased}} + \left(1 - \delta_j^{+1\text{-biased}}\right) \delta_i^j \right], 
\]
where $\delta_i^c=\exp(-\lambda_c a_i^c)$. 
Player $i$ chooses $\boldsymbol{a}_i \in A_i$ to minimize this disruption probability. Equivalently, he solves  
\begin{align*}
\min_{(a_i^c)_{c \in C_i}} \quad 
& -a_i^{+1\text{-biased}}+
\sum_{j \in I \setminus \{i\}} \log\left( \delta_j^{+1\text{-biased}} + 
\left(1 - \delta_j^{+1\text{-biased}}\right) \exp(-\lambda_j a_i^j) \right) \\
\text{s.t.} \quad 
& a_i^c \geq 0 \ \text{for all } c \in C_i, \quad 
\sum_{c \in C_i} a_i^c \leq \tau_i,
\end{align*}
where the objective is the log-disruption probability and the constraint set is $A_i$.

The program is convex in $\boldsymbol{a}_i$. Let $\gamma_i \geq 0$ denote the Lagrange multiplier on the bandwidth constraint. For each $c \in C_i$, let $\eta_i^c \geq 0$ denote the Lagrange multiplier on the nonnegative constraint $a_i^c \geq 0$ . The first-order conditions are: 
\begin{align*}
\tag{$\mathrm{FOC}_{a_i^{+1\text{-biased}}}$}
 1  = \gamma_i - \eta_i^{+1\text{-biased}} , \\
\tag{$\mathrm{FOC}_{a_i^{-1\text{-biased}}}$}  0  = \gamma_i -\eta_i^{-1\text{-biased}},\\
\tag{$\mathrm{FOC}_{a_i^j}$}
 \frac{ \lambda_j \left(1 - \delta_j^{+1\text{-biased}}\right) \delta_i^j }{ \delta_j^{+1\text{-biased}} + \left(1 - \delta_j^{+1\text{-biased}}\right) \delta_i^j }   =\gamma_i - \eta_i^j.
\end{align*}
 Complementary slackness requires that 
 \[\gamma_i(\sum_{c} a_i^c-\tau_i)=0\quad \text{and}\quad  \eta_i^c a_i^c=0 \quad \forall c.\]
 
$\mathrm{FOC}_{a_i^{+1\text{-biased}}}$ implies $\gamma_i=1+\eta_i^{+1\text{-biased}}\geq1$. Substituting into
$\mathrm{FOC}_{a_i^{-1\text{-biased}}}$ yields $\eta_i^{-1\text{-biased}}=\gamma_i>0$, and complementary slackness implies  
$a_i^{-1\text{-biased}}=0$. By symmetry,
$a_j^{+1\text{-biased}}=0$ or, equivalently, 
$\delta_j^{+1\text{-biased}}=1$ for every $j\in S_-$. Substituting into $\mathrm{FOC}_{a_i^j}$ gives $\eta_i^j=\gamma_i>0$, and complementary slackness implies $a_i^j=0$. Thus, $i$ restricts attention to the
$+1$-biased primary source and other members of $S_+$. By symmetry, the opposite holds for members of $S_-$, which establishes Lemma~\ref{lem_partition}.

Inside $S_+$, let $x_i$ to denote $i$'s attention to the $+1$-biased primary source. 
If $a_i^j>0$, then $\eta_i^j=0$. Simplifying 
$\mathrm{FOC}_{a_i^j}$ accordingly gives 
\[
a_i^j=\frac{1}{\lambda_j}
\log\left[
\left(\frac{\lambda_j}{\gamma_i}-1\right)(e^{x_j}-1)
\right],
\]
where the right-hand side is positive if and only if 
$\lambda_j/\gamma_i>1$ and
$x_j>\phi(\lambda_j/\gamma_i)$. It follows that 
\[
a_i^j=\frac{1}{\gamma_i}
h^+\left(x_j,\frac{\lambda_j}{\gamma_i}\right)
\]
and substituting this into the log-disruption probability for $i$ yields
\[-\tau-\sum_{j \in S_+\setminus \{i\}}\left(\left[x_j^{+}-\phi\left(\frac{\lambda_j}{\gamma_i}\right)\right]^+- \frac{1}{\gamma_i}h^+\left(x_j^{+},\frac{\lambda_j}{\gamma_i}\right)\right).\]
 This establishes Lemma \ref{lem_general}. The special case $x_i>0$ considered in the main text obtains when $\gamma_i=1$.\qed

\vspace{5pt}

We make another useful observation about $h$: 
\begin{obs}\label{obs2}
For fixed $x>0$ and $\lambda>1$, the function  
$\gamma\mapsto\frac{1}{\gamma}h\left(x,\frac{\lambda}{\gamma}\right)$, $\gamma \in (0,\lambda)$ is strictly positive and strictly decreasing,  diverges as $\gamma\downarrow 0$, and vanishes as $\gamma \uparrow \lambda$.
\end{obs}

\vspace{-10pt}
\paragraph{Proof of Lemma \ref{lem_peersymmetry}.} By symmetry, it suffices to prove the lemma for $S_+$, whose size we denote by $n$.  In Section \ref{sec_proofsketch}, we already proved the case $n=2$. We now consider the case $n\geq 3$. 

We claim that $\gamma_i \equiv 1$ for all $i \in S_+$. The proof proceeds by fixing a pair  $(i, j)$ in $S_+$ and eliminating the following possibilities.

\vspace{-10pt}
\subparagraph{Case 1: $\gamma_i=1$, $\gamma_j>1$.} By Lemma \ref{lem_general}, player $i$ pays $x_i$ units of attention to the primary source, zero attention to player $j$, and $h^+(x_k,\lambda)$ units of attention to each $k \in S_+\setminus \{i,j\}$. Player $j$ pays zero attention to the primary source, $\frac{1}{\gamma_j}h^+(x_i, \frac{\lambda}{\gamma_j})$ units of attention to player $i$, and $\frac{1}{\gamma_j}h^+(x_k,\frac{\lambda}{\gamma_j})$ units of attention to each $k \in S_+ \setminus \{i,j\}$. 
The following is a contradiction: 
\begin{align*}
\tag{$\because$ $j$'s bandwidth constraint binds}\tau&=\dfrac{1}{\gamma_j} h^+\left(x_i,\dfrac{\lambda}{\gamma_j}\right)+\dfrac{1}{\gamma_j} \sum_{k \in S_+ \setminus \{i,j\}}h^+\left(x_k,\dfrac{\lambda}{\gamma_j}\right)\\
\tag{$\because$ Observation \ref{obs2}}& < h^+\left(x_i, \lambda\right)+\sum_{ k \in S_+ \setminus \{i,j\}} h^+\left(x_k, \lambda\right)\\
\tag{$\because$ Observation \ref{obs_function}}& \leq  x_i + \sum_{k \in S_+ \setminus \{i,j\}} h^+\left(x_k, \lambda\right)\\
\tag{$\because$ $i$'s bandwidth constraint binds}&=\tau.
\end{align*}

\vspace{-10pt}
 \subparagraph{Case 2: $\gamma_i, \gamma_j>1$.} In this case, both $i$ and $j$ pay no attention to the primary source and attract no peer attention. Their bandwidth constraints are 
\[\frac{1}{\gamma_i}\sum_{k \in S_+\setminus\{i, j\}} h^+\left(x_k, \frac{\lambda}{\gamma_i}\right)=\tau, \quad \frac{1}{\gamma_j}\sum_{k \in S_+\setminus\{i, j\}} h^+\left(x_k, \frac{\lambda}{\gamma_j}\right)=\tau.\]
Observation \ref{obs2} implies that $\gamma_i=\gamma_j>1$, and that 
\[\sum_{k \in S_+\setminus\{i, j\}}h^+(x_k,\lambda)>\tau.\]
Therefore, there exists 
$l\in S_+\setminus\{i,j\}$ such that $x_l>\phi(\lambda)$;  
$\gamma_l=1$ by Lemma \ref{lem_general}.
Expanding $l$'s bandwidth constraint yields a contradiction:
\begin{align*}
\tau&=x_l+\sum_{k \in S_+\setminus \{i,j,l\}} h^+(x_k,\lambda)\\
&>\phi(\lambda)+h(x_l,\lambda) + \sum_{k \in S_+\setminus \{i,j,l\}} h^+(x_k,\lambda)\\
&=\phi(\lambda)+\sum_{k \in S_+\setminus\{i,j\}} h^+(x_k,\lambda)\\
&> \phi(\lambda)+\tau,
\end{align*}
where the second line holds by Observation \ref{obs_function}. 

Therefore, $\gamma_i=\gamma_j=1$ remains as the only possibility. Between $i$ and $j$, 
\[x_i+h^+(x_j,\lambda)=\tilde{\tau}, \quad x_j+h^+(x_i,\lambda)=\tilde{\tau},\quad \text{where} \quad  \tilde{\tau}=\tau-\sum_{k \in S_+ \setminus\{i,j\}} h^+(x_k,\lambda).\]
This is precisely the two-player case considered in Section \ref{sec_proofsketch}, with an effective bandwidth of $\tilde\tau$. If $\tilde\tau>\phi(\lambda)$, then the solution is as depicted in Figure \ref{fig:best_response_diagram} and satisfies $x_i=x_j$. If $\tilde\tau\le\phi(\lambda)$,  then $x_i=x_j=\tilde\tau$. In both cases, $x_i=x_j$.

Repeating the argument for all pairs of $(i,j)$ gives $x_i \equiv x(n)$, where $x(n)$ solves $x+(n-1)h^+(x,\lambda)=\tau.$
If $x(n)\leq \phi(\lambda)$, then $h^+(x(n),\lambda)=0$ and $x(n)=\tau-(n-1) h^+(x(n),\lambda)=\tau>\phi(\lambda)$, a contradiction. Therefore, $x(n)>\phi(\lambda)$ and  $x(n)+(n-1)h(x(n),\lambda)=\tau.$ Substituting this into Lemma \ref{lem_general} yields $-\tau-(n-1)\xi(n)$ as the log-disruption probability, where $\xi(n)=x(n)-\phi(\lambda)-h(x(n),\lambda)>0$. \qed 

\vspace{-10pt}
\paragraph{Proof of Lemma \ref{lem_typesymmetry}.}
In the proof of Lemma \ref{lem_peersymmetry}, we solved $x(n)$ and $\xi(n)$ for $n \geq 2$.  For $n=1$, let $x(1)=\tau$ and $\xi(1)=\tau-\phi(\lambda)-h(\tau,\lambda)>0$. Then $n\xi(n)> (n-1)\xi(n)$ holds for all $n \in \mathbb{N}$. 

Suppose Lemma \ref{lem_typesymmetry} is false. 
One possibility is that one type is divided between the two
chambers. Suppose there exist $i \in I_+\cap S_+$ and $j \in I_+\cap S_-$. 
Currently $i$'s log-disruption probability equals 
$-\tau-(|S_+|-1) \xi(|S_+|)$. If $i$ leaves $S_+$ and joins $S_-$, then he optimally allocates attention between the $-1$-biased primary source and each current member of $S_-$. The latter pays $x(|S_-|)$ units of attention to the $-1$-biased primary source. Since 
\begin{align*}
\tau& =x(|S_-|)+(|S_-|-1)\,h(x(|S_-|),\lambda) >
|S_-|\,h(x(|S_-|),\lambda),
\end{align*}
$i$'s optimal attention allocation after the deviation is 
\[a_i^k\equiv h(x(|S_-|),\lambda) \quad \forall k \in S_-,\quad  a_i^{-1\text{-biased}}=\tau-|S_-|\,h(x(|S_-|),\lambda)>0.\] 
The second inequality implies that $\gamma_i=1$ after the deviation. Substituting into  Lemma \ref{lem_general}, $i$'s log-disruption probability after the deviation equals 
\[-\tau-\sum_{k \in S_-}\left(x(|S_-|)-\phi(\lambda)-h(x(|S_-|),\lambda)\right)=-\tau-|S_-| \xi(|S_-|).\]
Therefore, $i$ weakly prefers to stay in $S_+$ rather than joining $S_-$ if and only if 
\[\cancel{\log \frac1 2} +\log\beta -\tau - (|S_+|-1)\, \xi(|S_+|) \leq \cancel{\log\frac1 2}-\tau - |S_-| \, \xi(|S_-|).\]
The two sides represent the log-expected loss before and after the deviation. The constant $\log 1/2$ will subsequently be ignored whenever the context is clear.

Likewise, $j$ weakly prefers to stay in $S_-$ rather than joining $S_+$ if and only if 
\[-\tau - (|S_-|-1) \, \xi(|S_-|)\leq  \log \beta - \tau - |S_+|\, \xi(|S_+|).\]

Combining the inequalities for $i$ and $j$,
\begin{align*}
\log \beta-(|S_+|-1)\, \xi(|S_+|) \leq - |S_-| \, \xi(|S_-|)<
 - (|S_-|-1) \, \xi(|S_-|)\leq \log \beta - |S_+|\, \xi(|S_+|).
\end{align*}
which contradicts $\xi(|S_+|)>0$.
Therefore, either $i$ or $j$ strictly benefits from joining the other's chamber, which contradicts equilibrium.

Another possibility is that $S_+=I_-$ and $S_-=I_+$. Since $|S_+|=|S_-|=N$, the no-deviation condition for a member in $S_+$ is 
\[
-(N-1)\xi(N)\leq\log\beta-N\xi(N),
\quad \text{or equivalently} \quad
\xi(N)\leq\log\beta\]
which contradicts $\xi(N)>0>\log\beta$. \qed

\vspace{-10pt}
\paragraph{Proof of Lemma \ref{lem_ece}.} We already established the limit $N\rightarrow \infty$ in Section \ref{sec_proofsketch}. The limit $\tau\downarrow \phi(\lambda)$ obtains by replacing $1/n$ with $\Delta=\tau-\phi(\lambda)$ in the proof of $N\rightarrow \infty$. Below we establish the limit $\beta\downarrow 0$. 

If a player pays full attention to his own-biased source and chooses the default following no message, his log-expected loss equals $\log\beta-\tau+\log 1/2$. If he chooses the alternative following  no message, then the minimum log-expected loss is bounded below by $-\lambda\tau+\log 1/2$.  This is because, in a hypothetical situation where one of the peers has infinite bandwidth and so always knows the true state, paying full attention to that peer yields a log-disruption probability of $-\lambda\tau$.   Fixing $(\lambda, \tau)$ and letting $\beta\downarrow 0$, eventually $\log\beta-\tau<-\lambda \tau$. Under this condition, choosing the default following no message becomes a dominant strategy, hence every PSPBE is an ECE. The existence and uniqueness of the ECE follows from the construction in the proof of Lemma \ref{lem_peersymmetry}. \qed 

\subsection{Proof of Lemma \ref{lem_mse}} 
\subsubsection{Preliminaries}
\paragraph{Problem formulation.} A mixed strategy equilibrium of our game takes the following form: ex ante, a player specifies the decision $d$ he intends to make upon receiving no message ex post. Given $d$, the player seeks only $d$-rejecting messages, by allocating attention among the $d$-biased primary source and other players. Ex post, the player makes decision $d$ absent a message that rejects it and $-d$ otherwise. Randomization can only occur ex ante, and only if the player is indifferent between the two intended decisions. 

We therefore represent a mixed strategy equilibrium by $(p_i, x_i^+,x_i^{-})_{i \in I}$. The term $p_i \in [0,1]$ denotes the probability that $d_i=+1$. With the complementary probability,  $d_i=-1$.  Given $d_i=+1$ (resp. $d_i=-1$),  $x_i^+$ (resp. $x_i^-$) denotes the attention allocation to the $+1$-biased (resp. $-1$-biased) primary source. The superscripts indicate primary sources and should not be confused with positive- and negative-part operators.  

Attention allocation among peers can be solved analogously as earlier. Let  $\chi_{i}^+\coloneqq p_i (1-\exp(-x_{i}^+))$ denote the probability that player $i$ directly receives a message from the $+1$-biased primary source. When $i$ decides the attention allocation to $j$ given $d_i=+1$, he anticipates that $j$ directly receives a message from the  $+1$-biased primary source with probability $\chi_j^+$. This probability used to be $1-\exp(-x_j^+)$ in the PSPBE analysis and is now discounted by $p_j$ to reflect that $j$ may be randomizing between the two primary sources. Substituting $\chi_j^+$ into the previous analysis, we obtain 
\begin{align*}
\frac{1}{\lambda}
\log
\max\!\left\{
\left(
\frac{\lambda}{\gamma_{i}^+}-1
\right)
\left(
\frac{\chi_{j}^+}{1-\chi_{j}^+}
\right),
\,1
\right\},
\end{align*}
as $i$'s attention allocation to $j$. The term $\gamma_i^+$ denotes the Lagrange multiplier on $i$'s bandwidth constraint given $d_i=+1$. As before, $\gamma_i^+ \geq 1$, $\gamma_i^+>1$ implies $x_i^+=0$, and $x_i^+>0$ implies $\gamma_i^+=1$. Further manipulation yields
\begin{align*}
\min\!\left\{
\left(
\frac{\lambda/\gamma_{i}^+}
{\lambda/\gamma_{i}^+-1}
\right)
(1-\chi_{j}^+),
\,1
\right\}
\end{align*}
as the probability that $i$ fails to receive a $+1$-biased message from $j$.

The case $d_i=-1$ can be solved analogously by defining $\chi_i^-\coloneqq(1-p_i)(1-\exp(-x_i^-))$. Then 
\begin{align*}
\frac{1}{\lambda}
\log
\max\!\left\{
\left(
\frac{\lambda}{\gamma_{i}^-}-1
\right)
\left(
\frac{\chi_{j}^-}{1-\chi_{j}^-}
\right),
\,1
\right\}
\end{align*}
is  $i$'s attention allocation to $j$, and  
\begin{align*}
\min\!\left\{
\left(
\frac{\lambda/\gamma_{i}^-}
{\lambda/\gamma_{i}^--1}
\right)
(1-\chi_{j}^-),
\,1
\right\}.
\end{align*}
is the probability that he fails to receive a $-1$-biased message from  $j$.

We say that player $i$ \emph{strictly mixes} if $p_i \in (0,1)$, which happens only if he is ex-ante indifferent between the two intended decisions. The precise condition is
\begin{align*}
&\beta\exp\left(-x_{i}^+\right)\prod_{j \in I \setminus \{i\}}\min\left\{\left(\frac{\lambda/\gamma_{i}^+}{\lambda/\gamma_{i}^+-1}\right)(1-\chi_{j}^+), 1\right\}\\
&=\exp\left(-x_{i}^-\right)\prod_{j \in I \setminus \{i\}}\min\left\{\left(\frac{\lambda/\gamma_{i}^-}{\lambda/\gamma_{i}^- -1}\right)(1-\chi_{j}^-), 1\right\} \quad \text{if} \quad t_i=1,
\end{align*}
and
\begin{align*}
&\exp\left(-x_{i}^+\right)\prod_{j \in I \setminus \{i\}}\min\left\{\left(\frac{\lambda/\gamma_{i}^+}{\lambda/\gamma_{i}^+ -1}\right)(1-\chi_{j}^+), 1\right\}\\
&=\beta\exp\left(-x_{i}^-\right)\prod_{j \in I \setminus \{i\}}\min\left\{\left(\frac{\lambda/\gamma_{i}^-}{\lambda/\gamma_{i}^- -1}\right)(1-\chi_{j}^-), 1\right\} \quad \text{if}\quad  t_i=-1.
\end{align*}

\vspace{-10pt}
\paragraph{$\gamma_i^+\equiv \gamma_i^-\equiv 1$ in equilibrium.} We demonstrate that in every equilibrium,  $\gamma_i^+\equiv 1$ among all players who seek the $+1$-biased message with positive probability. 

Denote the set of such players by $A$. If our claim were false, then one possibility is that $\gamma_i^+=1$ and $\gamma_j^+>1$ for some $i,j \in A$. $\gamma_j^+>1$ implies $x_j^+=0$, so $i$ ignores $j$ when seeking the $+1$-biased message, giving rise to the following bandwidth constraint: 
\[\tau=x_i^+ + \sum_{k \in A \setminus \{i,j\}}\frac{1}{\lambda}\log\max\left\{(\lambda-1)\frac{\chi_k^+}{1-\chi_k^+},1\right\}.\]
By contrast, $j$'s bandwidth constraint when seeking the $+1$-biased message is 
\begin{align*}\tau&=\sum_{k\in A \setminus \{j\}}\frac{1}{\lambda}\log\max\left\{\left(\frac{\lambda}{\gamma_j^+}-1\right)\frac{\chi_k^+}{1-\chi_k^+}, 1\right\}, 
\end{align*}
where 
\begin{align*}
\mathrm{RHS}&<\sum_{k\in A \setminus \{j\}}\frac{1}{\lambda}\log\max\left\{\left(\lambda-1\right)\frac{\chi_k^+}{1-\chi_k^+}, 1\right\}\\
&\leq h^+(x_i^+,\lambda)+\sum_{k\in A \setminus \{i,j\}}\frac{1}{\lambda}\log\max\left\{\left(\lambda-1\right)\frac{\chi_k^+}{1-\chi_k^+}, 1\right\}\\
& \leq x_i^++\sum_{k\in A \setminus \{i,j\}}\frac{1}{\lambda}\log\max\left\{\left(\lambda-1\right)\frac{\chi_k^+}{1-\chi_k^+}, 1\right\}\\
&=\tau.
\end{align*}
The first inequality holds because lowering $\gamma_j^+$ to one strictly increases the summation term whose original value is strictly positive.  The second inequality uses the fact that $\chi/(1-\chi)\leq \exp(x)-1$ and, when the equality holds, the log-max term becomes $h^+(x,\lambda)$. The third inequality uses properties of $h$. The last equality holds because of $i$'s bandwidth constraint. Therefore, we have reached a contradiction.

Now suppose that $\gamma_i^+>1$ for all $i\in A$. Then
$x_i^+=0$ and hence $\chi_i^+=0$ for all $i \in A$, implying that when $i \in A$ seeks the $+1$-biased message, he ignores other members of $A$. Consequently, $x_i^+=\tau>0$, a contradiction.

Therefore, $\gamma_i^+\equiv 1$ remains as the only possibility. 
The proof for $\gamma_i^-\equiv 1$ among players seeking the $-1$-biased message with positive probability is analogous.

\begin{figure}[!h]
\centering
\begin{subfigure}[t]{0.48\textwidth}
\centering
\includegraphics[width=0.7\linewidth]{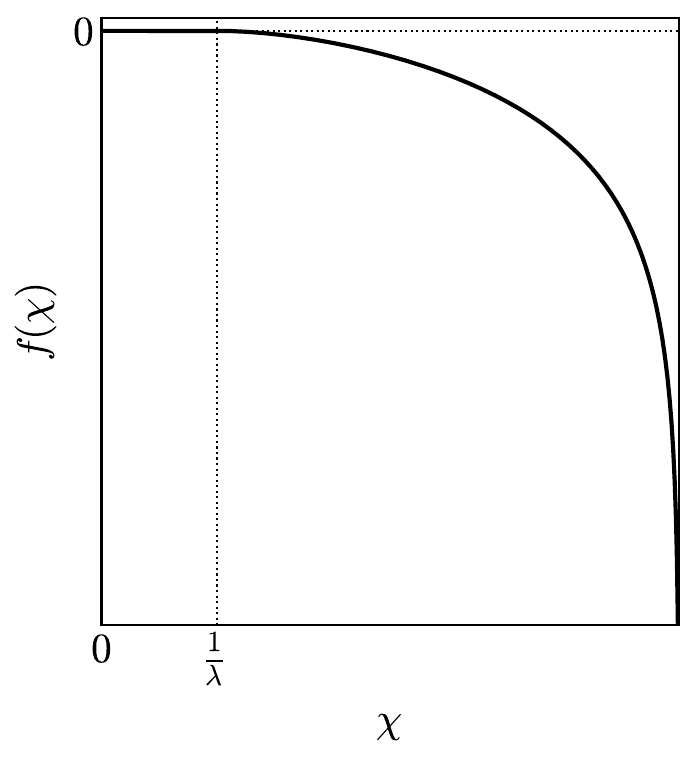}
\end{subfigure}
\hfill
\begin{subfigure}[t]{0.48\textwidth}
\centering
\includegraphics[width=0.7\linewidth]{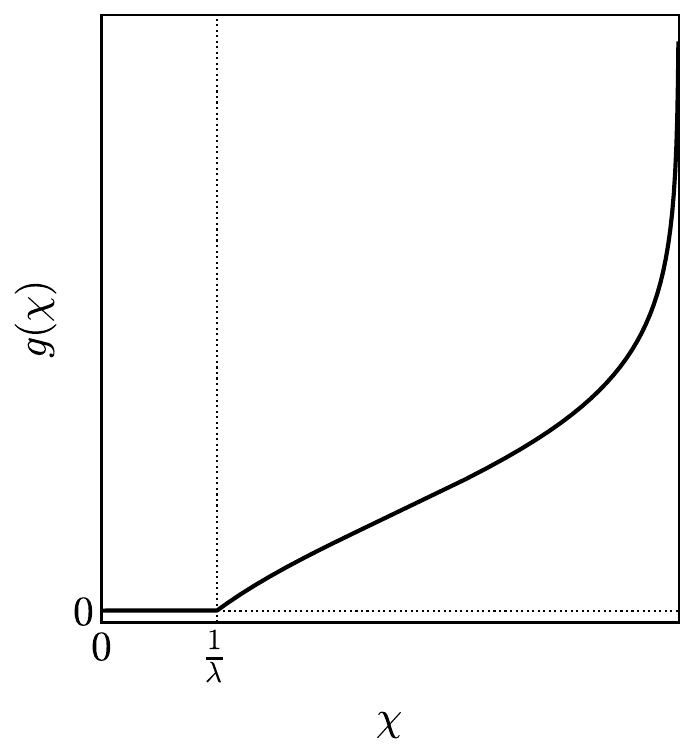}
\end{subfigure}
\caption{Plots of $f(\chi)$ and $g(\chi)$.}
\label{fig-fg}
\end{figure}

\vspace{-10pt}
\paragraph{Useful functions.} For $\chi \in [0, 1-\exp(-\tau)]$, define 
\[f(\chi)\coloneqq \frac{1}{\lambda}\log\max\left\{\frac{(\lambda-1)\chi}{1-\chi}, 1\right\}+\log\min\left\{\frac{\lambda(1-\chi)}{\lambda-1}, 1\right\}\]
and 
\[g(\chi)\coloneqq \frac{1}{\lambda}\log\max\left\{\frac{(\lambda-1)\chi}{1-\chi}, 1\right\}.\]
As illustrated in Figure \ref{fig-fg}: 

$\bullet$ $f=g=0$ on $[0, 1/\lambda]$. 

$\bullet$ On $\chi \in (1/\lambda, 1-\exp(-\tau)]$, $g$ is strictly increasing, $f$ is strictly decreasing, and $f(\chi)=g(\chi)+\log(1-\chi)+\phi(\lambda)$.

In the case where $d_i=+1$, $i$ pays $g(\chi_j^+)$ units of attention to $j$ and faces a log-disruption probability of  $-\tau+\sum_{j \neq i} f(\chi_j^+)$. The counterparts of these quantities in the PSPBE analysis can be obtained by letting  $p=1$ and evaluating $g$ and $f$ at $\chi=1-\exp(-x)$, which gives 
\[g(1-\exp(-x))=h^+(x,\lambda), \quad f(1-\exp(-x))=-[x-\phi(\lambda)-h^+(x,\lambda)]\mathbbm{1}_{x\geq \phi(\lambda)}.\]
For general $\chi \leq 1-\exp(-x)$, 
\[g(\chi) \leq h^+(x,\lambda), \quad f(\chi) \geq -[x-\phi(\lambda)-h^+(x,\lambda)]\mathbbm{1}_{x\geq \phi(\lambda)}.\]

\subsubsection{$\tau\downarrow \phi(\lambda)$} 
Fix $\beta \in (0,1)$, $\lambda>1$, $N\geq 2$, and a decreasing sequence $\tau\downarrow \phi(\lambda)$. For each $\tau$ and each equilibrium given $(\beta,\lambda,\tau, N)$---pure or mixed---define 
\[A=\{k \in I_+:  k \text{ strictly mixes}\}, \quad B=\{k \in I_-: k \text{ strictly mixes}\},\] 
\[C=\{k \in I: k\text{ seeks only ``+1''-biased messages}\},\]
\[D=\{k \in I: k \text{ seeks only ``$-1$''--biased messages}\}.\]
We claim that when $\tau$ is sufficiently near $\phi(\lambda)$, no player strictly mixes, or equivalently $A\cup B=\emptyset$.

Suppose this claim is false. Then there exists a subsequence $\{\tau_n, \text{equilibrium}_n\}$ along which $A_n\cup B_n \neq \emptyset$. We examine the asymptotic properties of this sequence. Whenever the context is clear, we omit   the subscript ``$n$'' from the notation of equilibrium objects.

Consider two cases. 

\vspace{-10pt}
\paragraph{Case 1:  $|A_n|, |B_n|\geq 1$.} For each $i \in A_n$, the following hold: 
\[
    \tau_n=x_{i}^++\sum_{k \in I \setminus \{i\}}g(\chi_k^+),\quad \tau_n=x_{i}^-+\sum_{k \in I \setminus \{i\}} g(\chi_k^-),
\]
and 
\[\log\beta=x_{i}^+-x_{i}^- -\sum_{k \in I \setminus \{i\}} \left[\log \min\left\{\frac{\lambda}{\lambda-1}(1-\chi_{k}^+),1\right\}-\log \min\left\{\frac{\lambda}{\lambda-1}(1-\chi_{k}^-),1\right\}\right].\]
The first two equations state that $i$'s bandwidth constraints bind when he seeks both types of messages, whereas the last equation states that he is indifferent ex ante. 
Solving these equations simultaneously gives 
\[\log\beta=-\sum_{k \in I \setminus \{i\}} \left[f(\chi_{k}^+)-f(\chi_{k}^-)\right].\] 

For each player $j \in B_n$, replacing $\log\beta$ with $-\log\beta$ in the above derivation yields
\[-\log\beta=-\sum_{k \in I\setminus \{j\}} \left[f(\chi_{k}^+)-f(\chi_{k}^-)\right].\]

Denoting
\begin{align*}
X_{k}= f(\chi_{k}^+)-f(\chi_{k}^-),\quad & S_{A_n}=\sum_{k \in A_n} X_{k}, \quad S_{B_n}=\sum_{k \in B_n} X_{k},\\
S_{C_n}=\sum_{k\in C_n} X_{k}, \quad & S_{D_n}=\sum_{k\in D_n} X_{k},
\end{align*}
we have 
\[
   \begin{cases}
   X_{i}-[S_{A_n}+S_{B_n}+S_{C_n}+S_{D_n}]=\log\beta & \forall i \in A_n,\\[1ex]
X_{j}-[S_{A_n}+S_{B_n}+S_{C_n}+S_{D_n}]=-\log\beta & \forall  j \in B_n.
\end{cases}
\]
Solving this system of equations yields 
\[\begin{cases}
X_{i}\equiv \dfrac{(2|B_n|-1)\log\beta-S_{C_n\cup D_n}}{|A_n|+|B_n|-1} & \forall i \in A_n, \\[3ex] 
X_{j}\equiv -\dfrac{(2|A_n|-1)\log\beta+S_{C_n\cup D_n}}{|A_n|+|B_n|-1} & \forall j \in B_n,
\end{cases}
\]
where $S_{C_n\cup D_n}=S_{C_n}+S_{D_n}$. In the next paragraph, we verify that $\lim_{n\rightarrow \infty} S_{C_n \cup D_n}=0$.  Consequently, the following holds uniformly over $i\in A_n$: 
\[
X_i\leq-\kappa_1+o(1)
\quad \text{where}\quad 
\kappa_1=\frac{-\log\beta}{2N-1}>0,
\]
which together with 
\[
 f(\chi_{i}^+)-f(\chi_{i}^-) \geq f(\chi_{i}^+)+0\geq -[x_{i}^+-\phi(\lambda)-h^+(x_i^+,\lambda)]\mathbbm{1}_{x_i^+\geq \phi(\lambda)}\]
 implies 
 \[
 x_{i}^+  \geq \phi(\lambda)+\kappa_2+o(1),\]
where $\kappa_2$ is a positive constant that depends only on $(\kappa_1,\lambda)$. However, this contradicts the assumption that $\lim_{n \rightarrow \infty}\tau_n=\phi(\lambda)$.

We verify that $\lim_{n\rightarrow \infty}S_{C_n\cup D_n}=0$. By definition, $f(\chi_k^-)=0$ and $X_k=f(\chi_k^+)$ for every $k \in C_n$, and $f(\chi_k^+)=0$ and $X_k=-f(\chi_k^-)$ for every $k \in D_n$. Since $0 \geq f(\chi)\geq f(1-\exp(-\tau_n))\to f(1/\lambda)=0$ and $|C_n\cup D_n|\leq 2N$, 
\[
\left|S_{C_n\cup D_n}\right|=\left\vert \sum_{k \in C_n\cup D_n} X_k\right\vert \leq \sum_{k \in C_n\cup D_n} |X_k|\leq 2N |f(1-\exp(-\tau_n))|
\to 0.
\] 
\vspace{-10pt}

\paragraph{Case 2: $|A_n|=0$ or $|B_n|=0$.} It suffices to treat the case $B_n=\emptyset$. The assumption $A_n\cup B_n \neq \emptyset$ implies that $A_n \neq \emptyset$. Solving $X_i$'s among  $i\in A_n$ gives 
\[(1-|A_n|)X_{i}\equiv \log\beta+S_{C_n\cup D_n}.\]
If $|A_n|=1$, this equation becomes
$0=\log\beta+S_{C_n\cup D_n}$, which cannot hold for sufficiently
large $n$ because $S_{C_n\cup D_n}\to0$ and $\log\beta<0$.
If $|A_n|\geq2$, then for sufficiently large $n$, similar manipulation as in Case 1 shows that $x_{i}^-  \geq \phi(\lambda)+\kappa_2$, 
which contradicts  $\lim_{n\rightarrow \infty}\tau_n=\phi(\lambda)$.

\subsubsection{$N\rightarrow \infty$} 
Fix $\beta \in (0,1)$, $\lambda>1$,  $\tau>\phi(\lambda)$, and a diverging sequence $\{N\}$ of population sizes. For each $N$ and each equilibrium given $(\beta,\lambda,\tau, N)$, define the sets $A_N$, $B_N$, $C_N$, and $D_N$ as in the previous subsection. We claim that for sufficiently large $N$, no player strictly mixes, or equivalently $A_N\cup B_N =\emptyset$. 

Suppose this claim is false. Then for each $N$, there exists $n(N)>N$ such that $A_{n(N)}\cup B_{n(N)} \neq \emptyset$. In what follows, we write $n(N)$ simply as $n$. Consider two cases. 

\vspace{-10pt}

\paragraph{Case 1: $|A_n|$, $|B_n|\geq 1$.}  We already solved $(X_{i}, X_j)$ for $(i,j)\in  A_n \times B_n$ in Case 1 of the previous section. An inspection of their expressions reveals that $X_{i}-X_{j}=2\log \beta$, so either $X_{i}\leq \log\beta$ or $-X_{j}\leq \log\beta$. To avoid repetition, we shall only treat the case where $X_{i}\leq \log\beta$.

For all $i \in A_n$,
\begin{align*}
\log\beta\geq X_i =f(\chi_{i}^+)-f(\chi_{i}^-) \geq f(\chi_{i}^+) \\
\implies \chi_i^+ \geq \frac{1}{\lambda}+ \text{ a positive constant depending on } (\lambda, \beta)\\
\implies g(\chi_i^+)\geq \text{ a positive constant depending on } (\lambda,\beta).
\end{align*}
Summing the bandwidth constraints across $i \in A_n$ gives 
\[|A_n|\tau\geq \sum_{i \in A_n}\sum_{k \in A_n \setminus \{i\}} g(\chi_k^+) \geq |A_n|(|A_n|-1) \cdot \text{ a positive constant depending on } (\lambda,\beta).\]
Consequently, $A_n$ remains bounded as $n\rightarrow \infty$. Denoting $\tilde{A}_n= I_+\setminus A_n$, $|\tilde{A_n}|\rightarrow \infty$ as $n\rightarrow \infty$. 
By definition, $\tilde{A}_n\subseteq C_n \cup D_n$, so either $|C_n|\to\infty$, or $|C_n|$ remains bounded and $|D_n|\to\infty$.

Suppose $|C_n|\rightarrow \infty$. A line-by-line replication of Lemma \ref{lem_peersymmetry} shows that members of $C_n$ must play symmetric strategies. Let $y_n>0$ denote the attention that each $k \in C_n$ pays to the primary source.
If $y_n>\phi(\lambda)$, then
$(|C_n|-1)h(y_n,\lambda)\leq\tau$, so
$h(y_n,\lambda)=O(1/|C_n|)$. The Taylor-expansion argument in the proof of
Lemma \ref{lem_ece} then gives
$$y_n-\phi(\lambda)=O(1/|C_n|)
\quad \text{and} \quad
y_n-\phi(\lambda)-h(y_n,\lambda)=O(1/|C_n|^2).$$
Writing $q_n=1-\exp(-y_n)$, $q_n\geq 1/\lambda$ if and only if $y_n\geq \phi(\lambda)$, and moreover 
\begin{align*}
f(q_n)= -[y_n-\phi(\lambda)-h(y_n,\lambda)]\mathbbm{1}_{y_n\geq \phi(\lambda)}=O\left(\frac{1}{|C_n|^2}\right).
\end{align*}
Summing over $k \in C_n$ yields
\[S_{C_n}=\sum_{k \in C_n} f(\chi_{k}^+)-f(\chi_{k}^-)=\sum_{k \in C_n} f(\chi_{k}^+)-0=|C_n|f(q_n)=O\left(\frac{1}{|C_n|}\right)=o(1).\]

Likewise, $S_{D_n}=o(1)$ for large $n$ if $|D_n|\rightarrow \infty$.
The remainder of the proof distinguishes between two cases: $|C_n|\rightarrow \infty$ vs. $|D_n|\rightarrow \infty$.

\vspace{-10pt}

\subparagraph{Subcase (a): $|C_n|\rightarrow \infty$.} 
Fix a pair $(i,j) \in A_n \times  C_n$. When $i$ seeks the $+1$-biased message, his bandwidth constraint is 
\[\tau=x_i^+ + \sum_{k \in A_n \cup B_n\cup C_n\setminus \{i\}} g(\chi_k^+).\]
Comparing it with $j$'s bandwidth constraint: 
\[\tau=y_n+ \sum_{k \in A_n \cup B_n \cup C_n \setminus \{j\} } g(\chi_k^+)\]
gives 
\begin{align*}
 g(\chi_i^+)-x_i^++\phi(\lambda)=g(q_n)-y_n+\phi(\lambda).
\end{align*}
In the above equation
\[\text{LHS}\leq g(\chi_i^+)+\log(1-\chi_i^+)+\phi(\lambda)=f(\chi_i^+),\]
where the inequality follows from $\chi_i^+\leq 1-\exp(-x_i^+)$, and the equality uses the identity of $f$ and $g$ and the fact that $\chi_i^+ >1/\lambda$. Meanwhile,
\[\text{RHS}=\begin{cases}
f(q_n)=O\left(1/|C_n|^2\right) &  \text{ if } y_n\geq \phi(\lambda),\\
\phi(\lambda)-y_n> 0 & \text{ if } y_n< \phi(\lambda).
\end{cases}
\]
Since $f\leq 0$,  $y_n\geq \phi(\lambda)$ is the only possible case, in which
\[0\geq f(\chi_i^+)\geq f(q_n)=O\left(\frac{1}{|C_n|^2}\right)=o(1). \]
However, this contradicts $f(\chi_i^+)\leq \log\beta<0$, hence subcase (a) is impossible.

\vspace{-10pt}
\subparagraph{Subcase (b): $|C_n|$ is bounded and $|D_n|\rightarrow \infty$.} Recall that 
\[X_{j}\equiv -\dfrac{(2|A_n|-1)\log\beta+S_{C_n}+S_{D_n}}{|A_n|+|B_n|-1}\quad \forall n \text{ and } j \in B_n. \]
As $n\rightarrow \infty$, $S_{D_n}$ vanishes, $S_{C_n}$ is bounded and weakly negative, and $A_n$ is nonempty and bounded. Therefore, the asymptotic behavior of $X_j$ is determined by that of $B_n$.

$\bullet$ If $|B_n|$ stays bounded, then $\liminf_{n \rightarrow \infty}X_j>$ a strictly positive constant. Repeating the argument for $(i,j) \in A_n \times C_n$ in Subcase (a) for a pair $(i,j) \in B_n \times D_n$ will give rise to a similar contradiction. 

$\bullet$ Suppose $|B_n|\rightarrow \infty$. Summing over the bandwidth constraints across $k \in B_n$ gives 
\begin{align*}
|B_n|\tau\geq \sum_{k \in B_n}\sum_{l \in B_n \setminus \{k\}} g(\chi_l^-)=\left(|B_n|-1\right)\sum_{k \in B_n}g(\chi_k^-) 
\geq |B_n|(|B_n|-1) \min_{k \in B_n} g(\chi_k^-),
\end{align*}
and consequently 
\[\min_{k \in B_n} g(\chi_k^-)=O\left(\frac{1}{|B_n|}\right)=o(1).\]
Let $i \in B_n$ attain this min. Then $f(\chi_i^-)=o(1)$. 

Since $|C_n|$ is bounded while $|\tilde A_n|\to\infty$ and
$\tilde A_n\subseteq C_n\cup D_n$, 
$|\tilde A_n\cap D_n|\to\infty$. Fix
$j\in\tilde A_n\cap D_n$.
For $k \in \{i,j\}$, let $L_k^+$ and $L_k^-$ denote the minimum log-disruption probabilities when player $k$ seeks the $+1$-biased message and the $-1$-biased message, respectively, holding his opponents' strategies as in equilibrium. Since $j$ is $+1$-biased and seeks the $-1$-biased message in equilibrium,
$
L_j^-\leq \log\beta+L_j^+.
$
Since $i$ is $-1$-biased and strictly mixes in equilibrium,  
$
L_i^+=\log\beta+L_i^-.
$

Compare the optimization problems defining $L_i^+$ and $L_j^+$. When seeking the $+1$-biased message, $i$ faces: $A_n$, $B_n \setminus\{i\}$, and $C_n$, as potential secondary sources. By contrast, $j$ faces $A_n$, $B_n$, and $C_n$. Therefore, $L_j^+\leq L_i^+$, and consequently 
\[
L_j^-
\leq
\log\beta+L_j^+
\leq
\log\beta+L_i^+
=
2\log\beta+L_i^-.\]

Next,  compare the optimization problems defining $L_i^-$ and
$L_j^-$. When seeking the $-1$-biased message, $i$ faces: $A_n$, $B_n\setminus\{i\}$, and $D_n$ as potential secondary sources. By contrast, $j$ faces $A_n$, $B_n$, and $D_n \setminus \{j\}$. Recall that 
$
f(\chi_i^-)=o(1)
$ and $f(\chi_j^-)=O(1/|D_n|)$. Therefore, $L_i^- - L_j^- =o(1)$, which contradicts the above conclusion that $L_i^- - L_j^- \geq -2\log\beta> 0$.  Therefore, subcase (b) is also impossible. 

\vspace{-10pt}
\paragraph{Case 2: $|A_n|=0$ or $|B_n|=0$.}
By symmetry, suppose that $B_n=\emptyset$. Since
$A_n\cup B_n\neq\emptyset$, we have $A_n\neq\emptyset$. Moreover,
all type-$-1$ players use pure strategies, so
$
I_-=(I_-\cap C_n)\cup(I_-\cap D_n).
$
Because $|I_-|=n$, at least one of the following holds:
$$
|I_-\cap C_n|\geq\frac n2
\qquad\text{or}\qquad
|I_-\cap D_n|\geq\frac n2.
$$
Consider two subcases.

\vspace{-10pt}
\subparagraph{Subcase (a): $|I_-\cap C_n|\geq n/2$.}
In this case, $|C_n|\to\infty$. The same manipulation as in Case 1-(a) shows that 
\[f(\chi_k^+)\equiv f(q_n)=O\left(\frac{1}{|C_n|^2}\right) \quad \forall k \in C_n \]
and 
\[0 \geq f(\chi_k^+)\geq f(q_n) \quad \forall k \in A_n.\]
Thus for each $i\in I_-\cap C_n$, we can bound his equilibrium log-disruption probability from seeking the $+1$-biased message  below as   
\begin{align*}
L_i^+=-\tau + \sum_{k \in A_n\cup C_n \setminus \{i\}} f(\chi_k^+)\geq -\tau+  (|A_n|+|C_n|-1)f(q_n)=-\tau +O\left(\frac{1}{|C_n|}\right),
\end{align*}
where the inequality uses the fact that $A_n\subseteq I_+$, hence $|A_n|\leq n\leq 2|C_n|$. 
If $i$ deviates to seeking the $-1$-biased message, then $L_i^-\leq -\tau$. For large $n$,
$L_i^+>L_i^-+\log\beta,$
hence the deviation strictly lowers $i$'s expected loss. Therefore, subcase (a) is impossible.

\vspace{-10pt}
\subparagraph{Subcase (b): $|I_-\cap D_n|\geq n/2$.}
In this case, $|D_n|\to\infty$. Relabeling, we obtain that 
\[f(\chi_k^-)\equiv f(q_n)=O\left(\frac{1}{|D_n|^2}\right) \quad \forall k \in D_n, \quad 0 \geq f(\chi_k^-)\geq f(q_n) \quad \forall k \in A_n.\]
In equilibrium, the log-disruption probability faced by each $i\in A_n$ from seeking the $-1$-biased message satisfies  
\begin{align*}
L_i^-&= -\tau + \sum_{k \in A_n\cup D_n \setminus \{i\}} f(\chi_k^-) \geq -\tau +O\left(\frac{1}{|D_n|}\right).
\end{align*}
If $i$ deviates to seeking the $+1$-biased message, then $L_i^+\leq -\tau$. For large $n$,
$L_i^++\log \beta < L_i^-,$
hence the deviation strictly lowers $i$'s expected loss. Therefore, subcase (b) is also impossible.

\vspace{8pt}
By now, we have exhausted all possible cases and demonstrated that none of them can arise asymptotically as $n\rightarrow \infty$. This completes the proof.

\section{Proof of Theorem \ref{thm_tau}}\label{sec_proof_thm2}

\paragraph{Preliminaries.} Throughout, we fix the players' defaults and parameterize the game by $\theta \coloneqq (\theta_i)_{i \in I}$, where $\theta_i\coloneqq (\beta_i,\lambda_i,\tau_i)$. Let the parameter space $\Theta$ be a  topological space in $\mathbb{R}^{3|I|}$, and denote the set of equilibria at $\theta$---pure or mixed---by $\mathcal{E}(\theta)$. Theorem \ref{thm_ece} essentially identifies parameters $\theta_0 \in int(\Theta)$ such that $\mathcal{E}(\theta_0)=\{\text{ECE}(\theta_0)\}$ and moreover $\mathrm{ECE}(\theta_0)$ is a strict equilibrium. Parts~A and B of Theorem \ref{thm_tau} examine properties of the set $\mathcal{E}(\theta)$ for $\theta$ near $\theta_0$.

It is well known that in games like ours (strategy spaces are compact and convex and continuous in the parameter, payoffs are continuous in strategies), the set of equilibria is upper-hemicontinuous (uhc) in the parameter. That is, for every neighborhood $U$ of $\mathcal{E}(\theta_0)$, there exists a neighborhood $V$ of $\theta_0$ such that $\mathcal E(\theta)\subseteq U$ for every $\theta\in V$.
See, for example, the textbook treatment in \cite{aliprantis2006infinite}. Building on this result, we demonstrate that for parameters sufficiently near $\theta_0$, the game has a unique equilibrium, which remains an ECE. 

\vspace{-10pt}
\paragraph{Step 1. Show that every equilibrium is an ECE.} For $d \in \{-1, +1\}$, let $C_i^d\coloneqq \{d\text{-biased}\}\cup  I \setminus \{i\}$  denote the set of sources that player $i$ may benefit from attending to, given $d$ as the intended decision upon receiving no message.    $A_i^d(\theta)\coloneqq \{\boldsymbol{a}_i \in \mathbb{R}_+^{|C_i^d|}: \sum_{c \in C_i^d} a_i^c \leq \tau_i\}$ is the simplex of feasible pure attention allocations associated with $d$. 
Let $V_i^d(\theta, \sigma_{-i})$ denote $i$'s minimum expected loss from choosing $d$, given the opponents' (possibly mixed) strategies $\sigma_{-i}$, i.e., 
\[V_i^d(\theta, \sigma_{-i})\coloneqq \min_{\boldsymbol{a}_i \in A_i^d(\theta)}L_i^d\left(\boldsymbol{a}_i; \theta, \sigma_{-i}\right),\] 
Since $\mathrm{ECE}(\theta_0)$ is a strict pure-strategy equilibrium, 
\[V_i^{d_i^*}(\theta_0, \delta_{\boldsymbol{a}_{-i}^*})<V_i^{-d_i^*}(\theta_0, \delta_{\boldsymbol{a}_{-i}^*}).\]
Moreover, $V_i^d$ is continuous in $(\theta,\sigma_{-i})$ by Berge's maximum theorem, and $A_i^d(\theta)$ is continuous in $\theta$ by construction. Thus for $(\theta,\sigma)$ sufficiently near $(\theta_0, \delta_{\boldsymbol{a}^*_{-i}})$, 
\[V_i^{d_i^*}(\theta,\sigma_{-i})<V_i^{-d_i^*}(\theta, \sigma_{-i}).\]
The uhc of $\mathcal{E}(\theta)$ then implies that for $\theta$ sufficiently near $\theta_0$, $d_i^*$ remains the unique optimal decision in every equilibrium. Repeating this argument for all players, we conclude that for $\theta$ sufficiently near $\theta_0$, every equilibrium is an ECE and, consequently, a pure-strategy equilibrium.

\vspace{-10pt}
\paragraph{Step 2. Show that the ECE is unique.} We analyze the set of pure-strategy equilibria among type $+1$ players, denoting it by   $\mathcal{E}(\theta)$ with a slight abuse of notation. 

By Lemma \ref{lem_general}, $\mathcal{E}(\theta)$ consists of the fixed points of $\psi(\cdot; \theta): \prod_{i \in I_+}[0,\tau_i] \rightarrow \prod_{i \in I_+}[0,\tau_i]$, where 
\[\psi_i(\boldsymbol{x};\theta)=[\tau_i-\sum_{j \in I_+\setminus \{i\}} h^+(\boldsymbol{x}_j,\lambda_j)]^+.\]
In words, $\psi(\cdot;\theta)$ maps each vector $\boldsymbol{x}$ of primary-source attention to the residual capacities after deducting the optimal peer attention given $\boldsymbol{x}$ from the bandwidths. Denoting $\Psi(\boldsymbol{x};\theta)\coloneqq \psi(\boldsymbol{x};\theta)-\boldsymbol{x}$, 
\[\mathcal{E}(\theta)=\{\boldsymbol{x} \in \prod_{i \in I_+}[0,\tau_i]: \Psi(\boldsymbol{x}; \theta)=0\},\]
which is nonempty by Brouwer's fixed point theorem.

The uhc of $\mathcal{E}(\theta)$ in $\theta$ says that for every neighborhood $U$ of $\mathcal{E}(\theta_0)=\{\boldsymbol{x}_0\}$, there exists a neighborhood $V(U)$ of $\theta_0$ such that $\mathcal{E}(V(U)) \subseteq U$. On the other hand, we have shown that $\boldsymbol{x}_0=x(N)\boldsymbol{1}_N>  \phi(\lambda)\boldsymbol{1}_{N}$, so $\Psi$ is smooth in $(\boldsymbol{x},\theta)$ in a small neighborhood of $(\boldsymbol{x}_0,\theta_0)$, with $\nabla_{\boldsymbol{x}}\Psi=-M$ non-singular.  By the implicit function theorem, there exist neighborhoods $W$ of $\boldsymbol{x}_0$ and $V'$ of $\theta_0$ such that for all $\theta\in V'$, $\Psi(\boldsymbol{x};\theta)=0$ has a unique solution in $W$, denoted by $\boldsymbol{x}(\theta)$. Let $U \subseteq W$, it follows that $\mathcal{E}(\theta)=\{\bm x(\theta)\}$ for all $\theta \in V(U) \cap V'$.

\vspace{-10pt}
\paragraph{Comparative statics of visibility.} In Section \ref{sec_proofsketch2}, we already examined the equilibrium and welfare consequences of perturbing $\tau_i$ for a single $i\in I_+$. Here, we consider a small perturbation of $\lambda_i$. 

In a small neighborhood of $\theta_0$, the vector $\boldsymbol{x}$ of equilibrium primary-source attention satisfies  
\[
x_i + \sum_{j \in I_+ \setminus \{i\}} h(x_j, \lambda_j) = \tau_i \quad \forall i \in I_+.
\]
We suppress the dependence of $\boldsymbol{x}$ on $\theta$ because the context is clear. Taking derivatives w.r.t. $\lambda_1$ yields, in the matrix form: 
\[
\frac{\partial \boldsymbol{x}} {\partial \lambda_1} = -h_\lambda(x_1, \lambda_1) M^{-1}(\boldsymbol{1}_N - \boldsymbol{e}_1),
\]
where $M^{-1}$ is as solved in Section \ref{sec_proofsketch2}, and $\boldsymbol{e}_1$ is a basis vector. Further algebraic manipulation yields
\[
M^{-1}(\boldsymbol{1}_N - \boldsymbol{e}_1) = \diag\left(\frac{1}{1 - g_1}, \dots, \frac{1}{1 - g_N}\right) \left( \frac{\displaystyle \frac{g_1}{1 - g_1}}{1 + \displaystyle \sum_{i=1}^N \frac{g_i}{1 - g_i}} \boldsymbol{1}_N - \boldsymbol{e}_1 \right),
\]
where the first coordinate of the right-hand-side vector is negative, whereas the remaining coordinates are positive. It follows that 
\[\sgn\left(\frac{\partial x_1}{\partial\lambda_1}\right)=\sgn \left(h_{\lambda}(x_1,\lambda_1)\right), \quad \sgn\left(\frac{\partial x_i}{\partial \lambda_1}\right)=-\sgn \left(h_{\lambda}(x_1,\lambda_1)\right) \quad \forall i \in I_+\setminus \{1\},\]
where $h_{\lambda}(x_1,\lambda_1)$ can be either positive, negative, or zero by Observation \ref{obs_function}. 

Substituting into peer attention gives
\[
\frac{d}{d\lambda_1}h(x_1,\lambda_1)
=
h_x(x_1,\lambda_1)\frac{\partial x_1}{\partial\lambda_1}
+h_\lambda(x_1,\lambda_1), \quad \frac{d}{d\lambda_1}h(x_i,\lambda_i)=h_x(x_i,\lambda_i)\frac{\partial x_i}{\partial\lambda_1} \quad \forall i \neq 1,
\]
where the first term has the same sign as $h_{\lambda}(x_1,\lambda_1)$, and the second term has the opposite sign to $h_{\lambda}(x_1,\lambda_1)$. Taken together, we obtain the three scenarios stated in Part B of the theorem. 

So far we have not imposed symmetry on either the environment or the equilibrium before the perturbation. The remainder of the proof imposes symmetry on both and examines the change in the total expected loss inside $I_+$ following the perturbation.

As shown in Section \ref{sec_proofsketch2}, the log-disruption probability for each $i \in I_+$ equals 
\[-X+\sum_{j \in I_+\setminus \{i\}} \phi(\lambda_j) \quad \text{where}\quad X=\sum_{j \in I_+} x_j.\]
  At the symmetric ECE, $g_i\equiv g$, hence  
\[\frac{\partial X}{\partial \lambda_1} = \boldsymbol{1}_N^{\top} \frac{\partial \bm x}{\partial \lambda_1}=
-\frac{(N-1)h_\lambda(x,\lambda)}{1+(N-1)g}.
\]
Let $L_i$ denote $i$'s expected loss, 
\begin{align*}
L_i
&=\frac{\beta}{2}
\exp\left(
-X+\sum_{j\in I_+\setminus\{i\}}\phi(\lambda_j)
\right),\\
\frac{\partial L_i}{\partial\lambda_1}
&=-\frac{\beta}{2}
\exp\left(
-X+\sum_{j\in I_+\setminus\{i\}}\phi(\lambda_j)
\right)
\left[
\frac{\partial X}{\partial\lambda_1}
-\mathbbm{1}_{\{i\ne1\}}\phi'(\lambda_1)
\right],\\
\left.
\frac{\partial}{\partial\lambda_1}
\sum_{i\in I_+}L_i
\right|_\text{sym ECE}
&=\frac{\beta}{2}e^{-Nx+(N-1)\phi(\lambda)}(N-1)
\left[
\frac{Nh_\lambda(x,\lambda)}{1+(N-1)g}
+\phi'(\lambda)
\right].
\end{align*}
Tedious algebra shows that 
\[\frac{Nh_\lambda(x,\lambda)}{1+(N-1)g}+
\phi'(\lambda)=
\frac{(N-1)(1-g)-N(\lambda-1)h(x,\lambda)}
{\lambda(\lambda-1)[1+(N-1)g]}<0,\]
hence
\[\left.
\frac{\partial}{\partial\lambda_1}
\sum_{i\in I_+}L_i
\right|_\text{sym ECE}<0.\]

\vspace{-10pt}
\paragraph{Proof of Proposition \ref{prop_population}.} Inside a chamber of size $n$, the log-disruption probability for each member equals \[-nx(n)+(n-1)\phi(\lambda)=-n\left[x(n)-\phi(\lambda)\right]+\phi(\lambda).\]
Thus the proposition holds if for sufficiently large $n$, the derivative of the log-disruption probability w.r.t. $n$, 
$-[x(n)-\phi(\lambda)-n x'(n)]$, is positive. 

Since
$
x(n)+(n-1)h(x(n),\lambda)=\tau,
$
taking the total derivative with respect to $n$ yields
\[
\left[1+(n-1)h_x(x(n),\lambda)\right]x'(n)
+h(x(n),\lambda)=0,
\]
and hence
\[
x'(n)=-\frac{h(x(n),\lambda)}
{1+(n-1)h_x(x(n),\lambda)}.
\]
Write $\Delta(n)=x(n)-\phi(\lambda)$ and recall that
$\Delta(n)=O(1/n)$ as $n\to\infty$. Taylor expansion gives
\begin{align*}
h(x(n),\lambda)
&=\Delta(n)
+\frac12 h_{xx}(\phi(\lambda),\lambda)\Delta^2(n)
+o(\Delta^2(n)),\\
h_x(x(n),\lambda)
&=1+h_{xx}(\phi(\lambda),\lambda)\Delta(n)
+o(\Delta(n)),
\end{align*}
where we use $h(\phi(\lambda),\lambda)=0$ and
$h_x(\phi(\lambda),\lambda)=1$. Therefore, for sufficiently large $n$, 
\begin{align*}
x(n)-\phi(\lambda)+nx'(n)
&=\frac{
\Delta(n)\left[1+(n-1)h_x(x(n),\lambda)\right]
-nh(x(n),\lambda)}
{1+(n-1)h_x(x(n),\lambda)}\\
&=\frac{
\frac{n-2}{2}h_{xx}(\phi(\lambda),\lambda)\Delta^2(n)
+o(n\Delta^2(n))}
{n[1+o(1)]}\\
&=\frac12 h_{xx}(\phi(\lambda),\lambda)\Delta^2(n)
+o(\Delta^2(n))
<0,
\end{align*}
where the last line uses $h_{xx}(\phi(\lambda),\lambda)<0$. The desired claim follows. \qed 

\vspace{-10pt}

\paragraph{Proof of Corollary \ref{cor_merger}.} With a mega source, $C_i=\{\text{mega}\}\cup I\setminus \{i\}$, $\bm a_i=(a_i^c)_{c \in C_i}$, and $A_i=\{\bm{a}_i \in \mathbb{R}_+^{2N}: \sum_{c \in C_i} a_i^c \leq \tau\}$. For a fixed $\bm{a}$, let $D$ denote the probability that all the information transmission paths from the mega source to player $i$ are disrupted. Following no message, $i$'s posterior remains at the prior: 
\[\mathbb{P}(\omega=1 \mid \text{no message}, \bm a)=\frac{\frac{1}{2} D}{\frac{1}{2} D+\frac{1}{2}D}=\frac{1}{2}.\]
Therefore, he chooses the default as in the ECE. 

Let $x_i$ denote $i$'s attention to the mega source. Since this attention enables $i$ to disseminate both types of messages, it follows from Lemma \ref{lem_general} that every other player $j \in I \setminus \{i\}$ directs 
$\frac{1}{\gamma_j
} h^+(x_i, \lambda/\gamma_j)$ units of attention to $i$. The vector $(x_i)_{i \in I}$ of primary-source attention can be obtained by solving the ECE primary-source attention in a chamber of $|I|=2N$ members. By Lemmas \ref{lem_general} and \ref{lem_peersymmetry}, every player pays $x(2N)$ units of attention to the mega source, $h(x(2N),\lambda)$ units of attention to every other player, and faces a log-disruption probability of $-\tau-(2N-1)\xi(2N)$. Now apply Proposition \ref{prop_population}. \qed

\section{Proof of Theorem \ref{thm:efficiency}}\label{sec_proof_thm3}
\paragraph{Dual representation.} We develop a dual representation of the efficiency program. 

As noted in Section \ref{sec_efficient}, attention to other players generates only private benefits but no social benefit. Fixing primary-source attention $(x_i^+, x_i^-)_{i \in I}$ (the superscripts indicate primary sources and should not be confused with positive and negative parts), the efficiency program reduces to the minimization problem solved in the proof of Lemma \ref{lem_general}. Let $\gamma_i$ denote the Lagrange multiplier on player $i$'s bandwidth constraint. 

In the case where $d_i=+1$ following no message, $i$'s attention to each  $j \in I\setminus \{i\}$ equals 
\[\frac{1}{\gamma_i}h^+\left(x_j^{+},\frac{\lambda}{\gamma_i}\right).
\]
His bandwidth constraint can be expressed as 
\[
\frac{1}{\gamma_i}\sum_{j \in I\setminus \{i\}} h^+\left(x_j^{+},\frac{\lambda}{\gamma_i}\right)= \tau_i-x_i^{+}-x_i^{-},\]
where the left-hand side denotes the total attention to other players, and the right-hand side denotes the residual capacity after deducting primary-source attention from the bandwidth. 

$\bullet$ When $\tau-x_i^+-x_i^->0$, Observation~\ref{obs2} implies that the equation holds for a unique
\[\gamma_i<
\frac{\lambda}{\phi^{-1}\!\left(\max_{j \in I \setminus \{i\}} x_j^+\right)}.\]


$\bullet$ When $\tau-x_i^+-x_i^- =0$, the equation holds for an interval of $\gamma_i$: \
$$
\left[
\frac{\lambda}
{\phi^{-1}\left(\max_{j\in I\setminus\{i\}}x_j^+\right)},
\lambda
\right].
$$
In such cases, we select the minimum $\gamma_i$ in this interval.

In both cases, $\phi(\lambda/\gamma_i)\leq \max_{j \in I \setminus \{i\}} x_j^+$ holds by construction.

Substituting the above equation into Part (ii) of Lemma \ref{lem_general} yields player $i$'s log-disruption probability as 
\[-(\tau-x_i^{-})-\sum_{j \in I\setminus \{i\}}\left(\left[x_j^{+}-\phi\left(\frac{\lambda}{\gamma_i}\right)\right]^+ - \frac{1}{\gamma_i}h^+\left(x_j^{+},\frac{\lambda}{\gamma_i}\right)\right). \]
 The term $-(\tau-x_i^{-})$ represents the log-disruption probability in a benchmark where a  social planner deducts  $x_i^{-}$  from $i$'s bandwidth and directs the residual capacity to the $+1$-biased primary source only. If, instead, the planner optimally allocates the residual capacity among the $+1$-biased primary source and $j \neq i$, the resulting reduction in log-disruption probability is captured by the summation term. 

The social planner chooses a vector $\left(x_i^{+},x_i^{-}\right)_{i \in I}$ of primary-source attention to minimize the total expected loss across players. Given this vector, $(\gamma_i)_{i \in I}$ is determined as above and in turn determines the  attention network among players.

\vspace{-10pt}

\paragraph{Proof of Part (i).} Denote $\tau-\phi(\lambda)=\Delta$, and suppose that $d_i=+1$ under an efficient allocation.  Parameterize $\gamma_i$ by $\Delta$, and recall that 
$\max_{j \neq i} x_j^{+} \geq \phi\left(\frac{\lambda}{\gamma_i(\Delta)}\right)$ holds by construction. Together with the upper bound $x_j^+\leq \tau=\phi(\lambda)+\Delta$, this implies $\gamma_i(\Delta)=1+O(\Delta)$, and consequently
\begin{align*}
&\left[x_j^{+}-\phi\left(\frac{\lambda}{\gamma_i(\Delta)}\right)\right]^+ - \frac{1}{\gamma_i(\Delta)}h^+\left(x_j^{+}, \frac{\lambda}{\gamma_i(\Delta)}\right)   \\ 
&=\left[x_j^{+}-\phi\left(\frac{\lambda}{\gamma_i(\Delta)}\right)\right]^+-\frac{1}{1+O(\Delta)}\left\{\left[x_j^{+}-\phi\left(\frac{\lambda}{\gamma_i(\Delta)}\right)\right]^++O(\Delta^2)\right\},\\
&=O(\Delta^2).
\end{align*}
Summing this result over $j \neq i$ gives $-(\tau-x_i^{-})-O(\Delta^2)$ as the log-disruption probability for player $i$. Compared to attending to the $+1$-biased source alone, $i$ benefits $O(\Delta^2)$ from other players' attention to the $+1$-biased source, while paying a constant marginal cost for every unit of attention directed to the $-1$-biased source. By symmetry, the rest of society benefits $O(\Delta^2)$ from $x_i^-$, and efficiency requires $x_i^-=O(\Delta^2)$. The remainder of the proof follows routinely. \qed

\vspace{-10pt}
\paragraph{Proof of Part (ii).} The argument in Section \ref{sec_efficient} didn't solve for the fully symmetric efficient allocation. We now complete this omitted step.

We begin by defining several notions of symmetry. We adopt the convention that negating a primary source reverses its bias. Label \(I_+=\{1,\cdots,N\}\) and \(I_-=\{-N,\cdots,-1\}\). Define $C_+\coloneqq \{+1\text{-biased}\}\cup I_+$, $C_-\coloneqq \{-1\text{-biased}\}\cup I_-$, and $C\coloneqq C_+\cup C_-$.

\begin{defn}
    An attention allocation is: 
\begin{description}
\item[Symmetric across types] if \( x_i^c = x_{-i}^{-c} \) for all \( i \in I \) and \( c \in C \setminus \{i\} \).
\item[Symmetric within types] if all players of the same type allocate identical amounts of attention to their own-biased primary source, the opposite-biased primary source, each other same-type player, and each opposite-type player.
\item[Fully symmetric] if it is symmetric both across and within types. 
\end{description} 
\end{defn}

The proof proceeds in two steps.
\vspace{-10pt}

\subparagraph{Step 1.} Show that every cross-type-symmetric efficient allocation exhibits echo chambers. 

\begin{lem}\label{lem_efficient_mirror}
   Every cross-type-symmetric efficient allocation satisfies $x_{i}^{c}=0$ for all $(i,c) \in \left(I_+\times C_-\right) \cup \left(I_-\times C_+\right)$. 
\end{lem}

\begin{proof}
Under cross-type symmetry, efficiency requires that each player choose his default given no message. 
The expected loss faced by player $i \in I_+$ therefore equals
\[\frac{\beta}{2}\;\exp\left(-x_i^{+}-\sum_{j \in I \setminus \{i\}} \left[x_j^{+}-\phi\left(\frac{\lambda}{\gamma_i}\right)\right]^+\right).\]
We claim that $x_i^{-}=0$ holds under every cross-type-symmetric efficient allocation. 

Suppose, toward a contradiction, that $x_i^{-}>0$ under such an allocation.  Consider a perturbation that reduces $x_i^{-}$ by a small $\epsilon>0$ and increases $x_i^{+}$ by $\epsilon$. We examine the impact of this perturbation on each player.

Consider first the impact on player $i$. Since $i$'s total attention allocation to the primary sources is unaffected by the perturbation, $\gamma_i$ and hence $\phi(\lambda/\gamma_i)$ remain unchanged. The change in $i$'s expected loss is simply
$L_i\epsilon,$
where $L_i$ denotes his expected loss before the perturbation.

Next, consider \( j \in I_+ \setminus \{i\} \) such that \( x_i^{+} > \phi(\lambda/\gamma_j) \). That is, player \( j \) allocates positive attention to \( i \) before the perturbation. 
If we were to fix the original $\gamma_j$, then the increase in $x_i^+$ would raise $j$'s attention to $i$. 
To continue satisfying \( j \)’s bandwidth constraint, $\gamma_j$ must change by $\Delta_{\gamma_j}$ such that 
 \[
h_x\!\left(x_i^{+}, \frac{\lambda}{\gamma_j}\right)\,\epsilon 
+ \sum_{\substack{k \neq j: \\ x_k^{+} > \phi(\lambda/\gamma_j)}} \left[
   \frac{d}{d\gamma_j} h\!\left(x_k^{+}, \frac{\lambda}{\gamma_j}\right) 
   - \frac{1}{\gamma_j} h\!\left(x_k^{+}, \frac{\lambda}{\gamma_j}\right) 
  \right] 
\Delta_{\gamma_j} = 0.
\]
Straightforward algebra shows that for each $k$ as above,
\[\frac{d}{d\gamma_j} h\!\left(x_k^{+}, \frac{\lambda}{\gamma_j}\right) =\frac{1}{\gamma_j}h\left(x_k^{+}, \dfrac{\lambda}{\gamma_j}\right)-\dfrac{d}{ d \gamma_j }\phi\left(\dfrac{\lambda}{\gamma_j}\right). \]
Therefore,
\[
h_x\left(x_i^{+}, \frac{\lambda}{\gamma_j} \right)\; \epsilon = n_j \cdot \frac{d}{d\gamma_j} \phi\left(\frac{\lambda}{\gamma_j}\right)\cdot \Delta_{\gamma_j}, 
\]
where \( n_j \coloneqq \#\{k \neq j : x_k^{+} > \phi(\lambda/\gamma_j)\} \). Thus \( \Delta_{\gamma_j} > 0 \), with an intuitive interpretation: as \( i \) becomes more informed, $j$ raises  attention to $i$, as captured by the left-hand-side expression. To continue satisfying $j$'s bandwidth constraint, $\gamma_j$ must increase, which leads $j$ to raise the threshold $\phi(\lambda/\gamma_j)$ used to screen for secondary sources. For sufficiently small \( \epsilon \), the set \( \{k \neq j : x_k^{+} > \phi(\lambda/\gamma_j)\} \) remains unchanged after the perturbation. The impact on \( j \)’s expected loss  equals 
\[
L_j \;\left[ \epsilon - n_j \cdot \frac{d}{d\gamma_j} \phi\left( \frac{\lambda}{\gamma_j} \right) \cdot \Delta_{\gamma_j} \right] = L_j \; \left[1 - h_x\left(x_i^{+}, \frac{\lambda}{\gamma_j}\right) \right] \; \epsilon,
\]
where the coefficient in front of \( \epsilon \) on the right-hand side is positive but less than \( L_j \).

Next, consider $j \in I_+\setminus \{i\}$ such that $x_i^{+}\leq \phi(\lambda/\gamma_j)$. That is, $j$ ignores $i$ before the perturbation. After the perturbation, $j$ may or may not ignore $i$. Regardless, $j$'s expected loss decreases weakly.

Finally, consider $j \in I_-$. If $x_i^{-}\leq \phi(\lambda/\gamma_j)$ before the perturbation, then this remains true afterwards. If $x_i^{-} >\phi(\lambda/\gamma_j)$, then for sufficiently small $\epsilon$, the perturbation does not change the set $\{k \neq j: x_k^{-}>\phi(\lambda/\gamma_j)\}$, whose size is again denoted by $n_j$. Repeating the argument above yields
 \[L_j \;\left[1-h_x\left(x_i^{-}, \frac{\lambda}{\gamma_j}\right)\right] \; (-\epsilon)\]
as the change in $j$'s expected loss.

Taking the sum over all players gives $\epsilon$ times 
\begin{align*}
L_i+\sum_{j \in I_+\setminus \{i\}: \ x_i^{+}>\phi(\lambda/\gamma_j)} L_j \left[1-h_x\left(x_i^{+}, \frac{\lambda}{\gamma_j}\right)\right]\\
-\sum_{j \in I_-: \ x_i^{-}>\phi(\lambda/\gamma_j)} L_j \left[1-h_x\left(x_i^{-}, \frac{\lambda}{\gamma_j}\right)\right] 
\end{align*}
as an upper bound for the change in the total expected loss. Under cross-type symmetry, 
$\gamma_j=\gamma_{-j}$ and $L_j=L_{-j}$ hold for all $j$, and $x_j^+\geq x_j^-$ hold for all $j \in I_+$. Therefore, the preceding expression is strictly less than
\begin{align*}
L_i-L_{-i}+ \sum_{j \in I_+\setminus \{i\}: \ x_i^{+}>\phi(\lambda/\gamma_j)} L_j \left[1-h_x\left(x_i^{+}, \frac{\lambda}{\gamma_j}\right)\right]\\
-\sum_{j \in I_+\setminus \{i\}: \ x_i^{-}>\phi(\lambda/\gamma_j)} L_j \left[1-h_x\left(x_i^{-}, \frac{\lambda}{\gamma_j}\right)\right] \\
\leq L_i-L_{-i}+ \sum_{j \in I_+\setminus \{i\}: \ x_i^{-}>\phi(\lambda/\gamma_j)} L_j  \left[1-h_x\left(x_i^{-}, \frac{\lambda}{\gamma_j}\right)\right]\\
-\sum_{j \in I_+ \setminus \{i\}: \ x_i^{-}>\phi(\lambda/\gamma_j)} L_j \left[1-h_x\left(x_i^{-}, \frac{\lambda}{\gamma_j}\right)\right]\\
= 0.
\end{align*}
Thus the perturbation strictly reduces the total expected loss, which contradicts efficiency before perturbation. 
\end{proof}

\vspace{-10pt}
\subparagraph{Step 2.} Solve the fully-symmetric efficient allocation in the large $N$ limit.

\begin{lem}\label{lem_efficient_full}
As $N \rightarrow \infty$, the log-disruption probability faced by each player converges to  
\[
-\gamma^{\infty} \left[\tau-\phi\left(\frac{\lambda}{\gamma^{\infty}}\right)\right] - \phi\left(\frac{\lambda}{\gamma^{\infty}}\right).
\]
The term $\gamma^{\infty}$ denotes the limiting value of the Lagrange multiplier on each  player’s bandwidth constraint. It solves  
\[
\frac{\gamma - 1}{\lambda - \gamma} + \phi\left(\frac{\lambda}{\gamma}\right) = \tau
\]
and belongs to $(1,\lambda(1-e^{-\tau})]$.
\end{lem}

\begin{proof}
By Lemma \ref{lem_efficient_mirror}, we can parameterize a candidate solution by $(x,y)$ denoting a representative player's attention to his own-biased source and each same-type peer.  We minimize the log-disruption probability faced by a representative player, i.e., 
\begin{align*}
\min_{(x,\, y)} \quad & -x +
(N - 1) \log\big( \exp(-x) + \left(1 - \exp(-x)\right) \exp(-\lambda y) \big) \\
\text{s.t.} \quad & x \geq 0, \quad y \geq 0, \quad x + (N - 1)y \leq \tau.
\end{align*}

Denoting the Lagrange multipliers by $(\eta_x, \eta_y, \gamma)$, the first-order conditions are:
\begin{align*}
\tag{FOC$_x$}
1 + (N - 1) \frac{ \exp(-x) \left[1 - \exp(-\lambda y)\right] }
{ \exp(-x) + \left[1 - \exp(-x)\right] \exp(-\lambda y) }
&= \gamma - \eta_x, \\
\tag{FOC$_y$}
\lambda (N - 1) \frac{ \left[1 - \exp(-x)\right] \exp(-\lambda y) }
{ \exp(-x) + \left[1 - \exp(-x)\right] \exp(-\lambda y) }
&= (N - 1)\gamma - \eta_y.
\end{align*}
Supposing that $x,y>0$, so that $\eta_x=\eta_y=0$. FOC$_x$ then implies  $\gamma>1$, while FOC$_y$ implies $\gamma<\lambda$. 
Solving the FOCs yields
\[
\lambda y = -\log\left(1 - \frac{(\gamma - 1)\lambda}{(N - 1)(\lambda - \gamma)}\right), \quad
x = \log\left(1 + \frac{\gamma}{\lambda - \gamma - \dfrac{(\gamma - 1)\lambda}{N - 1}} \right).
\]
Substituting, the bandwidth constraint reduces to 
\[
\log\left(1 + \frac{\gamma}{\lambda - \gamma - \dfrac{(\gamma - 1)\lambda}{N - 1}} \right)
- \frac{N - 1}{\lambda} \log\left(1 - \frac{(\gamma - 1)\lambda}{(N - 1)(\lambda - \gamma)} \right)
= \tau.
\]

The left-hand side is strictly increasing in $\gamma$ for given $(\tau,\lambda,N)$. Let $\gamma(N)$ denote the unique solution, $\gamma(N)$ is strictly increasing in $N$. Moreover, it is  uniformly bounded above by $\lambda(1-e^{-\tau})<\lambda$, since 
\[\tau\geq x(N)>\log\left(1+\frac{\gamma(N)}{\lambda-\gamma(N)}\right)\implies \gamma(N)<\lambda(1-e^{-\tau}).\]
Therefore, $\gamma(N)\rightarrow \gamma^{\infty}\leq \lambda(1-e^{-\tau})$ as $N\rightarrow \infty$, where $\gamma^{\infty}$ is the unique solution to the limiting equation 
\[
\phi\left(\frac{\lambda}{\gamma}\right)+\frac{\gamma-1}{\lambda-\gamma}
=\tau.
\]
Since the left-hand side tends to $\phi(\lambda)<\tau$ as $\gamma\downarrow 1$, $\gamma^{\infty}>1$. 

For sufficiently large $N$, 
\begin{align*}
x(N)-\phi\left(\frac{\lambda}{\gamma(N)}\right)&=\log\left(1+\frac{\gamma(N)}{\lambda-\gamma(N)-\dfrac{(\gamma(N)-1)\lambda}{N-1}}\right)-\log\left(1+\frac{\gamma(N)}{\lambda-\gamma(N)}\right)\\
&=\left.\frac{d}{d\Delta}\log\left(1+\frac{\gamma (N)}{\lambda-\gamma(N)-\Delta}\right)\right\vert_{\Delta=0}\cdot \frac{(\gamma(N)-1)\lambda}{N-1}+O\left(\frac{1}{N^2}\right)\\
&=\frac{\gamma(N)(\gamma(N)-1)}{\lambda-\gamma(N)}\cdot \frac{1}{N-1}+O\left(\frac{1}{N^2}\right).
\end{align*}
As $N\rightarrow \infty$, 
\begin{align*}
\text{log-disruption probability} & =-N\left[x(N)-\phi\left(\frac{\lambda}{\gamma(N)}\right)\right]-\phi\left(\frac{\lambda}{\gamma(N)}\right)\\
&\rightarrow -\frac{\gamma^{\infty}(\gamma^{\infty}-1)}{\lambda-\gamma^{\infty}}-\phi\left(\frac{\lambda}{\gamma^{\infty}}\right) \\
&=-\gamma^{\infty}\left[\tau-\phi\left(\frac{\lambda}{\gamma^{\infty}}\right)\right]-\phi\left(\frac{\lambda}{\gamma^{\infty}}\right).
\end{align*}
Denote the last expression by $D^{\infty}$: 

$\bullet$ Since $\gamma^{\infty}>1$, 
\[D^{\infty}-(-\tau)=-(\gamma^{\infty}-1)\left[\tau-\phi\left(\frac{\lambda}{\gamma^{\infty}}\right)\right]<0.\]
By contrast, the boundary point $(x,y)=(\tau,0)$ yields a log-disruption probability of $-\tau$ and hence is sub-optimal than $(x(N), y(N))$ for sufficiently large $N$. The other boundary point $(x,y)=(0,\tau/(N-1))$ is clearly sub-optimal. Thus for large $N$, assuming an  interior solution is without loss. 

$\bullet$ Since $\gamma^{\infty}\leq \lambda(1-e^{-\tau})$,  
\[D^{\infty}= -\gamma^{\infty}\tau+(\gamma^{\infty}-1)\phi\left(\frac{\lambda}{\gamma^{\infty}}\right)>-\gamma^{\infty}\tau \geq -\tau \lambda (1-e^{\tau}).\]
The right-hand side is the log-disruption probability per capita under the hypothetical benchmark constructed in Section \ref{sec_efficient}.  It follows that for sufficiently large $N$, the expected loss per capita is $\Theta(1)$-larger  under the fully symmetric efficient allocation than under the benchmark. 
\end{proof}

\section{Proofs of Propositions \ref{prop_ece} and \ref{prop_cp}} 
Proposition \ref{prop_cp} is immediate from the complete version of Lemma \ref{lem_general}.

For Proposition \ref{prop_ece}, we prove only the large-$N$ limit. Proofs of other two limits: $\max \beta_g \downarrow 0$ and $\max \tau_g-\phi(\lambda_g)\downarrow 0$, follow analogous arguments to those in Theorem \ref{thm_ece}. We will use the notation
\[\mathrm{COR}_+=\{i \in I_+: x_i>\phi(\lambda_i)\}, \quad \mathrm{PER}_+=I_+\setminus \{i\},\]
read as ``core-plus'' and ``periphery-plus.''  $\mathrm{COR}_-$ and $\mathrm{PER}_-$ can be defined analogously for $I_-$. 

We first establish an analog of Lemma \ref{lem_typesymmetry} for asymmetric societies. 

\begin{lem}\label{lem_typesymmetry_asymmetric}
    In an asymmetric society described in Section \ref{sec_asymmetric}, the following holds in any PSPBE: if $i \in g \cap S_+$ and $j \in g \cap S_-$, then $i \in \mathrm{PER}_+$ and $j \in \mathrm{PER}_-$. 
\end{lem}

\begin{proof}
Fix $i, j, g$ as above. Since $i,j$ belong to the same group, they share identical  $(\beta,\tau,\lambda)$. By Lemma \ref{lem_general} and Proposition \ref{prop_cp}, the log-disruption probabilities faced by $i$ and $j$ equal 
\[-\tau-\sum_{k \in \mathrm{COR}_+\setminus \{i\}}\left(\left[x_k^{+}-\phi\left(\frac{\lambda_k}{\gamma_i}\right)\right]^+- \frac{1}{\gamma_i}h^+\left(x_k^{+},\frac{\lambda_k}{\gamma_i}\right)\right)\]
and 
\[-\tau-\sum_{k \in \mathrm{COR}_-\setminus \{j\}}\left(\left[x_k^{-}-\phi\left(\frac{\lambda_k}{\gamma_j}\right)\right]^+ - \frac{1}{\gamma_j}h^+\left(x_k^{-},\frac{\lambda_k}{\gamma_j}\right)\right),\]
respectively.  Supposing that $g \subset I_+$ (the case $g \subset I_-$ is analogous), the following hold for $i$: 
\begin{align*}
    \log\beta &-\tau-\sum_{k \in \mathrm{COR}_+ \setminus \{i\}} \left(\left[x_k^{+} - \phi\left(\frac{\lambda_k}{\gamma_i}\right)\right]^+-\frac{1}{\gamma_i}h^+\left(x_k^{+}, \frac{\lambda_k}{\gamma_i}\right)\right)\\
    \leq &-\tau-\sum_{k \in \mathrm{COR}_-} \left(\left[x_k^{-} - \phi\left(\frac{\lambda_k}{\gamma_i'}\right)\right]^+-\frac{1}{\gamma_i'}h^+\left(x_k^{-}, \frac{\lambda_k}{\gamma_i'}\right)\right)\\
     \leq & -\tau-\sum_{k \in \mathrm{COR}_-\setminus \{j\}} \left(\left[x_k^{-} - \phi\left(\frac{\lambda_k}{\gamma_j}\right)\right]^+-\frac{1}{\gamma_j}h^+\left(x_k^{-}, \frac{\lambda_k}{\gamma_j}\right)\right).
\end{align*}
The first inequality says that $i$ weakly prefer to stay in $S_+$ rather than joining $S_-$. If $i$ were to join $S_-$, he can ignore $j$ and replicate $j$'s current position, or even do better. Denoting the Lagrange multiplier under the optimal allocation by $\gamma_i'$, the second inequality holds, with equality if and only if ignoring $j$ is optimal, i.e., $x_j^{-}\leq \phi(\lambda/\gamma_i')=\phi(\lambda/\gamma_j)$. 

Analogous argument for $j$ shows that 
\begin{align*}
    &-\tau-\sum_{k \in \mathrm{COR}_- \setminus \{j\}} \left(\left[x_k^{-} - \phi\left(\frac{\lambda_k}{\gamma_j}\right)\right]^+-\frac{1}{\gamma_j}h^+\left(x_k^{-}, \frac{\lambda_k}{\gamma_j}\right)\right)\\
    \leq \log\beta & -\tau-\sum_{k \in \mathrm{COR}_+\setminus \{i\}} \left(\left[x_k^{+} - \phi\left(\frac{\lambda_k}{\gamma_i}\right)\right]^+-\frac{1}{\gamma_i}h^+\left(x_k^{+}, \frac{\lambda_k}{\gamma_i}\right)\right).
\end{align*}
Equality requires that ignoring $i$ be optimal for $j$ after joining
$S_+$, i.e., $x_i^{+}\leq \phi(\lambda/\gamma_j')=\phi(\lambda/\gamma_i)$. 

Combining the inequalities for $i$ and $j$ yields 
\begin{align*}
    \log\beta &-\tau-\sum_{k \in \mathrm{COR}_+ \setminus \{i\}} \left(\left[x_k^{+} - \phi\left(\frac{\lambda_k}{\gamma_i}\right)\right]^+-\frac{1}{\gamma_i}h^+\left(x_k^{+}, \frac{\lambda_k}{\gamma_i}\right)\right) \\
    \leq & -\tau-\sum_{k \in \mathrm{COR}_-\setminus \{j\}} \left(\left[x_k^{-} - \phi\left(\frac{\lambda_k}{\gamma_j}\right)\right]^+-\frac{1}{\gamma_j}h^+\left(x_k^{-}, \frac{\lambda_k}{\gamma_j}\right)\right) \\
    \leq \log\beta & -\tau-\sum_{k \in \mathrm{COR}_+\setminus \{i\}} \left(\left[x_k^{+} - \phi\left(\frac{\lambda_k}{\gamma_i}\right)\right]^+-\frac{1}{\gamma_i}h^+\left(x_k^{+}, \frac{\lambda_k}{\gamma_i}\right)\right).
\end{align*}
Because the first and third terms coincide, equality holds throughout. Consequently, $x_i^{+}\leq \phi(\lambda/\gamma_i)$ and $x_j^{-}\leq \phi(\lambda/\gamma_j)$. However, if $i \in \mathrm{COR}_+$, then  $\gamma_i=1$ and $x_i^+>\phi(\lambda)= \phi(\lambda/\gamma_i)$, which leads to a contradiction. Therefore, $i \in \mathrm{PER}_+$. The proof for $j \in \mathrm{PER}_-$ is analogous. 
\end{proof}

We next establish the analog of Lemma \ref{lem_peersymmetry} for an asymmetric society. 

\begin{lem}\label{lem_peersymmetry_asymmetric}
    In any PSPBE, if $i,j \in g \cap S_+$ or if $i,j \in g \cap S_-$, then $i$ and $j$ must adopt identical attention allocation. 
\end{lem}

\begin{proof}
Suppose $i,j \in g \cap S_+$. A line-by-line replication of Lemma \ref{lem_peersymmetry} eliminates Case 1: $\gamma_i>1=\gamma_j$ or $\gamma_j>1=\gamma_i$. During the replication, we only use the fact that $i$ and $j$ belong to the same group and so share identical $(\beta,\lambda,\tau).$
The remaining cases are $\gamma_i=\gamma_j=1$ and $\gamma_i=\gamma_j>1$. 
Lemma \ref{lem_general} implies that their attention allocations to every $k \in S_+ \setminus \{i,j\}$  are identical. 
If $\gamma_i=\gamma_j>1$, then $x_i=x_j=0$. 
If $\gamma_i=\gamma_j=1$, the two-player argument in the proof of Lemma
\ref{lem_peersymmetry} gives $x_i=x_j$.
The proof for the case $i,j \in g\cap S_-$ is analogous.
\end{proof}

Combining Lemmas \ref{lem_typesymmetry_asymmetric} and \ref{lem_peersymmetry_asymmetric} yields:
\begin{lem}\label{lem_size}
In any PSPBE, either $\mathrm{COR}_+=\emptyset$ or $|\mathrm{COR}_+|=\Theta(N)$. The same is true for $\mathrm{COR}_-$. 
\end{lem}
\begin{proof}
By Lemma \ref{lem_typesymmetry_asymmetric}, 
we can partition all groups into three categories: those who are divided between $S_+$ and $S_-$, those who join $S_+$ exclusively, and those who join $S_-$ exclusively. By Lemma \ref{lem_peersymmetry_asymmetric}, if group $g$ joins $S_+$ exclusively, then either $g \subseteq \mathrm{COR}_+$ or $g \subseteq \mathrm{PER}_{+}$. Therefore, $\mathrm{COR}_+=\cup_{g \subseteq \mathrm{COR}_+}g$, with $|\mathrm{COR}_+|=\sum_{g \subseteq \mathrm{COR}_+}|g|=0$ or $\Theta(N)$. 
\end{proof}

Lemma \ref{lem_size} implies that for large $N$, the following hold for $i\in S_+$: 
\begin{align*}
&-\tau_i-\sum_{k \in \mathrm{COR}_+\setminus \{i\}}\left(\left[x_k^{+}-\phi\left(\frac{\lambda_k}{\gamma_i}\right)\right]^+- \frac{1}{\gamma_i}h^+\left(x_k^{+},\frac{\lambda_k}{\gamma_i}\right)\right)\\
 &\geq -\tau_i-\sum_{k \in \mathrm{COR}_+ \setminus \{i\}}\left(\left[x_k^{+}-\phi\left(\lambda_k\right)\right]^+- h^+\left(x_k^{+},\lambda_k\right)\right)\\
 &\geq -\tau_i-\sum_{k \in \mathrm{COR}_+ \setminus \{i\}}\left[x_k^{+}-\phi\left(\lambda_k\right)- h\left(x_k^{+},\lambda_k\right)\right]\\
&=-\tau_i-O\left(\frac{1}{N}\right).
\end{align*}
The first line is $i$'s log-disruption probability. The second line holds by the properties $h$ and $\phi$. The third line follows from the definition of $\mathrm{COR}_+$. If $\mathrm{COR}_+=\emptyset$, then the fourth line holds automatically.  If $|\mathrm{COR}_+|=\Theta(N)$, then by Lemma \ref{lem_size}, for all $g \subseteq  \mathrm{COR}_+$, $x_k^+ \equiv x_g$ across $k \in g$, with $h(x_g,\lambda_g)=O(1/N)$. A second-order Taylor expansion of $x_g$ around $\phi(\lambda_g)$ yields $x_g-\phi(\lambda_g)-h(x_g,\lambda_g)=O(1/N^2)$. Summing over $g \subseteq \mathrm{COR}_+$, the total remains  $O(1/N)$.


Inside a chamber, a PSPBE exists by Brouwer's fixed point theorem; see the proof of Theorem \ref{thm_tau}.\qed

\begingroup
\footnotesize
\setlength{\bibsep}{-.5pt}      
\setlength{\itemsep}{0pt}     
\setlength{\parsep}{0pt}      
\bibliographystyle{aer} 
\bibliography{references-echo.bib}
\endgroup

\cleardoublepage

    \vspace*{16em}
    \begin{center}
        \Huge{
        Online Appendix for \\ ``Rationally Inattentive Echo Chambers''\\ by Lin Hu, Anqi Li, and Xu Tan}
    \bigbreak
    \end{center}

\thispagestyle{empty}
\cleardoublepage

\appendix
\setcounter{section}{0}

\setcounter{footnote}{0}



\gdef\thesection{O.\arabic{section}}
\newtheorem{thmO}{Theorem}
\renewcommand{\thethmO}{O.\arabic{thmO}}
\newtheorem{defnO}{Definition}
\renewcommand{\thedefnO}{O.\arabic{defnO}}
\newtheorem{exmO}{Example}
\renewcommand{\theexmO}{O.\arabic{exmO}}
\newtheorem{propO}{Proposition}
\renewcommand{\thepropO}{O.\arabic{propO}}

\renewcommand{\theequation}{O.\arabic{equation}}
\setcounter{equation}{0}

\renewcommand{\thetable}{O.\arabic{table}}
\setcounter{table}{0}

\renewcommand{\thefigure}{O.\arabic{figure}}
\setcounter{figure}{0}

\renewcommand{\thefootnote}{O.\arabic{footnote}}
\setcounter{footnote}{0}


\setcounter{page}{1}

\section{Multiple periods with costly waiting}\label{sec_online_cm}
This appendix solves the two-player, two-period model developed in Section \ref{sec_cm}. 

\vspace{-10pt}
\paragraph{Preliminaries.} Redefine a player's default as the optimal decision under his prior. A player's own-biased source is biased toward his default, whereas the opposite-biased source is biased toward the alternative. For Alan, the default is $d=+1$, the own-biased source is $+1$-biased (or ``$-1$''-revealing), and the opposite-biased source is $-1$-biased (or ``$+1$''-revealing). The converse holds for Bob.

We list all candidate equilibrium attention strategies. 

\begin{defnO}
\begin{description}
    \item[One-period attention] The player allocates attention only in period one and decides immediately.
    
    \item[Own-biased attention] The player attends to his own-biased source in period one. If still uninformed, he continues to attend to the same source and possibly the other player in period two.
    
    \item[Opposite-biased attention] The player attends to his opposite-biased source in period one. If still uninformed, he continues to attend to the same source and possibly the other player in period two.
    
    \item[Alternating attention] The player attends to the opposite-biased source in period one. If still uninformed, he switches to the own-biased source and may also attend to the other player in period two.\footnote{The reverse form of alternating attention---first attending to the own-biased source and then switching to the opposite-biased source---is never optimal.}
\end{description}
\end{defnO}

\vspace{-10pt}

\paragraph{Lessons.} We begin by replicating the main lesson of CM19. 

\begin{lesson}
In the single-agent, two-period example studied in CM19, opposite-biased attention is optimal if 
        \begin{equation}
    \frac{\exp(-2\tau)}{1-\exp(-\tau)} \leq c \leq \frac{\exp(-\tau)\left[1-p_0\left(1+\exp(-\tau)\right)\right]}{1-p_0\left(1-\exp(-\tau)\right)}.
\label{eq-CM}
\end{equation}
\end{lesson}

\begin{proof}
With a single agent, say Alan, the expected loss equals 
\[
(1-p_0)\exp(-\tau)
\]
under one-period attention, 
\[
(1-p_0)\exp(-2\tau)+c\left[1-(1-p_0)(1-\exp(-\tau))\right]
\]
under own-biased attention, and 
\[
    p_0\exp(-2\tau)+c\left[1-p_0(1-\exp(-\tau))\right]
\]
under opposite-biased attention. For opposite-biased attention to be optimal, Alan's prior must be sufficiently neutral, so that his posterior favors state $-1$ over state $+1$  after one period without a message: 
\begin{equation}\label{eqn_posterior}
\frac{1-p_0}{1-p_0+p_0\exp(-\tau)}>\frac{1}{2} \Longleftrightarrow p_0<\frac{1}{1+\exp(-\tau)}.
\end{equation}
Note that the right-hand side of Condition \eqref{eq-CM} implies Condition \eqref{eqn_posterior}. 

To derive Condition \eqref{eq-CM}, consider the expected-loss expression under opposite-biased attention. The first term stems from making the wrong decision after two periods, which occurs if the true state is $+1$ and Alan's attention channel is disrupted in both periods. The probability of this event equals  $p_0\exp(-2\tau)$. The second term captures the cost of waiting, which is avoided if the state is $+1$ and Alan becomes informed after one period. The probability of this event equals $p_0(1 - \exp(-\tau))$. Taking the sum yields the expected total loss under opposite-biased attention.

The loss under own-biased learning can be derived analogously. The loss under one-period attention is easy to derive. Comparing them yields Condition~\eqref{eq-CM}.
\end{proof}

We next demonstrate that sustaining opposite-biased attention in equilibrium is harder in our two-player model than in CM19's framework. 

\begin{lesson}
  In the two-player, two-period game considered in this appendix, opposite-biased attention is an equilibrium if Condition \eqref{eq-CM} is strengthened to 
\begin{equation}
    \frac{p_0\exp(-2\tau)-(1-p_0)\exp(-2\tau-\Delta)}{(2p_0-1)(1-\exp(-\tau))} \leq c \leq \frac{\exp(-\tau)\left[1-p_0\left(1+\exp(-\tau)\right)\right]}{1-p_0\left(1-\exp(-\tau)\right)},
    \label{eq-cm1}
\end{equation}
and moreover if 
\begin{equation}\label{eq-cm2}
    \frac{p_0}{1-p_0} \leq \exp(\tau-\Delta), 
\end{equation}
where the term 
\[
\exp(\Delta) = 
\begin{cases}
    \exp\left(2 \tau - h(2\tau,\lambda)-\phi(\lambda)\right) &\text{ if } \tau-h(2\tau,\lambda)>0 \\
    \dfrac{\exp(-\tau)}{\exp(-2\tau) + \left(1-\exp(-2\tau)\right)\exp(-\lambda\tau)} & \text{ if } \tau-h(2\tau,\lambda)\leq 0
\end{cases}\]
quantifies the network effect introduced by multiple players. 
\end{lesson}

\begin{proof}
We wish to sustain opposite-biased attention in equilibrium, as illustrated in Table \ref{table0}. 
Starting from opposite-biased attention, Alan faces two potential deviations: own-biased attention and alternating attention, as the upper bound in Condition~\eqref{eq-cm1} rules out a deviation to
one-period attention.
\begin{table}[h!]
\centering
\begin{tabular}{lcc}
\hline
 & \textbf{Alan} & \textbf{Bob} \\
\hline
\textbf{Period 1} & $-1$-biased & $+1$-biased \\
\textbf{Period 2} & $-1$-biased & $+1$-biased \\
\textbf{Period-two decision if uninformed} & $-1$ & $+1$\\
\hline
\end{tabular}
\caption{Opposite-attention equilibrium.}\label{table0}
\end{table}

We demonstrate that the deviation to own-biased attention may become profitable, even when it is unprofitable in CM19. This can be seen from Table \ref{table1}:  
\begin{table}[h!]
\centering
\begin{tabular}{lcc}
\hline
 & \textbf{Alan} & \textbf{Bob} \\
\hline
\textbf{Period 1} & $-1$-biased $\rightarrow$ {\color{red}{$+1$-biased}} & $+1$-biased \\
\textbf{Period 2} & $-1$-biased $\rightarrow$ {\color{red}{$+1$-biased \& Bob}} & $+1$-biased \\
\textbf{Period-two decision if uninformed} & $-1$ $\rightarrow$ {\color{red}{$+1$}} & $+1$\\
\hline
\end{tabular}
\caption{Alan deviates from opposite-biased attention to own-biased attention.}\label{table1}
\end{table}

\noindent By period two, Bob has allocated $2\tau$ units of attention to the $+1$-biased source. From the analysis in the main text, it follows that if $\tau>h(2\tau,\lambda)$, then in period two,  Alan pays $h(2\tau,\lambda)$ units of attention to Bob and the remaining attention to the $+1$-biased source. The resulting disruption probability equals  
\[
\exp\left[-(\underbrace{\tau-h(2\tau,\lambda)}_\text{Alan's attention to primary source}+\underbrace{2\tau}_\text{Bob's attention to primary source})+\underbrace{\phi(\lambda)}_\text{threshold}\right]
\]
If, instead, $\tau<h(2\tau,\lambda)$, then Alan attends only to Bob in period two but not to the $+1$-biased source, and the expression for $\Delta$ changes accordingly. Compared to when Alan acts alone (as in CM19), the presence of Bob lowers the period-two disruption probability  from $\exp(-\tau)$ to $\exp(-\tau - \Delta)$. The expected total loss reduces to 
\[
(1-p_0)\exp(-2\tau-\Delta)+c\left[1-(1-p_0)(1-\exp(-\tau))\right]. 
\]
To continue deterring Alan's deviation to own-biased attention, we raise the left-hand side of Condition~\eqref{eq-CM} to that of Condition~\eqref{eq-cm1}.

Alternatively, Alan can deviate to alternating attention, as illustrated in Table \ref{table2}. 
\begin{table}[h!]
\centering
\begin{tabular}{lcc}
\hline
 & \textbf{Alan} & \textbf{Bob} \\
\hline
\textbf{Period 1} & $-1$-biased $\rightarrow$ {\color{red}{$-1$-biased}} & $+1$-biased \\
\textbf{Period 2} & $-1$-biased $\rightarrow$ {\color{red}{Bob \& $+1$-biased}} & $+1$-biased \\
\textbf{Period-two decision if uninformed} & $-1$ $\rightarrow$ {\color{red}{$+1$}} & $+1$\\
\hline
\end{tabular}
\caption{Alan deviates from opposite-biased attention to alternating attention.}\label{table2}
\end{table}
\noindent After the deviation, the period-two attention allocation is the same as under own-biased attention, whereas the period-one attention allocation (and hence the waiting cost) is the same as under opposite-biased attention. The period-two decision: $+1$, is wrong in state $-1$; the latter is never revealed by the period-one source.
The expected total loss equals 
\[(1-p_0) \cdot 1 \cdot \exp(-\tau-\Delta)+c\left[1-p_0\left(1-\exp(-\tau)\right)\right].\]
Compared to opposite-biased attention, this deviation is unprofitable if Condition \eqref{eq-cm2} holds. 
\end{proof}

We next give conditions for sustaining alternating attention as an equilibrium of our game.

\begin{lesson}
 In the game considered in this appendix, alternating attention is an equilibrium if 

\begin{equation}
\frac{1-p_0}{2p_0-1}\exp\left(-\tau-\Delta'\right) \leq c \leq \frac{\left(1-p_0\right)\exp\left(-\tau\right)\left(1-\exp(-\Delta')\right)}{1-p_0\left(1-\exp\left(-\tau\right)\right)},\label{eq-abba1}
\end{equation}
and if 
\begin{equation}
    \frac{p_0}{1-p_0} \geq \exp(\tau+\Delta''-\Delta')
\end{equation}
where the terms
\begin{align*}
\exp\left(\Delta'\right) 
&= \exp\left(\tau - h\left(\tau, \lambda\right) - \phi\left(\lambda\right)\right)  \text{ and } \\
\exp\left(\Delta''\right) 
&= \exp\left(\tau - h\left(\tau, \lambda\right) - h\left(\tau - h(\tau, \lambda), \lambda\right)-\phi(\lambda)\right) 
\end{align*}
quantify the network effects introduced by multiple players. 
\end{lesson}

\begin{proof}
Table \ref{table3} illustrates the alternating-attention equilibrium:
\begin{table}[h!]
\centering
\begin{tabular}{lcc}
\hline
 & \textbf{Alan} & \textbf{Bob} \\
\hline
\textbf{Period 1} & $-1$-biased  & $+1$-biased \\
\textbf{Period 2} & Bob \& $+1$-biased & Alan \& $-1$-biased \\
\textbf{Period two decision if uninformed} & $+1$ & $-1$\\
\hline
\end{tabular}
\caption{Alternating-attention equilibrium.}\label{table3}
\end{table}

\noindent Consider Alan's equilibrium loss. In period two, Alan views Bob as a secondary source with effective bandwidth $\tau$. Thus, he allocates $h(\tau,\lambda)$ units of attention to Bob and $\tau-h(\tau,\lambda)$ units of attention to $+1$-biased source. The resulting disruption probability equals 
\begin{align*}
&\exp\left[-(\underbrace{\tau-h(\tau,\lambda)}_\text{Alan's attention to primary source}+\underbrace{\tau}_\text{Bob's attention to primary source})+\underbrace{\phi(\lambda)}_\text{threshold}\right]\\ 
&=\exp(-\tau-\Delta').
\end{align*}
Alan's period-two decision: $+1$, is wrong in state $-1$; the latter is never revealed by his first-period source. The expected waiting cost can be derived as previously. Taken together, Alan's expected total loss in equilibrium equals 
\[(1-p_0)\cdot 1 \cdot \exp\left(-\tau-\Delta'\right)+c\left[1-p_0(1-\exp(-\tau))\right].\]

There are three potential deviations: one-period attention, own-biased learning, and opposite-biased learning. 
The loss under one-period attention is easy to derive. The loss under own-biased attention now becomes
\[
(1 - p_0) \cdot \exp(-\tau) \cdot \exp\left(-\tau-\Delta'\right) + c \left[1 - \left(1 - p_0\right)\left(1-\exp(-\tau)\right)\right].
\]
\begin{table}[h!]
\centering
\begin{tabular}{lcc}
\hline
 & \textbf{Alan} & \textbf{Bob} \\
\hline
\textbf{Period 1} & $-1$-biased $\rightarrow$ {\color{red} $+1$-biased } & $+1$-biased \\
\textbf{Period 2} & Bob \& $+1$-biased $\rightarrow$ {\color{red} Bob \& $+1$-biased} & Alan \& $-1$-biased \\
\textbf{Decision if uninformed} & $+1$ $\rightarrow$ {\color{red} $+1$} & $-1$\\
\hline
\end{tabular}
\caption{Alan deviates from alternating attention to own-biased attention.}\label{table4}
\end{table}
Compared to alternating attention, the difference lies in the first-period attention, which now informs the period-two decision but also raises the expected waiting cost. 

The loss under opposite-biased attention is the most delicate. 
\begin{table}[h!]
\centering
\begin{tabular}{lcc}
\hline
 & \textbf{Alan} & \textbf{Bob} \\
\hline
\textbf{Period 1} & $-1$-biased $\rightarrow$ {\color{red} $-1$-biased } & $+1$-biased \\
\textbf{Period 2} & Bob \& $+1$-biased $\rightarrow$ {\color{red} Bob \& $-1$-biased} & Alan \& $-1$-biased \\
\textbf{Decision if uninformed} & $+1$ $\rightarrow$ {\color{red} $-1$} & $-1$\\
\hline
\end{tabular}
\caption{Alan deviates from alternating attention to opposite-biased attention.}\label{table5}
\end{table}
In period two, Bob pays $h(\tau,\lambda)$ units of attention to Alan and the remaining attention $\tau-h(\tau,\lambda)$ to the $-1$-biased source. Upon deviating to opposite attention, Alan finds Bob's period-two attention to the $-1$-biased source useful. From the analysis in the main text, it follows that Alan pays $h\left(\tau-h \left(\tau,\lambda\right),\lambda\right)$ units of attention to Bob and the remaining attention $\tau-h\left(\tau-h \left(\tau,\lambda\right),\lambda\right)$ to the $-1$-biased. The resulting disruption probability equals 
\begin{align*}
&\exp\left[-(\underbrace{\tau-h(\tau,\lambda)}_\text{Bob's attention to primary source}+\underbrace{\tau-h\left(\tau-h \left(\tau,\lambda\right),\lambda\right)}_\text{Alan's attention to primary source})+\underbrace{\phi(\lambda)}_\text{threshold}\right]\\
&=\exp(-\tau-\Delta'').
\end{align*}
The remainder of the analysis follows routinely. The expected total loss equals 
\[p_0\cdot \exp(-\tau)\cdot \exp(-\tau-\Delta'')+c\left[1-p_0\left(1-\exp(-\tau)\right)\right].\]
Comparing all losses gives the desired condition. 
\end{proof}

Lastly, we note: 
\begin{lesson}
    The combination of two model elements: multiple agents and multiple periods with costly waiting, leads to a proliferation of cases, making a complete analysis challenging.
\end{lesson}

 For example, hybrid equilibria may arise in which one player uses own-biased attention and the other uses  opposite-biased attention.

\section{Richer decision and information environment}\label{sec_multi}
In this appendix, we analyze a variant of the baseline model with a richer decision and information environment.

The true state $\omega$ is uniformly distributed over $\Omega=\{1,\cdots,M\}$, where $M \geq 2$ and is finite. There are $M$ types of players, each with population $N\geq 2$. Type-$m$ players' default is $m$. Their  utility from making a decision $d \in \Omega$ is zero if $d=\omega$. If $d \neq \omega$, then the utility is $-\beta$ if $d=m$ and $-1$ if $d \neq m$. $\beta \in (0,1)$ captures the strength of the preference for default.

There are $M$ primary sources, one for each state. In state $\omega$, the $\omega$-revealing source broadcasts the message ``$\omega$,'' while all the other sources remain silent. Players attend to the primary sources and to one another as potential secondary sources as in the main text. All players share identical bandwidth $\tau>0$ and visibility $\lambda>0$.

A \emph{symmetric} PSPBE is summarized by four quantities: $\Delta^*$, $x^*$, $y^*$, and $z^*$. For  type-$m$ players, these quantities represent their attention to the $m$-revealing source, each other primary source, each other same-type player, and each different-type player, respectively. A symmetric PSPBE is a \emph{quasi-echo-chamber equilibrium} (QECE) if $\Delta^*=0$ and $y^*>z^*$. Namely, players seek information that tests rather than confirms their default and prioritize same-type players as secondary sources. If, in addition, $z^*=0$, then different-type players are ignored, and the equilibrium reduces to an ECE.

\begin{propO}\label{prop_finite}
Fix $(M, N, \tau,\lambda)$ such that $M, N \geq 2$, $\tilde{\lambda}\coloneqq (M-1)\lambda>1$, and $\tilde{\tau}\coloneqq \tau/(M-1)>\phi(\tilde{\lambda})$. There exists $\ul\beta>0$ such that for all $\beta \in (0, \ul\beta)$, the game in this appendix admits a unique symmetric $\mathrm{PSPBE}$, which is a  $\mathrm{QECE}$. Let  $N\rightarrow \infty$ or $\tilde{\tau}\downarrow \phi(\tilde{\lambda})$, holding all else constant, this equilibrium eventually reduces to an ECE. 
\end{propO}

\begin{proof}
For sufficiently small $\beta$, players choose their default given no message and ignore the primary source that confirms their default, i.e., $\Delta^*=0$. Fixing others' strategies as in equilibrium, the loss minimization problem faced by a  type-$m$ player is 
\begin{align*}
\min_{x,y,z} \quad & -x+ (N-1)\log\big(\exp(-x^*) + \left(1 - \exp(-x^*)\right)\exp(-\lambda y)\big) \\
& \quad + N(M-2) \log\big(\exp(-x^*) + \left(1 - \exp(-x^*)\right)\exp(-\lambda z)\big) \\
\text{s.t.} \quad & x, y, z \geq 0 \quad \text{and}\quad (M-1) x + (N - 1)y + (M - 1)N z \leq \tau.
\end{align*}
To understand the objective, fix state \(\omega\neq m\), in which the message “\(\omega\)” reaches the player directly from the \(\omega\)-revealing source or indirectly through secondary sources. The latter include \(N-1\) same-type players and each type-\(l\) players, \(l\neq m,\omega\). Summing the log-disruption probabilities across these channels yields the objective. Type-\(\omega\) players, by contrast, ignore the \(\omega\)-revealing source and therefore do not help transmit the message. 

The program is convex and can be solved using the Lagrangian method. Denoting the Lagrange multipliers on the constraints by $(\gamma, \eta_x, \eta_y, \eta_z)$, the first-order conditions are:
\begin{align*}
\tag{$\mathrm{FOC}_x$} 1-(M-1)\gamma+\eta_x=0\\
\tag{$\mathrm{FOC}_y$} \frac{\lambda(N-1)\left[1-\exp(-x^*)\right]\exp(-\lambda y)}{\exp(-x^*)+\left[1-\exp(-x^*)\right]\exp(-\lambda y)}-(N-1)\gamma+\eta_y=0\\
\tag{$\mathrm{FOC}_z$} \frac{\lambda N(M-2)\left[1-\exp(-x^*)\right]\exp(-\lambda z)}{\exp(-x^*)+\left[1-\exp(-x^*)\right]\exp(-\lambda z)}-N(M-1)\gamma+\eta_z=0
\end{align*}
At a symmetric equilibrium, $x^*=0$, together with $\Delta^*=0$, implies that no peer possesses any information, in which case a positive increase in $x$ would strictly reduce the loss. Hence $x^*>0$, so $\eta_x=0$. 
Solving the FOC's accordingly yields
\[
y^* = \frac{1}{\lambda} \log^+\left(\left(\frac{\lambda}{\gamma} - 1\right)\left(\exp(x^*) - 1\right)\right) = (M - 1) h^+\left(x^*, \tilde{\lambda}\right),
\]
and
\[
z^* = \frac{1}{\lambda} \log^+\left(\left(\frac{\lambda (M - 2)}{\gamma (M - 1)} - 1\right)\left(\exp(x^*) - 1\right)\right) = (M - 2) h^+(x^*, \breve{\lambda}).
\]
The terms $\tilde{\lambda}=(M-1)\lambda$ and $\breve{\lambda} = (M - 2)\lambda$ represent the effective visibility of a same-type player and a different-type player, respectively, as secondary sources. The former transmits $M-1$ types of messages, whereas the latter transmits $M-2$ types of messages. 

Clearly, $y^* > 0$ if and only if $x^* > \phi(\tilde{\lambda})$, $z^* >0$ if and only if $x^* > \phi(\breve{\lambda})$, and 
$\phi(\tilde{\lambda})<\phi(\breve{\lambda})$ because $\tilde{\lambda}>\breve{\lambda}$. Depending on whether 
\(x^*\leq\phi(\breve\lambda)\) or \(x^*>\phi(\breve\lambda)\), there are two possibilities. (1) If \(x^*\leq\phi(\breve\lambda)\), then \(z^*=0\), and \(x^*\) uniquely solves 
\[x^*+(N - 1)h^+(x^*, \tilde{\lambda})=\tilde{\tau}.\]
This is precisely the ECE-defining equation. Since $\tilde{\tau}>\phi(\tilde{\lambda})$ by assumption, we have  $x^*>\phi(\tilde{\lambda})$ and $y^*>0$. (2) If \(x^*>\phi(\breve\lambda)\), then both $y^*$ and $z^*$ are strictly positive, and $x^*$ uniquely solves 
\[x^* + (N - 1)h^+(x^*, \tilde{\lambda}) + (M - 2)N h^+(x^*,\breve{\lambda})= \tilde{\tau}.\]
The two left-hand sides join continuously at
$x^*=\phi(\breve{\lambda})$ and together are strictly increasing.
Hence exactly one of the two cases yields a valid solution, satisfying 
$x^*>0$,  $y^*>z^*$, and $\Delta^*=0$, and thus the equilibrium is a QECE.

To complete the proof, notice that $\tilde{\lambda}$ and $\breve{\lambda}$ depend only on $\lambda$ and $M$. Holding $\lambda$ and $M$ constant, taking $\tilde{\tau}\downarrow \phi(\tilde{\lambda})$ or $N \rightarrow \infty$ sends $x^*\rightarrow \phi(\tilde{\lambda})<\phi(\breve{\lambda})$. So eventually, $z^*=0$, and the QECE reduces to an ECE. 
\end{proof} 

\section{Noisy primary sources}\label{sec_noisy}
This appendix examines a variant of the baseline model featuring noisy primary sources. 

We model noise as erroneous, probabilistic activation of primary sources. Suppose each $\omega$-revealing source is activated with probability $\epsilon>0$ in state $-\omega$. In state $\omega$, the source is activated with probability one. When activated, it broadcasts the message ``$\omega$.'' 

Simple at first glance, such activation errors quickly turn the analysis intractable, because the same peer attention channel is now used to transmit both messages ``$\omega=+1$'' and ``$\omega=-1$'' in both states. Suppose player~$i$ receives ``$\omega=+1$'' but not ``$\omega=-1$'' from player~$j$. Then  $i$ infers that the channel between him and $j$ is connected, and that $j$ receives ``$\omega=+1$'' but not ``$\omega=-1$'' from the primary sources. Based on this inference, $i$ updates about the true state, which---without the product structure of the noiseless case---quickly turns into a tedious combinatorial exercise.  

Despite this challenge, we show that the ECE remains the unique PSPBE given strong preferences for default and small activation errors. 

\begin{propO}\label{prop_noisy}
Fix a sufficiently small $\beta>0$. There exists
$\bar\epsilon(\beta)>0$ such that the ECE is the unique PSPBE whenever
$\epsilon<\bar\epsilon(\beta)$.
\end{propO}

\begin{proof}
    We denote the messages received by a player by $(\cdot, \cdot)$, where the first coordinate equals $+1$ if at least one message ``$\omega=+1$'' is received, and $\emptyset$ otherwise. The second coordinate equals $-1$ if at least one message ``$\omega=-1$'' is received, and $\emptyset$ otherwise. There are four message types: $(+1, \emptyset)$, $(\emptyset, -1)$, $(\emptyset, \emptyset)$, and $(+1,-1)$. As in the main text, $D_{i}^{c}$ denotes the event in which all attention channels from primary source $c$ to player $i$, direct and indirect, are disrupted. $D_i^{c, \neg}$ denotes the complementary event in which at least one of these channels is connected.

Fix a joint attention allocation $\boldsymbol{a}$ and a small $\epsilon>0$. We derive player $i$'s decision loss as a function of his message type.

Given messages of type $(+1, \emptyset)$, the optimal decision is $d_i=+1$. This decision is inaccurate in state $-1$, in which the probability of receiving a message of type  $(+1,\emptyset)$ equals 
\[
\epsilon \cdot \mathbb{P}_{\boldsymbol{a}}\left(D_i^{+1\mathrm{-rev}, \neg}, \, D_i^{-1\mathrm{-rev}}\right).
\]
In the above expression, $\epsilon$ denotes the probability that the $+1$-revealing source is erroneously activated in state $-1$, $D_i^{+1-\mathrm{rev}, \neg}$ indicates that at least one channel from this source to $i$ is connected, and $D_i^{-1-\mathrm{rev}}$ indicates that all channels from the $-1$-revealing source to $i$ are disrupted. Importantly, the term $\mathbb{P}_{\boldsymbol{a}}(\cdot)$ depends only on $\boldsymbol{a}$ and not on $\epsilon$. Holding $\boldsymbol{a}$ fixed, $\epsilon \cdot \mathbb{P}_{\boldsymbol{a}}(\cdot)=O(\epsilon)$.

Likewise, the optimal decision is $d_i=-1$ given messages of type $(\emptyset, -1)$. This decision is wrong in state $+1$, in which the probability of receiving such messages equals 
\[\epsilon \cdot \mathbb{P}_{\boldsymbol{a}}\left(D_i^{+1-\mathrm{rev}}, D_i^{-1-\mathrm{rev}, \neg}\right).\]

The unconditional contribution of messages of type $(+1,-1)$ to expected decision loss is $O(\epsilon)$, because receiving both messages requires an erroneous activation.

For the null message $(\emptyset, \emptyset)$, suppose $d_i=+1$ is the optimal decision. This decision is wrong in state $-1$, in which $i$ receives $(\emptyset, \emptyset)$ with probability
\[
(1-\epsilon)\cdot \mathbb{P}_{\boldsymbol{a}}\left(D_i^{-1-\text{rev}}\right)
+\epsilon \cdot \mathbb{P}_{\boldsymbol{a}}\left(D_i^{+1-\text{rev}}, D_i^{-1-\text{rev}}\right).
\]
In the above expression, the first term corresponds to the event in which the $+1$-revealing source is inactivated in state $-1$, and so player $i$ receives no message from it; the $-1$-revealing source is activated, but all incoming channels to $i$ are disrupted. As in the main text, 
\[\mathbb{P}_{\boldsymbol{a}}\left(D_i^{-1-\text{rev}}\right)=\delta_i^{-1-\mathrm{rev}}\prod_{j \in I \setminus \{i\}} \left[\delta_j^{-1-\mathrm{rev}}+(1-\delta_j^{-1-\mathrm{rev}})\delta_i^j\right].\]
where $\delta_i^c=\exp(-\lambda_c a_i^c)$. The second term corresponds to the event in which both primary sources are activated, but all incoming channels to $i$ are disrupted. 

Multiplying the above probabilities by their respective decision losses  and taking the sum, we obtain 
\[
(1-\epsilon)\cdot \mathbb{P}_{\boldsymbol{a}}\left(D_i^{-1-\mathrm{rev}}\right) + \epsilon \cdot \text{a term that depends only on } \boldsymbol{a}
\]
as the expected decision loss. The ``term that depends only on $\boldsymbol{a}$'' can be derived as follows: 
\begin{enumerate}
    \item   $\boldsymbol{a}$ generates a distribution over attention networks. For each realized network, $a_i^c$ enters via $\delta_i^c$ if the channel from $c$ to $i$ is disrupted, and via $1-\delta_i^c$ if the channel is connected. Apply this rule to all $c \in C_i$.
    \item Sum over the realized networks. 
\end{enumerate}  
Although the term lacks a closed form, it suffices to note that both its value and its gradient w.r.t. $\boldsymbol{a}_i$ are bounded independently of $\epsilon$. 

For small $\epsilon$, taking logarithm of the expected loss and applying a Taylor expansion yields 
\begin{align*}
&-a_i^{-1-\mathrm{rev}} + \sum_{j \in I \setminus \{i\}} \log\!\left(\delta_j^{-1-\mathrm{rev}} + \left(1-\delta_j^{-1-\mathrm{rev}}\right)\delta_i^j\right)\\
&\quad + \epsilon \cdot \text{a term that depends only on } \boldsymbol{a}+O(\epsilon^2).
\end{align*}
Player $i$ chooses $\boldsymbol{a}_i$ to minimize this objective, taking opponents' $\boldsymbol{a}_{-i}$ as given. Denoting the Lagrange multipliers by $(\gamma, (\eta_i^c)_{c \in C_i})$, the first-order conditions are  
\begin{align*}
\tag{FOC$_{-1\text{-rev}}$} 1 + O(\epsilon) &= \gamma - \eta_{a_i^{-1-\mathrm{rev}}},\\
\tag{FOC$_{+1\text{-rev}}$} O(\epsilon) &= \gamma - \eta_{a_i^{+1-\mathrm{rev}}},\\
\tag{FOC$_{a_i^j}$} \frac{\lambda_j\left(1-\delta_j^{-1-\mathrm{rev}}\right)\delta_i^j}{\delta_j^{-1-\mathrm{rev}} + \left(1-\delta_j^{-1-\mathrm{rev}}\right)\delta_i^j} + O(\epsilon) &= \gamma - \eta_{a_i^j}.
\end{align*}
Compared to the FOCs derived in the proof of Lemma \ref{lem_general} Part (i), the above differ only $O(\epsilon)$. A line-by-line replication shows that player $i$---who chooses $d_i=+1$ given no message---ignores the $+1$-revealing source and those who choose $d=-1$ given no message. 

For sufficiently small $\beta$, choosing the default is a dominant strategy. Thus as $\beta,\epsilon\downarrow 0$, eventually the ECE emerges as the unique PSPBE of the game. 
\end{proof}

\end{document}

%% file: macros.tex
\def\supp{\mbox{supp}}
\DeclareMathOperator{\sgn}{sgn}
\DeclareMathOperator{\diag}{diag}

\def\ul{\underline}
\def\bar{\overline}

